\documentclass[12pt,letterpaper]{JHEP3}
\pdfoutput=1

\usepackage{amsmath,array,amssymb,amsfonts,amsxtra,makeidx,graphics,graphicx,amsthm,epsfig,epstopdf,multirow}

\newcommand{\be}{\begin{equation}}
\newcommand{\ee}{\end{equation}}
\newcommand{\beq}{\begin{equation}}
\newcommand{\beql}[1]{\begin{equation}\label{#1}}
\newcommand{\eeq}{\end{equation}}
\newcommand{\ba}{\begin{array}}
\newcommand{\ea}{\end{array}}
\newcommand{\bea}{\begin{eqnarray}}
\newcommand{\beal}[1]{\begin{eqnarray}\label{#1}}
\newcommand{\eea}{\end{eqnarray}}
\newcommand{\ben}{\begin{enumerate}}
\newcommand{\een}{\end{enumerate}}
\newcommand{\bean}{\begin{eqnarray*}}
\newcommand{\eean}{\end{eqnarray*}}
\newcommand{\eref}[1]{(\ref{#1})}
\newcommand{\sref}[1]{\S\ref{#1}}
\newcommand{\tref}[1]{Table~\ref{#1}}
\newcommand{\nn}{\nonumber}

\newcommand{\bvec}{\left(\ba{c}}
\newcommand{\evec}{\ea\right)}
\newcommand{\fref}[1]{Figure \ref{#1}}
\newcommand{\btab}[1]{\begin{tabular}{#1}}
\newcommand{\etab}{\end{tabular}}
\newcommand{\diag}{\mbox{diag}}

\def \Black {\pdfsetcolor{0 0 0 1}}

\newcommand{\BC}{\mathbb{C}}

\newcommand{\comment}[1]{}

 \title{Counting Orbifolds}
\author{John Davey, Amihay Hanany, Rak-Kyeong Seong\\
The Blackett Laboratory \\
Imperial College London, Prince Consort Road\\
London,  SW7 2AZ,  UK \\
{\tt jpdavey, a.hanany, rak-kyeong.seong@imperial.ac.uk}}

\preprint{Imperial/TP/10/AH/02}

\abstract{We present several methods of counting the orbifolds $\mathbb{C}^{D}/\Gamma$. A correspondence between counting orbifold actions on $\mathbb{C}^{D}$, brane tilings, and toric diagrams in $D-1$ dimensions is drawn. Barycentric coordinates and scaling mechanisms are introduced to characterize lattice simplices as toric diagrams. We count orbifolds of $\mathbb{C}^{3}$, $\mathbb{C}^{4}$, $\mathbb{C}^{5}$, $\mathbb{C}^{6}$ and $\mathbb{C}^{7}$. Some remarks are made on closed form formulas for the partition function that counts distinct orbifold actions.}


\begin{document}

%\pagestyle{plain}
%\setcounter{page}{1}
%\newcounter{bean}
%\baselineskip16pt

%\clearpage

%%%%%%%%%%%%%%%%%%%%%%%%%%%%%%%%%%%%%%%%%%%%%%%%%%%%%%%%%%%%%%%%%%%%%%%%%%%%%%%%%%%%%%%%%%%%%%%%%%%%%%%%%%%%%%%%%%%%%%%%%%%%

\section{Introduction}

D3-branes probing abelian orbifolds of $\mathbb{C}^{3}$ have been studied intensively in recent years \cite{DouglasMoore96,DouglasMoore97,DouglasGreeneMorrison97}. The world volume theory of these D3-branes is a $(3+1)$-dimensional quiver gauge theory \cite{Klebanov:1998hh,Acharya:1998db}. More recently, through the works by Bagger-Lambert \cite{BaggerLambert07,BaggerLambert08a,BaggerLambert08b}, Gustavsson \cite{Gustavsson07,Gustavsson08} and Aharony-Bergman-Jafferis-Maldacena (ABJM) \cite{ABJM08}, this idea has been extended to cover the description of M$2$-branes probing certain orbifolds of $\mathbb{C}^{4}$. It has been found that the world volume theory corresponding to a large class of these geometries is a $\mathcal{N}=2$ $(2+1)$-dimensional quiver Chern-Simons theory \cite{Martelli:2008si,Hanany:2008cd,Hanany:2008gx}.
\\

An orbifold is a quotient of a torus $M$ by a finite group $\Gamma$. In this work, we consider complex spaces $M=\mathbb{C}^{D}$ and finite abelian groups $\Gamma_{N}\subset SU(D)$ of order $N$ such that $\Gamma_{N}=\otimes_{i}{\mathbb{Z}_{n_i}}$ where $\prod_{i}{n_i}=N$. For $D \geq 2$, there is always at least one action of $\Gamma_N$ on the coordinates of the torus and for $D=2$ there is a \emph{unique} way in which $\Gamma_N$ can act on the coordinates of $\mathbb{C}^{2}$. For $D \geq 3$ this fact is no longer true and so it is interesting to count these orbifolds.\\

To illustrate the lack of uniqueness of orbifold actions for $D>2$ let us consider the orbifold $\mathbb{C}^{3}/\mathbb{Z}_{3}$ which is commonly referred to as the complex cone over the del Pezzo 0 ($dP_0$) surface. Here $\mathbb{Z}_{3}$ acts on all three coordinates of $\mathbb{C}^{3}$ non-trivially. However it should be noted that there is another action of $\mathbb{Z}_{3}$ on $\BC^3$ which acts trivially on one coordinate of $\mathbb{C}^{3}$. This orbifold singularity does not correspond to the complex cone over the $dP_0$ surface, and is often denoted as the $\mathbb{C}^{2}/\mathbb{Z}_{3}\times\mathbb{C}$ orbifold singularity.\\
 
As a second example, let us consider all orbifolds formed by acting $\mathbb{Z}_{7}$ on $\BC^3$. There are exactly three distinct orbifold actions: one that acts trivially on one of the complex coordinates ($\mathbb{C}^{2}/\mathbb{Z}_{7} \times \mathbb{C}$) and two others that act non-trivially on all three coordinates of $\mathbb{C}^{3}$. Therefore there is a naming ambiguity when we discuss $\mathbb{C}^{3}/\mathbb{Z}_{7}$ orbifolds with actions that act non-trivially on all of the $\mathbb{C}^{3}$ coordinates.\\

The three methods of counting orbifolds discussed here are:
\begin{itemize}
\item
We count orbifold actions as generators of irreducible representations of $\Gamma_N$. This method is described in great detail and readers may wish to skip the relevant sections (Sections \sref{sc2},  \sref{sc3p1}, \sref{sc3p2}, \sref{sc4p1}, \sref{sc4p2}, \sref{sc5p1} and \sref{sc5p2}).
\item
We make use of the toric description of orbifold singularities by counting $(D-1)$-simplices with hypervolume $N$. Orbifold actions of $\mathbb{C}^{2}/\Gamma_{N}$ correspond to lattice lines ($1$-simplices) of length $N$. Orbifold actions of $\mathbb{C}^{3}/\Gamma_N$ correspond to lattice triangles ($2$-simplices) of area $N$, while orbifold actions of $\mathbb{C}^{4}/\Gamma_N$ correspond to lattice tetrahedra ($3$-simplices) of volume $N$. By continuation, we expect that orbifold actions of $\mathbb{C}^{D}/\Gamma_{N}$ correspond to $(D-1)$-simplices of hypervolume $N$. Barycentric coordinates of relevant lattice points in toric diagrams are used to characterize lattice simplices (Sections \sref{sc3p4}, \sref{sc4p3} and \sref{sc5p3}).
\item
We use Brane Tilings of $\mathbb{C}^{3}$ orbifolds to illustrate counting of obifold actions (Section \sref{sc3p3}). Brane Tilings (or Dimers) are a graphical representation of the world volume theory of D$3$-brane probes \cite{HananyKen05,Hanany05,Yamazaki:2008bt,Davey:2009bp,Hanany:2005ss,Franco:2005sm,Kennaway:2007tq}.
\end{itemize}

Overall, the paper is set out as follows: Sections \sref{sc2}, \sref{sc3} and \sref{sc4} discuss orbifolds $\mathbb{C}^{2}/\Gamma_{N}$, $\mathbb{C}^{3}/\Gamma_{N}$ and $\mathbb{C}^{4}/\Gamma_{N}$ respectively. In Section \sref{scd}, we generalize the discussion to orbifolds $\mathbb{C}^{D}/\Gamma_{N}$ of any higher dimension $D$ and move on to Section \sref{scount} where we present an explicit closed form formula for the partition function of orbifold actions of the $\mathbb{C}^{3}$ orbifold \cite{HananyOrlando10}. Counting results of $\mathbb{C}^{3}$, $\mathbb{C}^{4}$, $\mathbb{C}^{5}$, $\mathbb{C}^{6}$ and $\mathbb{C}^{7}$ orbifolds are presented in Section \sref{scount} and in the Appendix. The main counting result is summarized in \tref{tmain}.\\

%%%%%%%%%%%%%%%%%%%%%%%%%%%%%%%%%%%%%%%%%%%%%%%%%%%%%%%%%%%%%%%%%%%%%%%%%%%%%%%%%%%%%%%%%%%%%%%%%%%%%%%%%%%%%%%%%%%%%%%%%%%%

\section{Orbifolds of $\mathbb{C}^{2}$ \label{sc2}}

In this section we introduce some mathematical language required to discuss orbifolds. In particular we consider in detail the orbifolds of $\mathbb{C}^{2}$ and discuss the concept of equivalence between two orbifold actions. We find that for each $N$ there is a unique action of $\Gamma_{N}$ on $\mathbb{C}^{2}$. This property is further illustrated in \sref{sc3p3}. The notation introduced in this section is used throughout the work.\\

%--------------------------------------------
\subsection{Orbifolds and Orbifold Actions \label{s11}}

Let us parameterize $\mathbb{C}^{2}$ by the complex coordinates $\{z_1,z_2\}$ and consider quotients $\mathbb{C}^{2}/\Gamma_{N}$ with $\Gamma_{N}$ being a discrete subgroup of $SU(2)$ and of order $N\in\mathbb{Z}^{+}$. We take $\Gamma_{N}=\mathbb{Z}_{N}$ where the cyclic group $\mathbb{Z}_{N}=\{e,\omega,\omega^2,\dots,\omega^{N-1}\}$ with $\omega^N=e$ being the identity element. \\

Let an irreducible representation of $\Gamma_{N}=\mathbb{Z}_{N}$ be called $R_{N}$ with elements $\omega^{(a_1,a_2)}\in R_{N}$ and $|R_{N}|=N$. The elements of the representation $\omega^{(a_1,a_2)}\in R_{N}$ are $2 \times 2$ matrices of the form
\beql{e1}
\omega^{(a_1,a_2)}=\diag
\bvec
e^{\frac{i2\pi a_{1}}{N}} \\
e^{\frac{i2\pi a_{2}}{N}}
\evec~~,
\eeq
with $a_1+a_2=0$ (mod $N$). The zero sum condition is a manifestation of the Calabi-Yau condition on the orbifold singularity. It can also be seen to come from the $\det=1$ property of $SU(2)$.\\

For the element $\omega^{(a_1,a_2)}$ to be a generator of the representation $R_{N}$, it has to fulfill a second condition which is that $\gcd{(N,a_1,a_2)}=1$. If $\gcd{(N,a_1,a_2)}=k > 1$ then $N=k N_0$ and so $\omega^{(a_1,a_2)}$ is a generator of a representation of $\Gamma_{N_{0}}=\mathbb{Z}_{N_0}$ and not $\Gamma_{N}=\mathbb{Z}_{N}$. For all elements $\omega^{(a_1,a_2)}\in R_{N}$ the identity element is given by $(\omega^{(a_1,a_2)})^{N}=1$, and $\mbox{det}(\omega^{(a_1,a_2)})=1$ is a result of the zero sum condition.\\

The generator $\omega^{(a_1,a_2)}$ of the representation $R_{N}$ of $\Gamma_{N}=\mathbb{Z}_{N}$ acts on the coordinates of $\mathbb{C}^{2}$ as
\beql{e2}
\omega^{(a_1,a_2)}~~:~~ \left(\ba{c} z_1 \\ z_2 \ea\right) ~\mapsto~ \omega^{(a_1,a_2)} \left(\ba{c} z_1 \\ z_2 \ea\right)
=\left(\ba{c} z_1~e^{\frac{i2\pi a_1}{N}} \\ z_2~~e^{\frac{i2\pi a_2}{N}} \ea\right)~.
\eeq
The identity action on the coordinates of $\mathbb{C}^{2}$ can be represented as $2$ consecutive mappings, $\omega^{(a_1,a_2)}\omega^{(a_2,a_1)}=1$, such that the inverse of the operator $\omega^{(a_1,a_2)}$ is identified as 
\beql{e3}
(\omega^{(a_1,a_2)})^{-1}=\omega^{(a_2,a_1)}~~.
\eeq\\

The dual of the generator $\omega^{(a_1,a_2)}$ of $R_{N}$ is the \textit{orbifold action} $(a_1,a_2)$ generating $\tilde{R}_{N}$ with $(a_1+a_2)\bmod{N}=0$ and $\gcd{(N,a_1,a_2)}=1$. %There is an isomorphism
%\beal{e4}
%\mathcal{R}~~:~~ R_N &\rightarrow& \tilde{R}_N \nn\\
%\omega^{(a_1,a_2)} &\mapsto& (a_1,a_2) ~~.
%\eea
Our convention is not to call $(a_1,a_2)$ for $\gcd{(N,a_1,a_2)}\neq 1$ an orbifold action of $\mathbb{C}^{2}/\Gamma_{N}$.\\

Let the set of all orbifold actions of $\Gamma_{N}=\mathbb{Z}_{N}$ be called $\mathcal{A}_{N}=\{A_{k}\}$ with $k=1,\dots,|\mathcal{A}_{N}|$. This set is defined as
\beql{e4b}
\mathcal{A}_{N}~~=~~\left\{
(a_1,a_2)~~|~~
(a_1+a_2)\bmod{N}=0~~,~~\gcd{(N,a_1,a_2)}=1
\right\}~~.
\eeq 
The set of orbifold actions $\mathcal{A}_{N}$ does not consist of distinct \textit{inequivalent} orbifold actions. An equivalence class of an orbifold action is defined as
\beql{e6}
[A_k]=\{A_l~~|~~A_l\sim A_k=(a_1,a_2)\}~~,
\eeq
such that two actions of the same equivalence class, $A_l\in[A_k]$ and $A_m\in[A_k]$, generate two irreducible representations of $\Gamma_{N}=\mathbb{Z}_{N}$ that are equivalent, $\tilde{R}_{N}(A_{l})\sim\tilde{R}_{N}(A_{m})$. We note that two representations are equivalent if they are the same up to a permutation of the complex coordinates of $\mathbb{C}^{2}$. Accordingly, the set of all orbifold actions can be written as
\beql{e7}
\bigcup_{i=1}^{N_E}{[A_i]}=\mathcal{A}_{N}~~,
\eeq
where $N_E$ is the number of equivalence classes in $\mathcal{A}_{N}$. Identifying and counting these equivalence classes is one of the main aims of this work.\\

\textbf{Example.} Let us consider an example of the conventions introduced in this section. Taking $\Gamma_{6}=\mathbb{Z}_{6}$, the set of all orbifold actions of $\mathbb{Z}_{6}$ is found as
\beql{e7b}
\mathcal{A}_{6}~~=~~\left\{
(1,5)~,~(5,1)
\right\}~~.
\eeq
We choose to take $(1,5)\in\mathcal{A}_{6}$ to generate the orbifold action representation of $\mathbb{Z}_{6}$ as
\beql{e7c}
\tilde{R}_{6}\left((1,5)\right)~~=~~
\left\{
(1,5)~,~(2,4)~,~(3,3)~,~(4,2)~,~(5,1)~,~(0,0)
\right\}~,
\eeq
where $(0,0)\in\tilde{R}_{6}$ is the identity element of the representation group. Considering $(a_1,a_2)=(3,3)\in\tilde{R}_{6}\left((1,5)\right)$, we note $\gcd{(N,a_1,a_2)}=\gcd{(6,3,3)}=k=3$. Accordingly, with $N=6=k N_{0}$, the representation generated by $(3,3)\in\tilde{R}_{6}\left((1,5)\right)$,
\beql{e7d}
\tilde{R}_{6}\left((3,3)\right)~~=~~
\left\{
(3,3)~,~(0,0)
\right\}
~~~\leftrightarrow~~~
\tilde{R}_{2}\left((1,1)\right)~~=~~
\left\{
(1,1)~,~(0,0)
\right\}~~,
\eeq
is a representation of $\mathbb{Z}_{N_{0}=2}$ and not a representation of $\mathbb{Z}_{N=6}$.

%---------------------------
\subsection{Equivalence of Orbifold Actions \label{s6}}

The choice of complex coordinates for $\mathbb{C}^{2}$, $\{z_1,z_2\}$, is an arbitrary choice. Any permutation of these lead to an equivalent parameterization. In $\mathbb{C}^{2}$, there are two equivalent coordinate sets $\{z_1,z_2\}$ and $\{z_2,z_1\}$. Important for the discussion is that two orbifold actions of $\mathbb{C}^{2}$ that differ only up to a permutation of complex coordinates are equivalent,
\beql{e8}
(a_1,a_2)~~\sim~~(a_2,a_1) ~~.
\eeq
From the findings on the inverse given by $(\omega^{(a_1,a_2)})^{-1}=\omega^{(a_2,a_1)}$ in \eqref{e2}, the inverse of any orbifold action in $\mathbb{C}^{2}$ is equivalent to the orbifold action itself.\\

A stronger equivalence condition can be found by considering two orbifold actions $A_{k}=(\tilde{a}_1,\tilde{a}_2)$ and $A_{l}=(a_1,a_2)$. Let $A_{l}$ be the generator of the representation $\tilde{R}_{N}(A_{l})$ of the orbifold group $\Gamma_{N}=\mathbb{Z}_{N}$, and $A_{k}$ be some generator of an unknown representation. We note that $A_{k}$ and $A_{l}$ are equivalent orbifold actions if and only if they are related by $m\in\mathbb{Z}$,
\beql{e9}
(\tilde{a}_1,\tilde{a}_2)~~=~~m(a_1,a_2)~\bmod{N}~~,
\eeq
with $\gcd{(N,m)}=1$. The implication of the condition $(\tilde{a}_1,\tilde{a}_2)=m(a_1,a_2)$ is that $A_{k}=(\tilde{a}_1,\tilde{a}_2)$ is an element of the representation $\tilde{R}_{N}(A_{l})$ generated by the action $A_{l}$. Moreover, the implication of the second condition $\gcd{(N,m)}=1$ is that $A_{k}$ is itself a generator of a representation $\tilde{R}_{N}(A_{k})$ of the same orbifold group $\Gamma_{N}=\mathbb{Z}_{N}$ of which $\tilde{R}_{N}(A_{l})$ is a representation. This means since $|\tilde{R}_{N}(A_{l})|=N$ the $\gcd{(N,m)}=1$ condition gives $|\tilde{R}_{N}(A_{k})|=N$. Because $A_{k}\in\tilde{R}_{N}(A_{l})$, the conclusion is $\tilde{R}_{N}(A_{l})\sim\tilde{R}_{N}(A_{k})$ and hence the generators of the respective representations are equivalent, $A_{l}\sim A_{k}$.\\

The importance of the coprime condition $\gcd{(m,N)}=1$ can also be highlighted in terms of orbifold operators $\omega^{(a_1,a_2)}$. Let $R_{N}\left(\omega^{(a_1,a_2)}\right)$ be the irreducible representation of $\Gamma_{N}=\mathbb{Z}_{N}$ generated by $\omega^{(a_1,a_2)}$. If $\gcd{(m,N)}=k\neq 1$ such that $N = k N_0$ with $N_0\in\mathbb{Z}$ then 
\beql{e11}
(\omega^{(a_1,a_2)})^{m}~=~\omega^{m(a_1,a_2)}=\diag\left(\ba{c} e^{\frac{i2\pi m a_1}{N}} \\ e^{\frac{i2\pi m a_2}{N}} \ea\right)
=\diag\left(\ba{c} e^{\frac{i2\pi a_1}{N_0}} \\ e^{\frac{i2\pi a_2}{N_0}} \ea\right)
\eeq
is not a generator of the representation $R_{N}\left(\omega^{(a_1,a_2)}\right)$ of $\Gamma_{N}=\mathbb{Z}_{N}$. This is a problem because two equivalent (or equal) operators $\omega^{(a_1,a_2)}$ and $\omega^{m(a_1,a_2)}$ need to be generators of two equivalent (or equal) representations of the same group $\Gamma_{N}$. For $\gcd{(m,N)}=k \neq 1$ with $N=k N_{0}$, $\omega^{m(a_1,a_2)}$ generates a representation of $\mathbb{Z}_{N_0}$ and not $\mathbb{Z}_{N}$ such that it cannot be equivalent to $\omega^{(a_1,a_2)}$.\\

As a result, all equivalence classes of orbifold actions are of the form,
\beql{e12}
[(a_1,a_2)]~=~\left\{(b_1,b_2)~~|~~
(b_1,b_2)=m(a_1,a_2),~\gcd{(m,N)}=1,~1<m\leq N 
\right\}~~.
\eeq
It is now possible to show that for all $N\in\mathbb{Z}^{+}$, there is only one equivalence class defined for $\mathbb{C}^{2}/\mathbb{Z}_{N}$ orbifold actions at a given order $N$ such that
\beql{e13}
\mathcal{A}_{N}\equiv [(1,N-1)]=\{(a_1,a_2)~~|~~(a_1,a_2) \sim (1,N-1)\}~~.
\eeq
A short demonstration of the proof is given in the discussion of orbifolds in $\mathbb{C}^{3}$ and their corresponding brane tiling representation (Section \sref{s2}).\\

\textbf{Example.} Re-using the example from Section \sref{s11} with $\Gamma_{6}=\mathbb{Z}_{6}$, we see that there are only two generators of representations of $\mathbb{Z}_{6}$, $(1,5),(5,1)\in\mathcal{A}_{6}$ which are equivalent under a permutation of the complex coordinates. Also, there is $m=5$ with $\gcd{(5,6)}=1$ such that $m(1,5)\bmod{N}=(5,1)$. Finally, $(5,1)$ generates the following representation of $\mathbb{Z}_{6}$
\beql{e13b}
\tilde{R}_{6}\left((5,1)\right)~~=~~
\left\{
(5,1)~,~(4,2)~,~(3,3)~,~(2,4)~,~(1,5)~,~(0,0)
\right\}~,
\eeq
which is equivalent to the representation $\tilde{R}_{6}\left((1,5)\right)$ in \eref{e7c} up to a permutation of the complex coordinates of $\mathbb{C}^{2}$, $\{z_1,z_2\}\rightarrow\{z_2,z_1\}$.\\

%%%%%%%%%%%%%%%%%%%%%%%%%%%%%%%%%%%%%%%%%%%%%%%%%%%%%%%%%%%%%%%%%%%%%%%%%%%%%%%%%%%%%%%%%%%%%%%%%%%%%%%%%%%%%%%%%%%%%%%%%%%%

\section{Orbifolds of $\mathbb{C}^{3}$ \label{sc3}}

In this section we draw a correspondence between orbifold actions, brane tilings and lattice triangles as toric diagrams. The concepts introduced for orbifold actions of $\mathbb{C}^{2}$ are extended to $\mathbb{C}^{3}$. A complication is introduced when proceeding from $\Gamma_{N}=\mathbb{Z}_{N}$ orbifold quotients to $\Gamma_{N}=\mathbb{Z}_{n_1}\times \mathbb{Z}_{n_2}$ quotients where $N=n_1 n_2$. In \sref{sc3p3}, the brane tiling and its correspondence to quiver gauge theories as world volume theories of D$3$-branes probing a non-compact Calabi-Yau orbifold singularity are introduced. We present the brane tilings corresponding to orbifolds of $\mathbb{C}^{3}$. Furthermore, we discuss equivalence in the context of brane tilings. In \sref{sc3p4}, the connection between $\mathbb{C}^{3}/\Gamma_{N}$ orbifolds and toric geometry is drawn, and the question of equivalent orbifold actions is re-defined in terms of lattice triangles of area $N$. We use barycentric coordinates of relevant lattice points in the toric diagram to characterize the lattice simplex. As a brief interlude, we show how brane tilings and toric diagrams in the context of $\mathbb{C}^{3}/\Gamma_{N}$ orbifolds can be used to represent orbifolds of the lower dimensional $\mathbb{C}^{2}/\Gamma_{N}$ orbifold. Using arguments based on the brane tiling representation, we illustrate how orbifolds $\mathbb{C}^{2}/\Gamma_{N}$ have always only one unique orbifold action at a given order $N$ as indicated in \eref{e13}.\\

%--------------------------------------------
\subsection{Orbifolds and Orbifold Actions \label{sc3p1}}

Let $\mathbb{C}^{3}$ be parameterized by $\{z_1,z_2,z_3\}$. We consider quotients $\mathbb{C}^{3}/\Gamma_{N}$ with $\Gamma_{N}\subset SU(3)$ and of order $N\in\mathbb{Z}^{+}$. In general, we consider orbifolds with $\Gamma_{N}=\mathbb{Z}_{n_1}\times\mathbb{Z}_{n_2}$ and order $n_1 n_2=N\in\mathbb{Z}^{+}$. Without loss of generality, it is assumed that $n_1\geq n_2$.\\

Let an irreducible representation of $\Gamma_{N}=\mathbb{Z}_{n_1}\times\mathbb{Z}_{n_2}$ be called $R_{(n_1,n_2)}$ with elements $\omega^{(\{a_i\},\{b_i\})}$, $i=1,\dots,3$ and $|R_{(n_1,n_2)}|=N$. The elements of the representation $\omega^{(\{a_i\},\{b_i\})}\in R_{(n_1,n_2)}$ are of the form
\beql{e14}
\omega^{(\{a_i\},\{b_i\})}~~=~~
\diag\bvec
e^{\frac{i2\pi a_1}{n_1}} \\ 
e^{\frac{i2\pi a_2}{n_1}} \\ 
e^{\frac{i2\pi a_3}{n_1}} 
\evec
\diag\bvec
e^{\frac{i2\pi b_1}{n_2}} \\ 
e^{\frac{i2\pi b_2}{n_2}} \\ 
e^{\frac{i2\pi b_3}{n_2}}
\evec
~=~
\diag\bvec
e^{i2\pi (\frac{a_1}{n_1}+\frac{b_1}{n_2})} \\ 
e^{i2\pi (\frac{a_2}{n_1}+\frac{b_2}{n_2})} \\ 
e^{i2\pi (\frac{a_3}{n_1}+\frac{b_3}{n_2})} 
\evec~~,
\eeq
with $(a_1+a_2+a_3)\bmod{n_1}=0$ and $(b_1+b_2+b_3)\bmod{n_2}=0$. The zero sum conditions are a manifestation of the Calabi-Yau condition on the orbifold $\mathbb{C}^{3}/\Gamma_{N}$ and the $\det=1$ property of $SU(3)$. We introduce notation such that \eref{e14} can be expressed as
\beql{e14b}
\omega^{(\{a_i\},\{b_i\})}=\omega^{(a_1,a_2,a_3)}\omega^{(b_1,b_2,b_3)}=\omega^{((a_1,a_2,a_3),(b_1,b_2,b_3))}~~. 
\eeq\\

For the element $\omega^{(\{a_i\},\{b_i\})}\in R_{(n_1,n_2)}$ to be also a generator of the representation, it has to fulfill $\gcd{(n_1,\{a_i\})}=1$ and $\gcd{(n_2,\{b_i\})}=1$. In addition, the identity element of the representation is defined as $(\omega^{(\{a_i\},\{b_i\})})^{N}=1$. The Calabi-Yau condition also results in $\mbox{det}(\omega^{(\{a_i\},\{b_i\})})=1$.\\

The generator $\omega^{(\{a_i\},\{b_i\})}$ of the representation $R_{(n_1,n_2)}$ acts on the coordinates of $\mathbb{C}^{3}$ as
\beql{e15}
\omega^{(\{a_i\},\{b_i\})}~~:~~ \left(\ba{c} z_1 \\ z_2 \\ z_3  \ea\right) ~\mapsto~ \omega^{(\{a_i\},\{b_i\})}\left(\ba{c} z_1 \\ z_2 \\ z_3  \ea\right)
~=~ \left(\ba{c} z_1~e^{i2\pi(\frac{a_1}{n_1}+\frac{b_1}{n_2})} \\ z_2~e^{i2\pi(\frac{a_2}{n_1}+\frac{b_2}{n_2})} \\ z_3~e^{i2\pi(\frac{a_3}{n_1}+\frac{b_3}{n_2})}  \ea\right) ~~.
\eeq
%There is again the possibility of writing the identity operator $1$ as a product of operators, here in $\mathbb{C}^{3}$ as a product of $3$  %elements $\omega^{(\{a_i\},\{b_i\})}$. It is important to note that in $D=3$, the choice of products is not unique with the possibilities being
%\beal{e16}
%\omega^{((a_1,a_2,a_3),(b_1,b_2,b_3))}\omega^{((a_2,a_3,a_1),(b_2,b_3,b_1))}\omega^{((a_3,a_1,a_2),(b_3,b_1,b_2))}=1 \nn\\
%\omega^{((a_1,a_2,a_3),(b_1,b_2,b_3))}\omega^{((a_2,a_1,a_3),(b_2,b_3,b_1))}\omega^{((a_3,a_2,a_1),(b_3,b_1,b_2))}=1 \nn\\
%\omega^{((a_1,a_2,a_3),(b_1,b_2,b_3))}\omega^{((a_2,a_1,a_3),(b_2,b_1,b_3))}\omega^{((a_3,a_2,a_1),(b_2,b_1,b_3))}=1
%\eea
%up to permutations of the complex coordinates of $\mathbb{C}^{3}$, and swaps of $a_i$ and $b_i$. Accordingly, the inverse is found to have the form,
%\beal{e17}
%(\omega^{((a_1,a_2,a_3),(b_1,b_2,b_3))})^{-1}~~&=&~~\omega^{((a_2,a_3,a_1),(b_2,b_3,b_1))}\omega^{((a_3,a_1,a_2),(b_3,b_1,b_2))} \nn\\
%~~&=&~~\omega^{((a_2,a_1,a_3),(b_2,b_3,b_1))}\omega^{((a_3,a_2,a_1),(b_3,b_1,b_2))} \nn\\
%~~&=&~~\omega^{((a_2,a_1,a_3),(b_2,b_1,b_3))}\omega^{((a_3,a_2,a_1),(b_2,b_1,b_3))} ~~.
%\eea\\

The dual to the generator $\omega^{(\{a_i\},\{b_i\})}$ of the representation $R_{(n_1,n_2)}$ is now the $2\times 3$ matrix orbifold action $((a_1,a_2,a_3),(b_1,b_2,b_3))$ generating the representation $\tilde{R}_{(n_1,n_2)}$ with $\gcd{(n_1,\{a_i\})}=1$ and $\gcd{(n_2,\{b_i\})}=1$. %There is an isomorphism
%\beal{e18}
%\mathcal{R}~~:~~R_{(n_1,n_2)} ~~&\rightarrow&~~\tilde{R}_{(n_1,n_2)} \nn\\
%\omega^{(\{a_i\},\{b_i\})} ~~&\mapsto&~~((a_1,a_2,a_3),(b_1,b_2,b_3)) ~~.
%\eea
For $\gcd{(n_1,\{a_i\})}\neq 1$ and $\gcd{(n_2,\{b_i\})}\neq 1$, $((a_1,a_2,a_3),(b_1,b_2,b_3))$ is not an orbifold action of $\mathbb{C}^{3}/\Gamma_{n_1 n_2}$.\\

Let the set of all generators of representations $\{\tilde{R}_{(n_1,n_2)}\}$ of $\Gamma_{N}$ orbifold groups of order $N=n_1 n_2$ be called $\mathcal{A}_{N}=\{A_{k}\}$ with $k=1,\dots,|\mathcal{A}_{N}|$. This set is defined as
\beql{e19}
\mathcal{A}_{N=n_1 n_2}~~=~~
\left\{
\left(
\ba{ccccc}
( & a_1, & a_2, & a_3 & ) \\
( & b_1, & b_2, & b_3 & ) 
\ea
\right)~~\Bigg|~~
\ba{c}
(a_1+a_2+a_3)\bmod{n_1}=0~~~,\\
(b_1+b_2+b_3)\bmod{n_2}=0~~~,\\
\gcd{(n_1,\{a_i\})}=1~~~,~~~\gcd{(n_2,\{b_i\})}=1
\ea
\right\}~~.
\eeq
As for $\mathbb{C}^{2}$ orbifolds, the set of orbifold actions $\mathcal{A}_{N}$ does not consist of distinct inequivalent orbifold actions. The set of orbifold actions $\mathcal{A}_{N}$ at a given order $N=n_1 n_2$ can be re-expressed as the union of all orbifold action equivalence classes $[A_{k}]$. If two orbifold actions $A_{l}\in[A_k]$ and $A_{m}\in[A_{k}]$ are of the same equivalence class $[A_k]$ and are both generators of representations $\tilde{R}_{(n_1,n_2)}(A_{l})$ and $\tilde{R}_{(\tilde{n}_1,\tilde{n}_2)}(A_{k})$ respectively with $N=n_1 n_2=\tilde{n}_1 \tilde{n}_2$, then the two representations of $\Gamma_{N}$ are equivalent  $\tilde{R}_{(n_1,n_2)}(A_{l})\sim\tilde{R}_{(\tilde{n}_1,\tilde{n}_2)}(A_{k})$ up to a permutation of the complex coordinates of $\mathbb{C}^{3}$.\\

It is of use to consider an orbifold action in terms of its components. An orbifold action $A_{k}$ in $\mathbb{C}^{3}$ consists of two components corresponding to the two rows in the $2\times 3$ orbifold action matrix $A_{k}=((a_1,a_2,a_3),(b_1,b_2,b_3))$. We denote the two components as $A_{k}^{(n_1)}=(a_1,a_2,a_3)$ and $A_{k}^{(n_2)}=(b_1,b_2,b_3)$ such that the action can be written as $A_{k}=(A_{k}^{(n_1)},A_{k}^{(n_2)})$. The dual operator has the corresponding notation $\omega^{(\{a_i\},\{b_i\})}=(\omega^{(a_1,a_2,a_3)},\omega^{(b_1,b_2,b_3)})$.\\

For the case when $n_2=1$ with $n_1 > n_2$, the orbifold action and its dual are of the form $A_{k}=(A_{k}^{(n_1)},(0,0,0))$ and $\omega^{(\{a_i\},\{b_i\})}=(\omega^{(a_1,a_2,a_3)},1)$ respectively. In this case, it is beneficial to talk about the \textit{effective} component $A_{k}^{(n_1)}$ of the orbifold action instead of the orbifold action $A_{k}$ itself. In the context of representations, for $\gcd{(n_1,\{a_i\})}=1$, the component $A_{k}^{(n_1)}$ is the generator of the representation $\tilde{R}_{n_1}$ of the group $\Gamma_{N=n_1}=\mathbb{Z}_{n_1}$ with $n_2=1$.\\

%---------------------------
\subsection{Equivalence of Orbifold Actions\label{s1} \label{sc3p2}}

As in $2$-dimensional space, any two orbifold actions that are related by a permutation of the complex coordinates in $\mathbb{C}^{3}$, $\{z_1,z_2,z_3\}$, are equivalent. Permutations in the coordinates $\mbox{Perm}(\{z_1,z_2,z_3\})$ correspond to permutations of columns $\{(a_i,b_i)\}$ with $i=1,2,3$ in the orbifold action matrix $A_k$.\\

For orbifolds where $n_2=1$ such that the orbifold is $\mathbb{C}^{3}/\mathbb{Z}_{n_1}$ with $N=n_1$, equivalence between orbifold actions can be determined in terms of the corresponding effective actions, $A_{k}^{(n_1)}=(a_1,a_2,a_3)$. As with $2$-dimensional orbifolds, $2$ effective actions are said to be equivalent if there is $m\in\mathbb{Z}$, $1<m<n_1=N$ such that
\beql{e23}
(a_1,a_2,a_3)=m(\tilde{a}_{1},\tilde{a}_{2},\tilde{a}_{3})\bmod{N} \eeq
with $\gcd(m,N)=1$. As for the reasoning in \eref{e11}, if $\gcd(m,N)=k\neq 1$ with $n_1=k n_{1(0)}$, then $A_{k}^{(n_1)}=(a_1,a_2,a_3)$ is a generator of a representation of $\mathbb{Z}_{n_{1(0)}}$ and not $\mathbb{Z}_{n_{1}}$. Since two orbifold actions that are generators of orbifold groups that have not the same order $N$ cannot be equivalent, $\gcd(m,N)=1$ is a requirement for two orbifold actions to be equivalent.\\

For the case when $n_1\neq 1$ and $n_2\neq 1$, there are two non-trivial representations $\tilde{R}_{n_1}(A_{k}^{(n_1)})$ and $\tilde{R}_{n_2}(A_{k}^{(n_2)})$ generated by two orbifold action components $A_{k}^{(n_1)}$ and $A_{k}^{(n_2)}$ respectively with $\gcd{(n_1,\{a_i\})}=1$ and $\gcd{(n_2,\{b_i\})}=1$. But we note here that it is not the components we want to compare, but the complete orbifold actions $A_{k}=(A_{k}^{(n_1)},A_{k}^{(n_2)})$ of $\mathbb{C}^{3}/\Gamma_{N}$. For this purpose, the representation $\tilde{R}_{(n_1,n_2)}(A_{k})$ of the whole product group $\Gamma_{N}=\mathbb{Z}_{n_1}\times\mathbb{Z}_{n_2}$ generated by $A_{k}=(A_{k}^{(n_1)},A_{k}^{(n_2)})$ is required.\\

Let $\rho_{ab}^{n_1,n_2}$ map from an element $A_{k}^{(n_1)}\in \tilde{R}_{n_1}$ of the representation $\tilde{R}_{n_1}$ of $\mathbb{Z}_{n_1}$ and an element $A_{k}^{(n_2)}\in \tilde{R}_{n_2}$ of the representation $\tilde{R}_{n_2}$ of $\mathbb{Z}_{n_2}$ to an element $A_{k}^{(n_1,n_2)}\in\bar{R}_{(n_1,n_2)}$ of some representation $\bar{R}_{(n_1,n_2)}$ of $\Gamma_{N}=\mathbb{Z}_{n_1}\times\mathbb{Z}_{n_2}$. It is defined as
\beql{e25}
\rho_{ab}^{n_1,n_2}~~:~~(A_{k}^{(n_1)},A_{k}^{(n_2)})~\mapsto~A_{k}^{(n_1,n_2)}=(a n_2 A_{k}^{(n_1)}+b n_1 A_{k}^{(n_2)})\bmod{N} ~~,
\eeq
where $1\leq a \leq n_1$ and $1\leq b \leq n_2$ and $N=n_1 n_2$. The action of the map $\rho_{ab}^{n_1,n_2}$ in terms of components of the orbifold operator $\omega^{(\{a_i\},\{b_i\})}=(\omega^{(a_1,a_2,a_3)},\omega^{(b_1,b_2,b_3)})$ is multiplication, 
\beal{e27}
\rho_{ab}^{n_1,n_2}~~:~~(\omega^{(a_1,a_2,a_3)},\omega^{(b_1,b_2,b_3)}) ~\mapsto~ 
&& (\omega^{(a_1,a_2,a_3)})^{a}(\omega^{(b_1,b_2,b_3)})^{b}  \nn\\
&=&\diag\left(\ba{c} e^{i2\pi\frac{a n_2 a_1+b n_1 b_1}{N}} \\ e^{i2\pi\frac{a n_2 a_2+b n_1 b_2}{N}} \\ e^{i2\pi\frac{a n_2 a_3+b n_1 b_3}{N}} \ea\right) ~~=~~\omega^{(a(a_1,a_2,a_3),b(b_1,b_2,b_3))} ~~. \nn\\
\eea\\

Accordingly, the map $\rho_{ab}^{n_1,n_2}$ in the parameter range $1\leq a \leq n_1$ and $1\leq b \leq n_2$ maps all elements of the representation $\tilde{R}_{n_1}(A_{k}^{(n_1)})$ of $\mathbb{Z}_{n_1}$ generated by $A_{k}^{(n_1)}$ and all elements of the representation $\tilde{R}_{n_2}(A_{k}^{(n_2)})$ of $\mathbb{Z}_{n_2}$ generated by $A_{k}^{(n_2)}$ into the representation $\bar{R}_{(n_1,n_2)}(\rho_{11}^{n_1,n_2}(A_{k}^{(n_1)},A_{k}^{(n_2)}))$ of $\Gamma_{N}=\mathbb{Z}_{n_1}\times\mathbb{Z}_{n_2}$ generated by $\rho_{11}^{n_1,n_2}(A_{k}^{(n_1)},A_{k}^{(n_2)})$. In short,
\beql{e27b}
\rho_{ab}^{n_1,n_2}~~:~~
\left(
\tilde{R}_{n_1}(A^{(n_1)}_{k}),\tilde{R}_{n_2}(A^{(n_2)}_{k})
\right) 
~~\rightarrow~~ \bar{R}_{(n_1,n_2)}\left(\rho_{11}^{n_1,n_2}(A_{k}^{(n_1)},A_{k}^{(n_2)})\right) ~~.
\eeq\\

The key observation is that for given two orbifold actions $A_k$ and $A_l$ if their corresponding representation of $\Gamma_{N}$, $\bar{R}_{(n_1,n_2)}(\rho_{11}^{n_1,n_2}(A_{k}))$ and $\bar{R}_{(\tilde{n}_1,\tilde{n}_2)}(\rho_{11}^{\tilde{n}_1,\tilde{n}_2}(A_{l}))$ with $N=n_1 n_2 =\tilde{n}_{1}\tilde{n}_{2}$, are equal up to permutations of the complex coordinates of $\mathbb{C}^{3}$, then $A_k$ and $A_l$ are equivalent. In converse, two orbifold actions $A_k$ and $A_l$ generate under the map $\rho_{ab}^{n_1,n_2}$ two equivalent representations of $\Gamma_{N}$. Accordingly, we define the equivalence class of an orbifold action as
\beql{e29}
[A_k]~~=~~
\left\{ 
A_{l} ~~\Big|~~ 
A_{l}\sim A_{k}~~\Leftrightarrow~~ 
\bar{R}_{(n_1,n_2)}(\rho_{11}^{n_1,n_2}(A_l))
~\sim~
\bar{R}_{(\tilde{n}_1,\tilde{n}_2)}(\rho_{11}^{\tilde{n}_1,\tilde{n}_2}(A_k)) 
\right\} ~~,
\eeq
with $N=n_1 n_2= \tilde{n}_{1}\tilde{n}_{2}$. The set of all orbifold actions at a given orbifold group order $N$ can be re-expressed as $\bigcup_{i=1}^{N_E}{[A_i]}=\mathcal{A}_{N}$ where $N_{E}$ is the number of equivalence classes in $\mathcal{A}_{N}$.\\

An alternative representation map to $\rho_{ab}^{n_1,n_2}$ is
\beal{e29b}
\tilde{\rho}^{n_1,n_2}_{ab}~~:~~
(\tilde{R}_{n_1}(A^{(n_1)}_{k}),\tilde{R}_{n_2}(A^{(n_2)}_{k}))
~~&\rightarrow&~~ \tilde{R}_{(n_1,n_2)}(A_k) \nn\\
(A_{k}^{(n_1)},A_{k}^{(n_2)})
~~&\mapsto&~~
\bvec
(a A_{k}^{(n_1)})\bmod{n_1}\\
(b A_{k}^{(n_2)})\bmod{n_2}
\evec ~~,
\eea
where the representation $\tilde{R}_{(n_1,n_2)}(A_{k})$ of $\Gamma_{N}=\mathbb{Z}_{n_1}\times\mathbb{Z}_{n_{2}}$ is generated by the orbifold action $A_{k}=(A_{k}^{(n_1)},A_{k}^{(n_2)})$ and not by $\rho_{11}^{n_1,n_2}(A_{k})$. The representation $\tilde{R}_{(n_1,n_2)}(A_{k})$ is isomorphic to the representation $\bar{R}_{(n_1,n_2)}(\rho_{11}^{n_1,n_2}(A_{k}))$, but is not used for testing equivalence between orbifold actions.\\

\textbf{Example.} It is instructive to consider an example with the following actions
\beql{e30}
A_{1}=\bvec (1,2,3)\\(0,0,0) \evec
~~~,~~~
A_{2}=\bvec (1,0,2)\\(0,1,1) \evec
\eeq
of the orbifolds $\mathbb{C}^{3}/\mathbb{Z}_{6}$ and $\mathbb{C}^{3}/\mathbb{Z}_{3}\times\mathbb{Z}_{2}$ respectively. It can be shown that $A_{1}$ of $\mathbb{C}^{3}/\mathbb{Z}_{6}$ is equivalent to $A_{2}$ of $\mathbb{C}^{3}/\mathbb{Z}_{3}\times\mathbb{Z}_{2}$ by first obtaining
\beal{e31}
\bar{R}_{(6,1)}(\rho_{11}^{6,1}(A_{1})) &=&\left\{
\underbrace{(1,2,3)}_{\rho_{11}^{6,1}},
\underbrace{(2,4,0)}_{\rho_{21}^{6,1}},
\underbrace{(3,0,3)}_{\rho_{31}^{6,1}},
\underbrace{(4,2,0)}_{\rho_{41}^{6,1}},
\underbrace{(5,4,3)}_{\rho_{51}^{6,1}},
\underbrace{(0,0,0)}_{\rho_{61}^{6,1}}
\right\} \nn\\
\bar{R}_{(3,2)}(\rho_{11}^{3,2}(A_{2})) &=&\left\{
\underbrace{(2,3,1)}_{\rho_{11}^{3,2}},
\underbrace{(4,3,5)}_{\rho_{21}^{3,2}},
\underbrace{(0,3,3)}_{\rho_{31}^{3,2}},
\underbrace{(2,0,4)}_{\rho_{12}^{3,2}},
\underbrace{(4,0,2)}_{\rho_{22}^{3,2}},
\underbrace{(0,0,0)}_{\rho_{32}^{3,2}}
\right\}
\eea 
where the labels $\rho_{ab}^{n_1,n_2}$ indicate the maps used to obtain each representation element. Under permutations of the complex coordinates of $\mathbb{C}^{3}$ or equivalently the columns of $A_{1}$ and $A_{2}$, $\{(a_i,b_i)\}$, it can be seen that $\bar{R}_{(6,1)}(\rho_{11}^{6,1}(A_{1}))\sim\bar{R}_{(3,2))}(\rho_{11}^{3,2}(A_{2}))$ indicating $A_{2}\in [A_{1}]$. As a convention in discussions, the action $A_{1}$ of $\mathbb{C}^{3}/\mathbb{Z}_{6}$ is favored over $A_{2}$ of $\mathbb{C}^{3}/\mathbb{Z}_{3}\times\mathbb{Z}_{2}$ keeping $(n_1-n_2)\rightarrow\mbox{max}(n_1-n_2)$. \\

Similarly, using the alternative map $\tilde{\rho}_{ab}^{n_1,n_2}$, one can find the representations
\beal{e31bb}
\tilde{R}_{(6,1)}(A_{1}) &=& \left\{
\underbrace{\bvec(1,2,3)\\(0,0,0)\evec}_{\tilde{\rho}_{11}^{6,1}},
\underbrace{\bvec(2,4,0)\\(0,0,0)\evec}_{\tilde{\rho}_{21}^{6,1}},
\underbrace{\bvec(3,0,3)\\(0,0,0)\evec}_{\tilde{\rho}_{31}^{6,1}},
\underbrace{\bvec(4,2,0)\\(0,0,0)\evec}_{\tilde{\rho}_{41}^{6,1}},
\underbrace{\bvec(5,4,3)\\(0,0,0)\evec}_{\tilde{\rho}_{51}^{6,1}},
\underbrace{\bvec(0,0,0)\\(0,0,0)\evec}_{\tilde{\rho}_{61}^{6,1}}
\right\} \nn\\
\tilde{R}_{(3,2)}(A_{2}) &=&\left\{
\underbrace{\bvec(1,0,2)\\(0,1,1)\evec}_{\tilde{\rho}_{11}^{3,2}},
\underbrace{\bvec(2,0,1)\\(0,1,1)\evec}_{\tilde{\rho}_{21}^{3,2}},
\underbrace{\bvec(0,0,0)\\(0,1,1)\evec}_{\tilde{\rho}_{31}^{3,2}},
\underbrace{\bvec(1,0,2)\\(0,0,0)\evec}_{\tilde{\rho}_{12}^{3,2}},
\underbrace{\bvec(2,0,1)\\(0,0,0)\evec}_{\tilde{\rho}_{22}^{3,2}},
\underbrace{\bvec(0,0,0)\\(0,0,0)\evec}_{\tilde{\rho}_{32}^{3,2}}
\right\} ~.\nn\\
\eea 
It can be seen that although isomorphic to the representations in \eref{e31}, the representations in \eref{e31bb} do not indicate equivalence.\\

\textbf{From $\mathbb{C}^{3}$ to $\mathbb{C}^{2}$ orbifolds.} As a final comment on the orbifold actions of $\mathbb{C}^{3}$, one can consider equivalences of orbifold actions in complex spaces of different dimension. Returning to the notion of operators $\omega^{(\{a_i\},\{b_i\})}\in R_{(n_1,n_2)}$ acting on the coordinates $\{z_1,z_2,z_3\}$ of $\mathbb{C}^{3}$, the orbifold operator $\omega^{(\{a_i\},\{b_i\})}$ can only be equivalent to a $2$-dimensional operator if and only if it acts only on $2$ coordinates $\{z_1,z_2,z_3\}$ of $\mathbb{C}^{3}$. That means, if $\{(a_m,b_m)\}=\{(0,0)\}$ for some $m\in\{1,2,3\}$ such that there is a trivial action on the $z_m$ coordinate of $\mathbb{C}^{3}$,
\beql{e31b}
(\omega^{(\{a_i\},\{b_i\})})_m ~~:~~ z_m ~\mapsto~ z_m~ (\omega^{(\{a_i\},\{b_i\})})_m = z_m ~~,
\eeq
then 
\beql{e31c}
\omega^{((a_i,a_j,0),(b_i,b_j,0))}=\bvec e^{i2\pi (\frac{a_i}{n_1}+\frac{b_i}{n_2})} \\ e^{i2\pi (\frac{a_j}{n_1}+\frac{b_j}{n_2})} \\ 1 \evec ~~\sim~~
\omega^{(n_2 a_i+n_1 b_i,n_2 a_j+n_1 b_j)} = \omega^{(\tilde{a}_1,\tilde{a}_2)} ~~, 
\eeq
where $i,j\neq m$, $\omega^{((a_i,a_j,0),(b_i,b_j,0))}$ is an operator in $\mathbb{C}^{3}/\mathbb{Z}_{n_1}\times\mathbb{Z}_{n_2}$ and $\omega^{(\tilde{a}_1,\tilde{a}_2)}$ is an operator in $\mathbb{C}^{2}/\mathbb{Z}_{n_1 n_2}$. In terms of orbifold actions, the above equivalence relation can be rewritten in a concise form as
\beql{e31d}
\bvec (a_i,a_j,0)\\ (b_i,b_j,0) \evec
\sim 
\bvec n_2 a_i+n_1 b_i,n_2 a_j+n_1 b_j\evec~~.
\eeq\\

%---------------------------
\subsection{Equivalence of Brane Tilings \label{s2} \label{sc3p3}}

Equivalence of two orbifold actions can be illustrated in the setting of brane tilings. In the context of brane boxes and brane configurations, this has been illustrated in \cite{Hanany:1997tb,Hanany:1998ru,HananyUranga98}. Brane tilings are periodic bipartite graphs on the $2$-dimensional torus that are dual descriptions of a Quiver Gauge Theory. \\

The world volume gauge theories that arise when a collection of D-branes probe a non-compact toric Calabi-Yau (CY) singularity, the CY $3$-fold, are quiver gauge theories. In $10$-dimensional Type IIB String Theory, the configuration of the probe D$3$-branes on the cone over the CY $3$-fold is T-dualised to a configuration of D$5$-branes suspended between $\mbox{NS}5$-branes. The resulting so called brane box configurations of NS$5$ and D$5$-branes, their corresponding T-dual configuration of D$3$-branes probing a non-compact Calabi-Yau singularity, and the $(3+1)$-dimensional D-brane world volume gauge theories have a combined description in the form of a brane tiling \cite{HananyUranga98,Hanany05}.\\

The configuration of $n_1$ NS$5$-branes and $n_2$ NS$5'$-branes, the $n_1 \times n_2$ brane box configuration, is T-dual to the orbifold $\mathbb{C}^{3}/\mathbb{Z}_{n_1}\times\mathbb{Z}_{n_2}$. The orbifold action can be considered as a labeling of distinct $n_1 \times n_2$ brane box configurations. Accordingly, under the brane tiling description of brane box configurations, two inequivalent orbifold actions correspond to two distinct brane tilings.\\

\textbf{Brane Tiling Dictionary.} A brane tiling is a bipartite graph embedded into the two torus. It is made of faces $\{\mathsf{F}_i\}$, edges $\{\mathsf{E}_{i}\}$ and nodes $\{\mathsf{N}_i\}$ that come in two colors (black and white) due to bipatiteness, such that each edge is connecting nodes of different coloring. The String Theory and Gauge Theory interpretations of the different components of the brane tiling are shown in \tref{t0}.\\

\begin{table}[th]
\begin{tabular}{l|l|l}
\hline
{\bf Brane tiling} & {\bf String theory} & {\bf Gauge theory} \\ \hline \hline 
$2n$-sided face & D5-branes & Gauge group with $n$ flavors \\ \hline
Edge between two & String stretched between D5-  & Bifundamental chiral multiplet\\
 faces $\mathsf{F}_{i}$, $\mathsf{F}_{j}$               &branes suspended between the  &  between gauge groups $i$ and $j$; \\ 
 & NS$5$ branes. & We orient the arrow such that \\
 & & the white node is to the right. \\ \hline 
$k$-valent vertex & Region where $k$ strings & Interaction between $k$ chiral \\
 & interact locally.&multiplets, i.e.~order $k$ term in\\
 && the superpotential. The signs for \\
      &         & the superpotential terms are\\
      && assigned such that white and \\
         &      & black nodes correspond to plus \\
&&         and minus signs respectively. \\ \hline
\end{tabular}
\caption{Dictionary for translating between brane tiling, string theory and gauge theory objects \cite{Hanany05}.}\label{t0}
\end{table}

\begin{figure}[ht!]
\begin{center}
\includegraphics[totalheight=5.5cm]{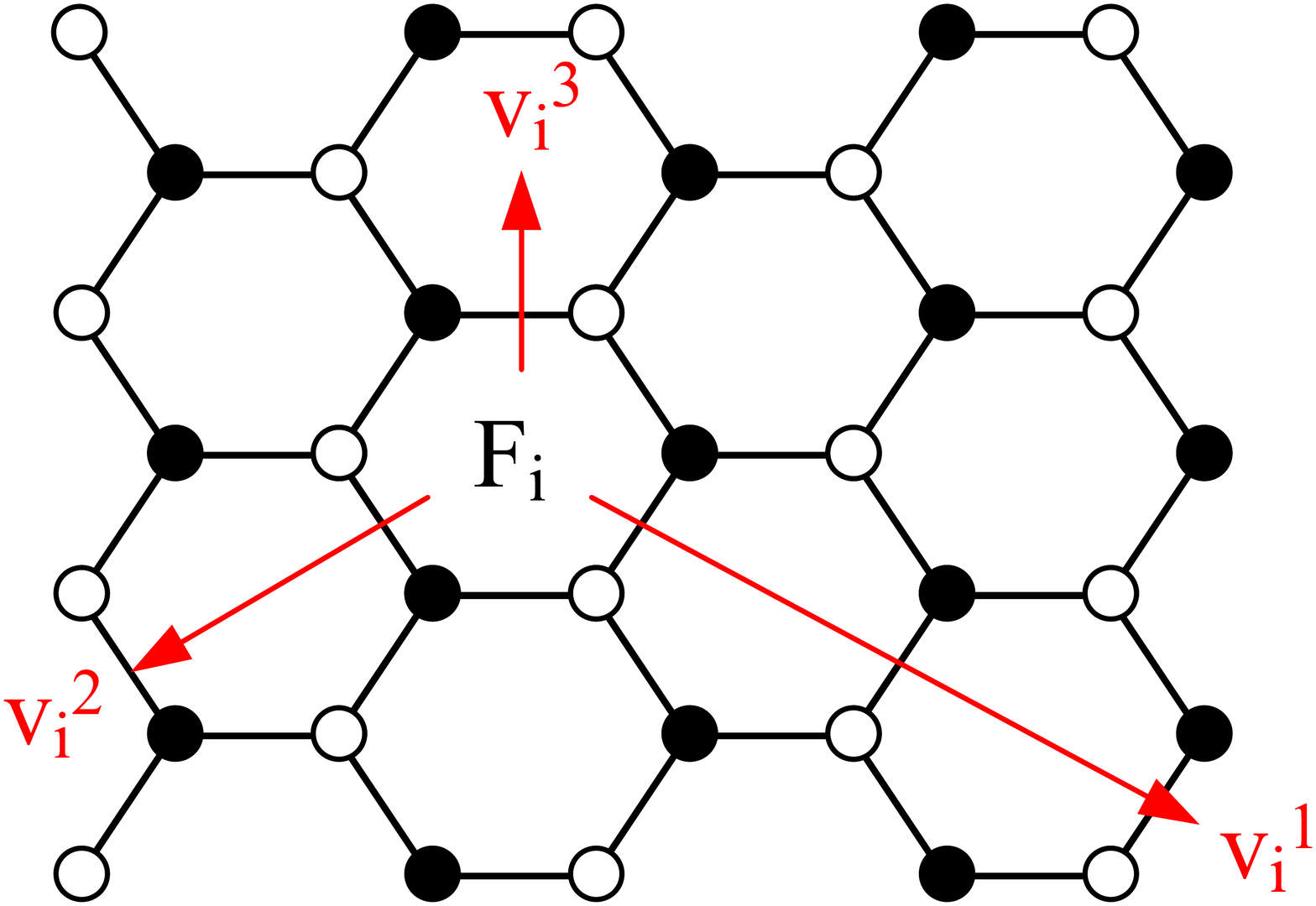}
\caption{The fundamental directions $\{v_{i}^{1},v_{i}^{2},v_{i}^{3}\}$ at a given face $\mathsf{F}_{i}$ in the brane tiling of $\mathbb{C}^{3}$.}
  \label{f1}
 \end{center}
\end{figure}

The order of the orbifold, $N=n_1 n_2$, is the number of faces in the fundamental domain of the tiling corresponding to the gauge groups $U(1)^{N}$ of the $(3+1)$-dimensional world volume gauge theory. Faces in the tiling for $\mathbb{C}^{3}$ orbifolds are hexagonal such that the tiling has $3$ symmetry axes corresponding to $3$ fundamental directions 
\beql{e31e}
\{v_{i}^{1},v_{i}^{2},v_{i}^{3}\}
\eeq
crossing at a face $\mathsf{F}_i$ in the tiling, with $i=1,\dots,N$, as shown in \fref{f1}. Note that the directions $\{v_{i}^{1},v_{i}^{2},v_{i}^{3}\}$ at a given face $\mathsf{F}_i$ are isomorphic to the complex coordinates $\{z_1,z_2,z_3\}$ of $\mathbb{C}^{3}$,
\beql{e31f}
\mathnormal{B}~~:~~\{z_1,z_2,z_3\} ~\rightarrow~ \{v_{i}^{1},v_{i}^{2},v_{i}^{3}\}~~.
\eeq
Moreover, these correspond to the generators $\sigma$ for a convex polyhedral cone \cite{Fulton} as shown in the discussion on toric geometry in Section \sref{s3}.\\

To represent the action $A_k$ in the brane tiling setup of the orbifold action $\mathbb{C}^{3}/\mathbb{Z}_{n_1}\times\mathbb{Z}_{n_2}$, it is useful to specify the face labels $\mathsf{F}_{i}$ as a pair of two positive integer numbers $\mathsf{F}_{i}=(f_{i1},f_{i2})$ with $f_{ij}\in\mathbb{N}_{0}$. Then the orbifold action can be visualized as acting on the face labels of the tiling in a chosen direction $v_{i}^{m}$,
\beql{e29}
A_{k}^{m}=\bvec a_m\\b_m\evec~~:~~\mathsf{F}_{i}=(f_{i1},f_{i2})~~\mapsto~~((f_{i1}+a_m)\bmod{n_1},(f_{i2}+b_m)\bmod{n_2}) ~~,
\eeq
where $A_{k}^{m}$ is a column of the orbifold action matrix $A_{k}$ such that $A_{k}=(A_{k}^{1},A_{k}^{2},A_{k}^{3})^{\top}$.
\\

\begin{figure}[ht!]
\begin{center}
\includegraphics[totalheight=5.5cm]{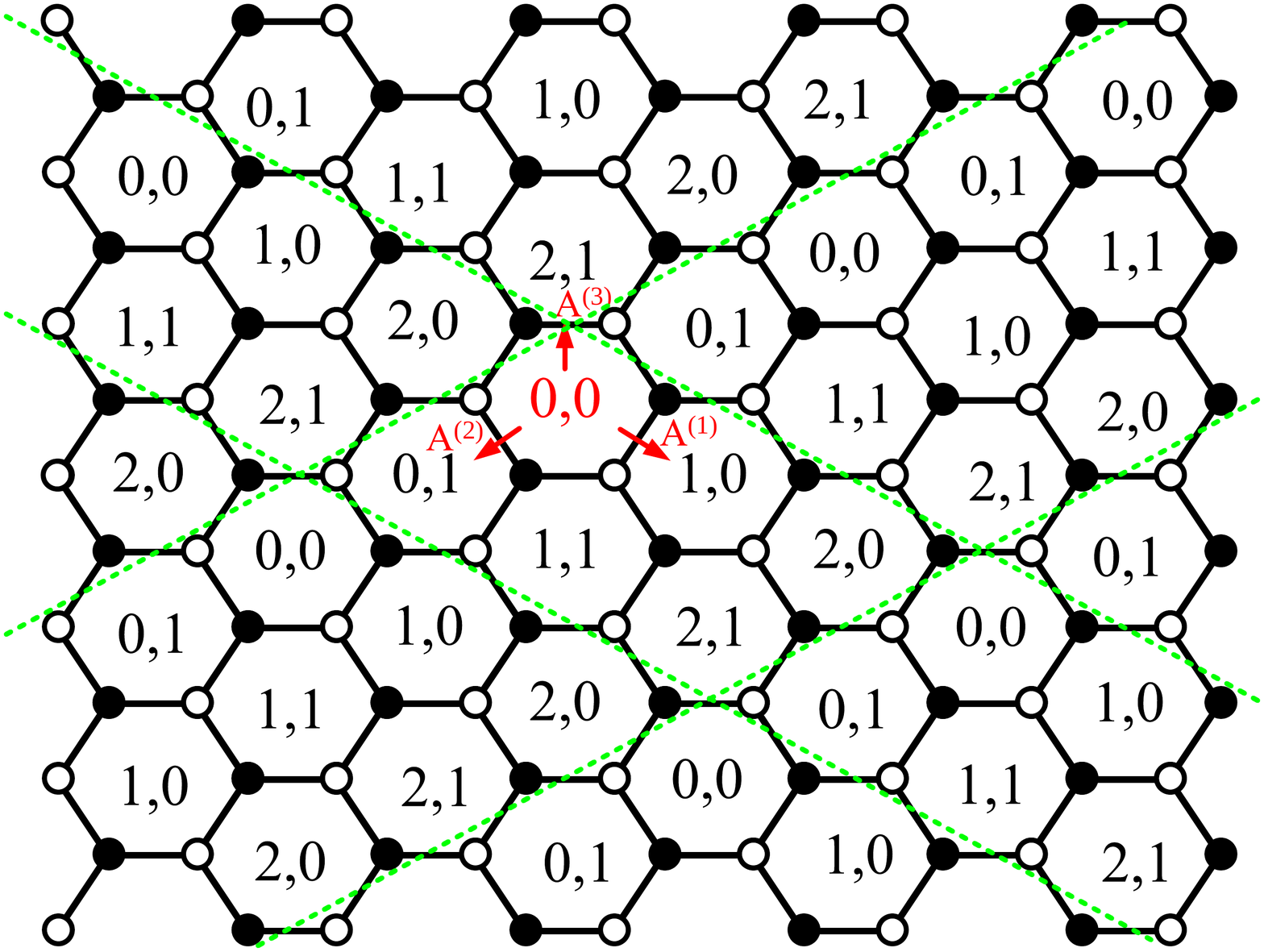}
\caption{The brane tiling for the orbifold $\mathbb{C}^{3}/\mathbb{Z}_{3}\times\mathbb{Z}_{2}$ with action $A_{2}=((1,0,2),(0,1,1))$.}
  \label{f2}
 \end{center}
\end{figure}

As an example, the orbifold used in Section \sref{s1}, $\mathbb{C}^{3}/\mathbb{Z}_{3}\times\mathbb{Z}_{2}$ with action $A_{2}=((1,0,2),(0,1,1))$ has a brane tiling as shown in \fref{f2} with an arbitrarily chosen reference face $\mathsf{F}_{1}=(f_{11},f_{12})=(0,0)$ that has $3$ direct neighbors along the fundamental directions $\{v_{1}^{1},v_{1}^{2},v_{1}^{3}\}$. These direct neighbors share with $\mathsf{F}_{1}$ a unique edge in the tiling and have labels given by
\beal{e30}
A_{1}^{1}~~&:&~~ (0,0)~~\mapsto~~(1,0) \nn\\
A_{1}^{2}~~&:&~~ (0,0)~~\mapsto~~(0,1) \nn\\
A_{1}^{3}~~&:&~~ (0,0)~~\mapsto~~(2,1)  ~~.
\eea
The entire brane tiling structure can be constructed by finding recursively the face labels of neighboring faces of all faces $\{\mathsf{F}_{i}\}$ in the brane tiling.\\

\begin{figure}[ht!]
\begin{center}
\includegraphics[totalheight=5.5cm]{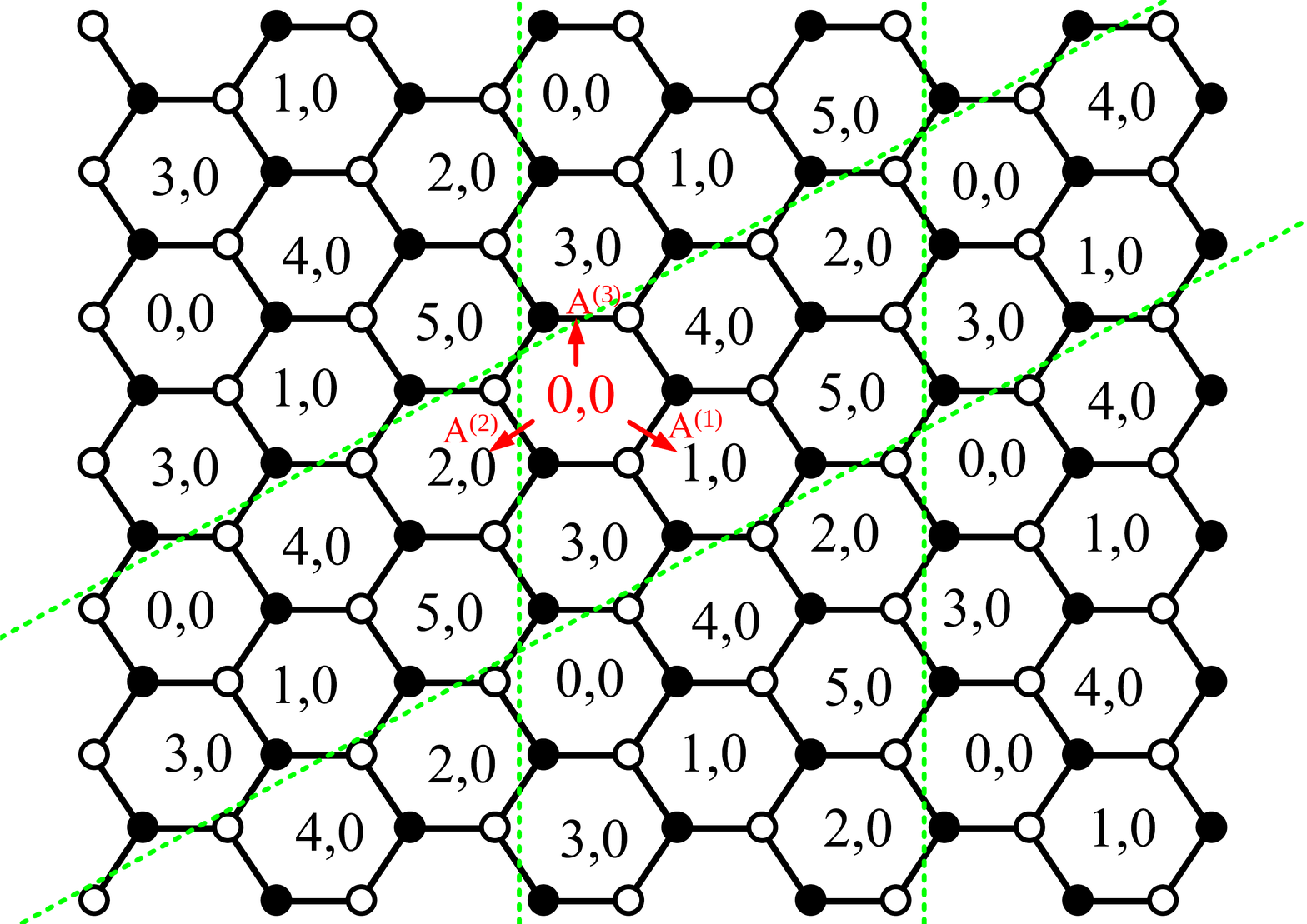}
\caption{The brane tiling for the orbifold $\mathbb{C}^{3}/\mathbb{Z}_{6}$ with action $A_{1}=((1,2,3),(0,0,0))$.}
  \label{f3}
 \end{center}
\end{figure}

\textbf{Equivalence of Brane Tilings.} It is now instructive to see how the brane tiling conveys equivalence between orbifold actions. Continuing using the example from the discussion in Section \sref{s1} with actions \eref{e27}, the brane tiling for the orbifold action $A_{1}=((1,2,3),(0,0,0))$ of $\mathbb{C}^{3}/\mathbb{Z}_{6}$ can be drawn as shown in \fref{f3}.
For any brane tiling with face labels $\mathsf{F}_{i}=(f_{i1},f_{i2})$, there is a consistent relabeling of faces $\rho$ such that
\beql{e31a}
\rho~~:~~\mathsf{F}_{i}=(f_{i1},f_{i2})~~\mapsto~~\bar{f}_{l}\in\mathbb{N}_{0} ~~,
\eeq
where $l=1,\dots,N$ and $\bar{f}_{l}\neq\bar{f}_{k}$ if $l\neq k$. For the tiling corresponding to $A_{1}=((1,2,3),(0,0,0))$ with faces $\{\mathsf{F}^{~A_{2}}_{i}\}$, a straightforward relabeling choice is \beql{e31aa}
\rho^{A_{1}}~:~(f_{i1}^{~A_{1}},f_{i2}^{~A_{1}})~\mapsto~\bar{f}_{l}=f_{i1}^{~A_{1}}\eeq
since $f_{i2}=0~\forall i$.
It can be now shown that there is a consistent relabeling $\rho^{A_{2}}$ such that it maps the face labels $\{\mathsf{F}_{i}^{~A_{2}}\}$ of the tiling for $A_{2}$ in the following way,
\beql{e31f}
\rho^{A_{2}} ~~:~~ \{\mathsf{F}_{i}^{~A_{2}}\}=\{(f_{i1}^{~A_{2}},f_{i2}^{~A_{2}})\}~~\rightarrow~~ \{\bar{f}_{l}\}=\rho^{A_{1}}(\{\mathsf{F}_{i}^{~A_{1}}\})~~,
\eeq
where $\rho^{A_2}$ is the map on the face labels of the $A_{2}$ action tiling as shown in \eref{e31aa}.\\

In fact, in general if the relation in \eref{e31f} holds for two brane tilings of orbifold actions $A_{1}\in\tilde{R}_{(n_1,n_2)}$ and $A_{2}\in\tilde{R}_{(n'_{1},n'_{2})}$ with $n_1 n_2=n'_{1}n'_{2}=N$, then $A_{1}\sim A_{2}$. For the above two example actions $A_{1}$ and $A_{2}$, the relabeling map on $\{\mathsf{F}_{i}^{~A_{2}}\}$ can be chosen as
\beal{e32}
\rho^{A_{2}} ~~:~~
(0,0)~~&\mapsto&~~0=\rho^{A_{1}}((0,0)) \nn\\
(1,0)~~&\mapsto&~~1=\rho^{A_{1}}((4,0)) \nn\\
(2,0)~~&\mapsto&~~2=\rho^{A_{1}}((2,0)) \nn\\
(0,1)~~&\mapsto&~~3=\rho^{A_{1}}((3,0)) \nn\\
(1,1)~~&\mapsto&~~4=\rho^{A_{1}}((1,0)) \nn\\
(2,1)~~&\mapsto&~~5=\rho^{A_{1}}((5,0))
\eea 
verifying that $A_{1}\sim A_{2}$ where $A_{1}\in\tilde{R}_{(6,1)}$ and $A_{2}\in\tilde{R}_{(3,2)}$. Accordingly, we have shown that $A_{1}\sim A_{2}$ in the context of brane tilings verifying the result in Section \sref{s1}. \\

Another correspondence can be identified between equivalent brane tilings and orbifold actions that are equivalent up to a permutation of the complex coordinates of $\mathbb{C}^{3}$, $\{z_1,z_2,z_3\}$. By the correspondence between the coordinates $\{z_1,z_2,z_3\}$ and the fundamental directions $\{v_{i}^{1},v_{i}^{2},v_{i}^{3}\}$ of a face $\mathsf{F}_{i}$ in the tiling, orbifold equivalence up to a permutation of coordinates corresponds to tiling equivalence due to permutations of $\{v_{i}^{1},v_{i}^{2},v_{i}^{3}\}$ that are interpreted as reflections or rotations around a face $\mathsf{F}_{i}$ in the tiling. Accordingly, orbifold action equivalence can be identified as a \textit{symmetry} on the brane tiling.\\

\textbf{The Calabi-Yau condition in the Brane Tiling.} The Calabi-Yau condition, which is manifested in the definition of the orbifold action and the $\det=1$ property of $SU(3)$, is naturally visualized in the brane tiling setup. Defining the minimal path element as the orbifold action along a given fundamental direction $v_{i}^{m}$ in the tiling under the function
\beql{e33}
f_{P}(A_{k}^{m},\mathsf{F}_{i})~~=~~\left\{ (f_{i1},f_{i2})~~,~~A_{k}^{m}(f_{i1},f_{i2})=((f_{i1}+a_m)\bmod{n_1},(f_{i2}+b_m)\bmod{n_2})  \right\} ~~,
\eeq
the Calabi-Yau condition requires that any path defined as 
\beql{e34}
f_{P}(A_{k}^{1},\mathsf{F_i})\cup f_{P}(A_{k}^{2},A_{k}^{1}\mathsf{F_i}) \cup f_{P}(A_{k}^{3},A_{k}^{2}A_{k}^{1}\mathsf{F_i})
\eeq
is closed by $A_{k}^{3}A_{k}^{2}A_{k}^{1}\mathsf{F}_{i}=\mathsf{F}_{i}$. Since this is true for any starting face $\mathsf{F}_{i}$ with $i=1,\dots,N$, every node in the tiling can be considered as a puncture on the two torus on which the brane tiling is drawn.\\

\textbf{Brane Tilings for $\mathbb{C}^{2}$ Orbifolds.} As a final comment on brane tilings, it is instructive to consider a possible representation of $\mathbb{C}^{2}$ orbifold actions by a brane tiling. As seen in \eref{e31d}, two orbifold actions from different dimensional orbifolds can be equivalent. Accordingly, a brane tiling of a $\mathbb{C}^{3}$ orbifold can represent an action on an orbifold of $\mathbb{C}^{2}$. These brane tilings are called linear by the property that the $\mathbb{C}^{3}$ orbifold action of the form $A_{k}=((a_i,a_j,0),(b_i,b_j,0))$ has a unity action component $A_{k}^{m}=\{(a_m,b_m)\}=\{(0,0)\}$ along the fundamental direction $v_{i}^{m}$ for $i,j\neq m$. This means that $A_{k}^{m}~:~\mathsf{F}_{i}\mapsto\mathsf{F}_{i}$ such that face labels along the $v_{i}^{m}$ direction stay invariant. Accordingly, it is sufficient to consider the tiling as a single row of faces with face labels generated either by $A_{k}^{i}$ or $A_{k}^{j}$. The corresponding brane construction was studied in \cite{HananyWitten,Karch:1998yv}.\\

From this point of view, and from the observation that actions of orbifolds $\mathbb{C}^{2}/\mathbb{Z}_{N}$, $(\bar{a}_1,\bar{a}_2)=(\bar{a}_1,N-\bar{a}_1)\in\tilde{R}_{N}$ are effectively labeled by a single parameter $\bar{a}_1\in\mathbb{Z}^{+}$ deduced from the Calabi-Yau condition $(\bar{a}_{1}+\bar{a}_{2})\mod{N}=0$, it can be shown that all orbifold actions in $\mathbb{C}^{2}$ are equivalent to the action $(1,N-1)\in\tilde{R}_{N}$ as indicated in the above discussion on $\mathbb{C}^{2}$ orbifolds in the context of \eref{e13} in Section \sref{s6}.\\

It is noted that the rows of faces of the linear brane tiling of the original $\mathbb{C}^{3}$ orbifold with $(a_m,b_m)=(0,0)$ are labeled by either $A_{k}^{i}=(a_i,b_i)$ or $A_{k}^{j}=(a_j,b_j)$ corresponding to selecting either a row in the $v^{i}$ or $v^{j}$ tiling direction respectively. From the fact that $A_{k}^{i}$ and $A_{k}^{j}$ are transpose components of a single $\mathbb{C}^{3}$ action $A_{k}=(A_{k}^{i},A_{k}^{j},A_{k}^{m})^{\top}$, the row of faces labeled by $A_{k}^{i}$ and the row labeled by $A_{k}^{j}$ are equivalent up to a relabeling. This can be seen when using the equivalence relation in \eref{e31d} which gives
\beql{e35}
\bar{a}_1=n_2 a_i+n_1 b_i~~,~~\bar{a}_2=n_2 a_j+n_1 b_j ~~~,
\eeq
where the $\mathbb{C}^{3}$ tiling face labeling by $A_{k}^{i}$ corresponds to the $\mathbb{C}^{2}$ action $(\bar{a}_1,N-\bar{a}_1)$ and the $\mathbb{C}^{3}$ tiling face labeling by $A_{k}^{j}$ corresponds to the $\mathbb{C}^{2}$ action $(N-\bar{a}_2,\bar{a}_2)$. By the Calabi-Yau condition in the context of the $\mathbb{C}^{2}$ orbifold, $(\bar{a}_i+\bar{a}_j)\bmod{N}=0$. This can be verified by the fact that $(a_{i}+a_{j}+a_{m}) \mod{n_1}=0$ and $(b_{i}+b_{j}+b_{m}) \mod{n_2}=0$ with $(a_m,b_m)=(0,0)$ in the current linear $\mathbb{C}^{3}$ tiling example. Accordingly, from these results, it can be easily shown that $(\bar{a}_1,\bar{a}_2)=(\bar{a}_1,N-\bar{a}_1)=(N-\bar{a}_2,\bar{a}_2)$.\\

To show now that $(\bar{a}_1,\bar{a}_2)\sim (1,N-1)$, we first choose the $\mathbb{C}^{3}$ tiling face labeling by $A_{k}^{i}$ and note that it is $N$-periodic, $(A_{k}^{i})^{N}=(0,0)$, since $N(a_1,b_2)\mod{N}=(0,0)$. By a relabeling map $\rho$, the face labels can be chosen as
\beal{e36}
\rho~~:~~~~~~~~~~~~~ 
A_{k}^{(i)} \mathsf{F} ~&\mapsto&~ (1,0) \nn\\
(A_{k}^{(i)})^{2} \mathsf{F} ~&\mapsto&~ (2,0) \nn\\
&\vdots& \nn\\
(A_{k}^{(i)})^{N-1} \mathsf{F} ~&\mapsto&~ (N-1,0) \nn\\
(A_{k}^{(i)})^{N} \mathsf{F} ~&\mapsto&~ (0,0) ~~,
\eea
which corresponds to a face row labeling by a new $\mathbb{C}^{3}$ action $\tilde{A}_{k}^{(i)'}=(a_i,b_i)=(1,0)$. Noting now that for $\mathbb{C}^{2}$ orbifolds $n_{2}=1$ and $n_{1}=N$, and using from the above relabeling $\rho$ the values $a_i=1$ and $b_i=0$ where $(a_i+a_j)\mod{n_1}=0$ and $(b_i+b_j)\mod{n_2}=0$, the relations in \eref{e35} can be solved to give $\bar{a}_{1}=1$ and $\bar{a}_{2}=N-1$ such that $(\bar{a}_{1}+\bar{a}_{2})\mod{N}=0$. Hence, it is shown that in general all $\mathbb{C}^{2}$ orbifold actions are equivalent to a single unique action $(1,N-1)$ at a given order $N$.

%---------------------------
\subsection{Equivalence of Toric Triangles \label{s3} \label{sc3p4}}

The brane tiling does not only encode the gauge groups, superpotential terms and chiral fields of the associated Quiver Gauge Theory arising from D$3$-branes probing the toric CY $3$-fold. The brane tiling also encodes information characterizing the toric geometry of the CY $3$-fold that is being probed by D$3$-branes. The algorithm that deciphers the encoding of the geometry data in the tiling is called the Forward Algorithm and has been described in the literature \cite{HananyKen05,Hanany05}. \\

The Forward Algorithm results in the toric diagram, a convex polyhedron embedded in the lattice $\mathbb{Z}^{2}$ for CY $3$-folds. For orbifold $\mathbb{C}^{3}/\Gamma_{N}$ dual to the Quiver Gauge Theory arising when D$3$-branes probe the CY $3$-fold, the corresponding toric diagram is a triangle with corner points $\{v_{i}\}$ where $i=1,\dots,3=D$. The Cartesian coordinates of the corner points are given by $v^{i}=(x_{1}^{i},x_{2}^{i},x_{3}^{i})$. The area of the triangle corresponds to the number of gauge groups $N$ in the Quiver Gauge Theory. \\

The Forward Algorithm acts as a map that links through the brane tiling setup orbifold actions $A_{k}$ of a given orbifold $\mathbb{C}^{3}/\Gamma_{N}$ to a toric diagram triangle $\sigma$ with corner points $\{v_{i}\}$. Accordingly, two equivalent orbifold actions need to be mapped by the Forward Algorithm to two equivalent toric diagram triangles. Conversely, two equivalent toric diagram triangles need to be mapped to two equivalent orbifold actions by the Inverse Algorithm \cite{HananyFengHe00}. \\

\textbf{Topology.} To understand equivalence between toric diagram triangles, it is important to identify the topological information a lattice triangle on a plane encodes. One feature is the area of the triangle $N$. It is conveniently expressed by Pick's Theorem formula of a lattice polygon,
\beql{e59}
N=2 N_I+N_B-2~~,
\eeq 
where $N_I$ is the number of lattice points enclosed by the perimeter of the polygon, and $N_B$ is the number of lattice points on the boundary edges of the polygon. Accordingly, internal and boundary lattice points of a toric diagram triangle do play a role in defining the parameterization of the topology of a toric diagram.\\

Let us define the $topology$ of a toric diagram corresponding to the orbifold $\mathbb{C}^{3}/\Gamma_{N}$. Calling the set of internal points $I$ and boundary points $B$, it is useful to introduce barycentric coordinates to express the defining points of the toric diagram, $w_k\in I\cup B$ with $k=1,\dots,|I\cup B|$. Knowing that every point $w_k\in I\cup B$ divides the toric triangle into $3$ subtriangles with areas $\lambda_{k1}$, $\lambda_{k2}$ and $\lambda_{k3}$, the barycentric coordinates of $w_k$ are defined as,
\beql{e60}
w_{k}=\frac{1}{N}\left(\lambda_{k1},\lambda_{k2},\lambda_{k3}\right) ~~,
\eeq
where $N=\sum_{i=1}^{3}{\lambda_{ki}}$. From \eref{e60}, we define the `topological character' of the toric diagram triangle $\sigma$ with corner points $\{v_i\}$ as the set of the barycentric coordinates of all the points $w_k\in I\cup B$,
\beql{e61}
\tau~~=~~\left\{\frac{1}{N}\left(\lambda_{k1},\lambda_{k2},\lambda_{k3}\right)~~|~~ w_{k}\in I\cup B\right\} ~~.
\eeq
The topological character $\tau$ can be considered as a $k\times 3$ matrix, where the $k$ rows represent the lattice points inside and on the boundary of the toric diagram triangle $\sigma$, and the $3$ columns represent the coordinate axes of the barycentric coordinates of the lattice points.\\

\textbf{Equivalence.} For two given toric diagram triangles $\sigma_1$ and $\sigma_2$ with the same area $N$ and topological characters $\tau_1=\tau(\sigma_1)$ and $\tau_2=\tau(\sigma_2)$ respectively, the toric diagram triangles $\sigma_1$ and $\sigma_2$ are said to be equivalent if the respective topological characters are the same, $\tau_1\sim\tau_2$, up to a permutation of the barycentric coordinate components, 
\beql{e62}
\mbox{Perm}\left(\left\{\frac{1}{N}(\lambda_{1i},\lambda_{2i},\dots,\lambda_{N_{IB}i})\right\}\right)
~~=~~\mbox{Perm}\left(\tau^{\top}\right)
~~,
\eeq
where $i=1,2,3$ and $N_{IB}=|I\cup B|$. It is important to note that the permutations of barycentric coordinate components in \eref{e62} correspond to permutation of the barycentric coordinate axes of a toric diagram triangle, $\{\hat{v}_1,\hat{v}_2,\hat{v}_3\}$.\\

To further evaluate the meaning of the equivalence condition based on the coordinate permutation in \eref{e62}, we consider the corner points $v_i$ defining the toric diagram triangle $\sigma$. The barycentric coordinates for the $3$ corner points are always of the form,
\beql{e63}
v_{1}=(0,0,1)~~~,~~~v_{2}=(0,1,0)~~~,~~~v_{3}=(1,0,0)~~~,
\eeq
independent of the toric diagram triangle area $N$. It is important to note that any permutation of the coordinate axes of the barycentric coordinates leaves the set of corner points $\{v_{i}\}$ in \eref{e63} invariant. Accordingly, we may identify the corner points in \eref{e63} as the $3$ unit vectors along the $3$ linearly independent barycentric coordinate axes, $\{\hat{v}_1,\hat{v}_2,\hat{v}_3\}$, of the toric diagram triangle.\\

\begin{figure}[ht!]
\begin{center}
\includegraphics[totalheight=6cm]{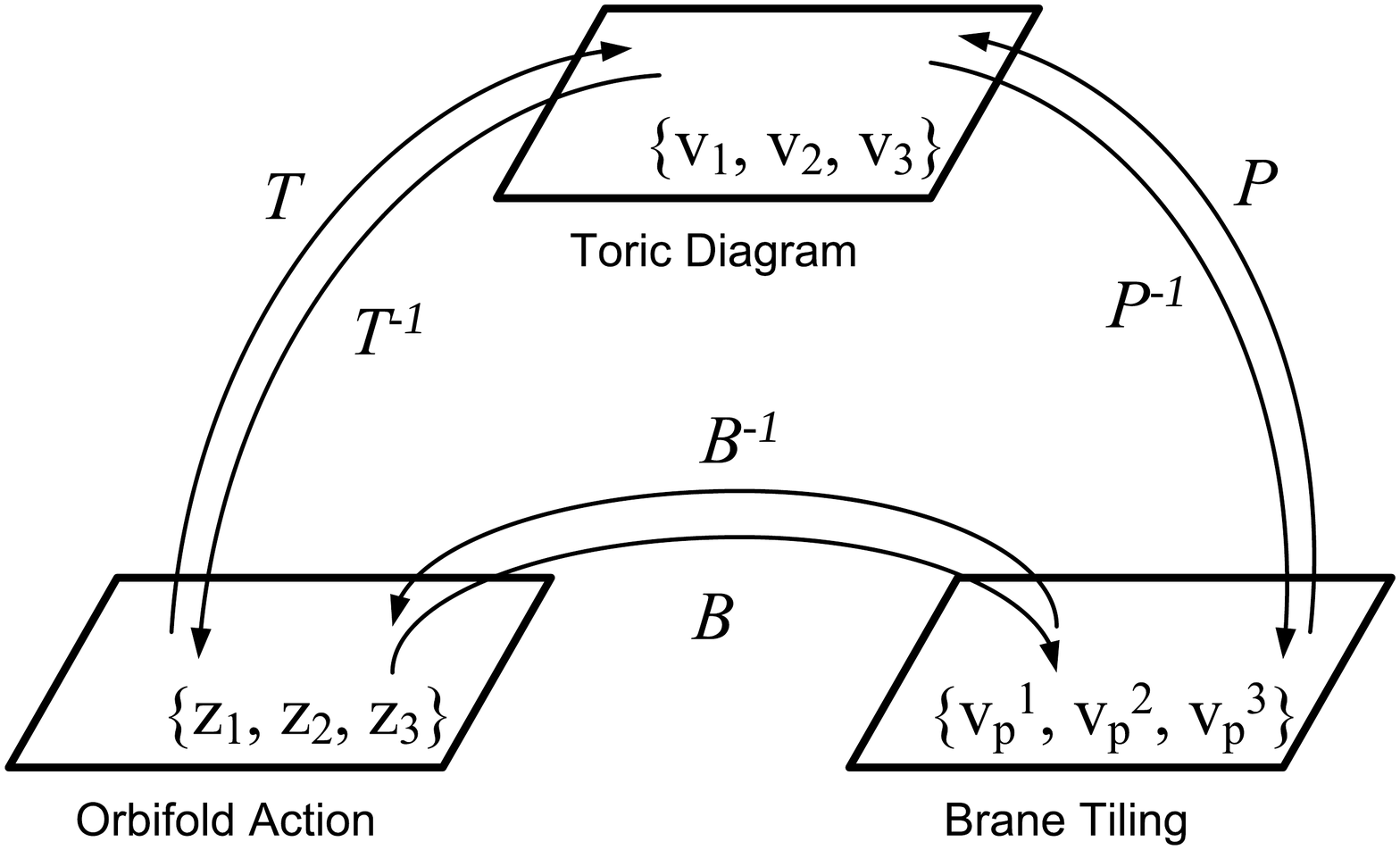}
\caption{Maps $\mathnormal{B}$, $\mathnormal{T}$ and $\mathnormal{P}$  and their corresponding inverses linking the coordinates of $\mathbb{C}^{3}$, $\{z_1,z_2,z_3\}$, the toric diagram $\{\hat{v}_1,\hat{v}_2,\hat{v}_3\}$, and the brane tiling $\{v_{p}^{1},v_{p}^{3},v_{p}^{3}\}$ at brane tiling face $\mathsf{F}_{p}$.}
  \label{f4}
 \end{center}
\end{figure}

It is now possible to show several links between the discussions on orbifold action equivalence in Section \sref{s1} and brane tiling equivalence in Section \sref{s2}. As pointed out in Section \sref{s1}, two orbifold actions are equivalent when they are related by a permutation of the coordinates $\{z_1,z_2,z_3\}$ on $\mathbb{C}^{3}$. It is important to note that there is an isomorphism between the coordinates of $\mathbb{C}^{3}$ and the barycentric coordinate axes $\{\hat{v}_1,\hat{v}_2,\hat{v}_3\}$ corresponding to the corner points of the toric diagram triangle $\sigma$,
\beql{e64}
\mathnormal{T}~~:~~\{z_1,z_2,z_3\} ~\rightarrow~ \{\hat{v}_1,\hat{v}_2,\hat{v}_3\}~~.
\eeq
Moreover, as noted in Section \sref{s2}, the fundamental directions on a brane tiling face $\mathsf{F}_{p}$, $\{v_{p}^{1},v_{p}^{2},v_{p}^{3}\}$, are isomorphic to the coordinates of $\mathbb{C}^{3}$, leading to the isomorphism between the barycentric coordinate axes of a toric diagram triangle and the fundamental directions $\{v_{p}^{1},v_{p}^{3},v_{p}^{3}\}$ on a brane tiling face $\mathsf{F}_{p}$,
\beql{e65}
\mathnormal{P}~~:~~\{v_{p}^{1},v_{p}^{3},v_{p}^{3}\}~\rightarrow~ \{\hat{v}_1,\hat{v}_2,\hat{v}_3\}~~.
\eeq
The maps $\mathnormal{B}$ in \eref{e31f}, $\mathnormal{T}$ in \eref{e64} and $\mathnormal{P}$ in \eref{e65} and their corresponding inverses $\mathnormal{B}^{-1}$, $\mathnormal{P}^{-1}$ and $\mathnormal{T}^{-1}$ link the coordinates of the orbifold, the toric diagram and the brane tiling as shown in \fref{f4}.\\

\begin{figure}[ht!]
\begin{center}
\includegraphics[totalheight=6cm]{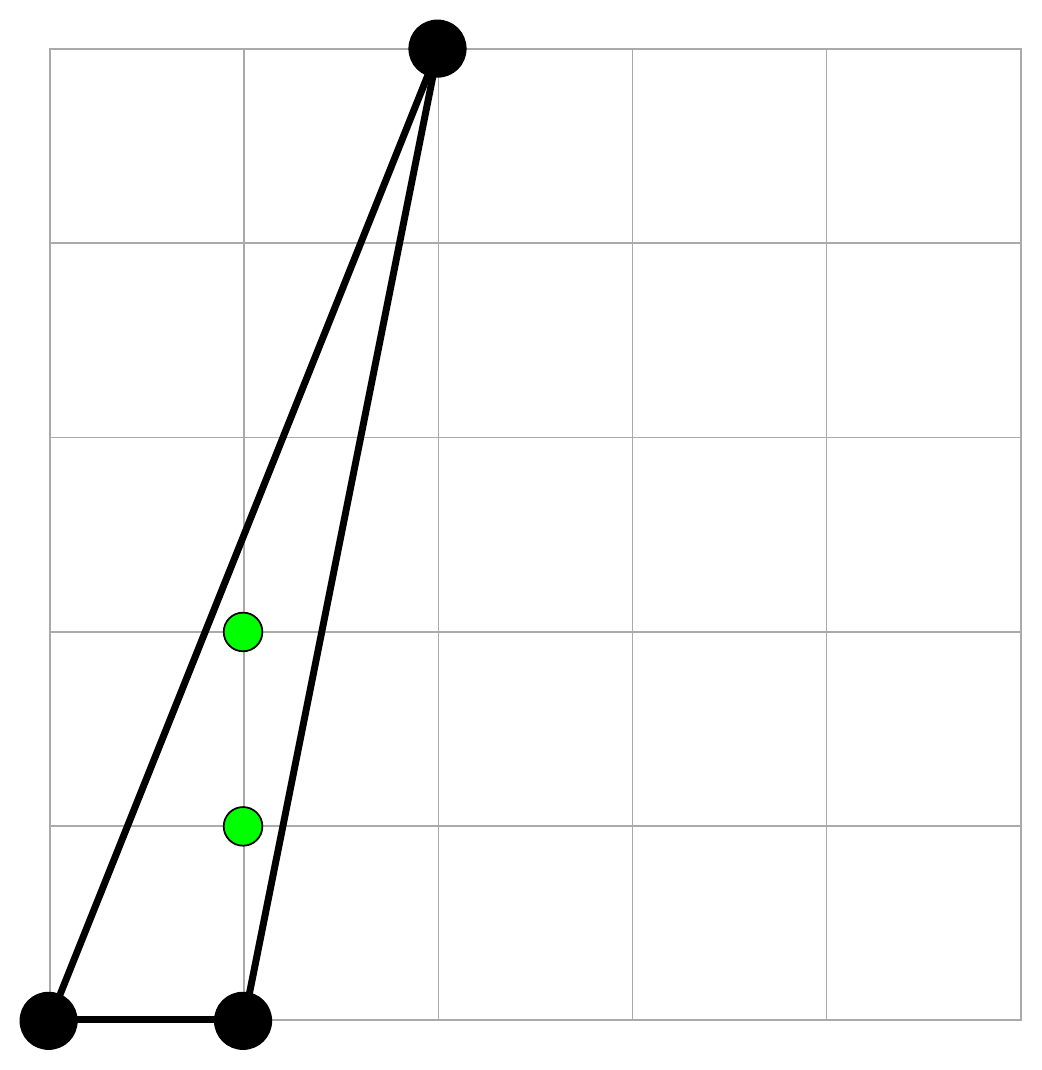}
\includegraphics[totalheight=6cm]{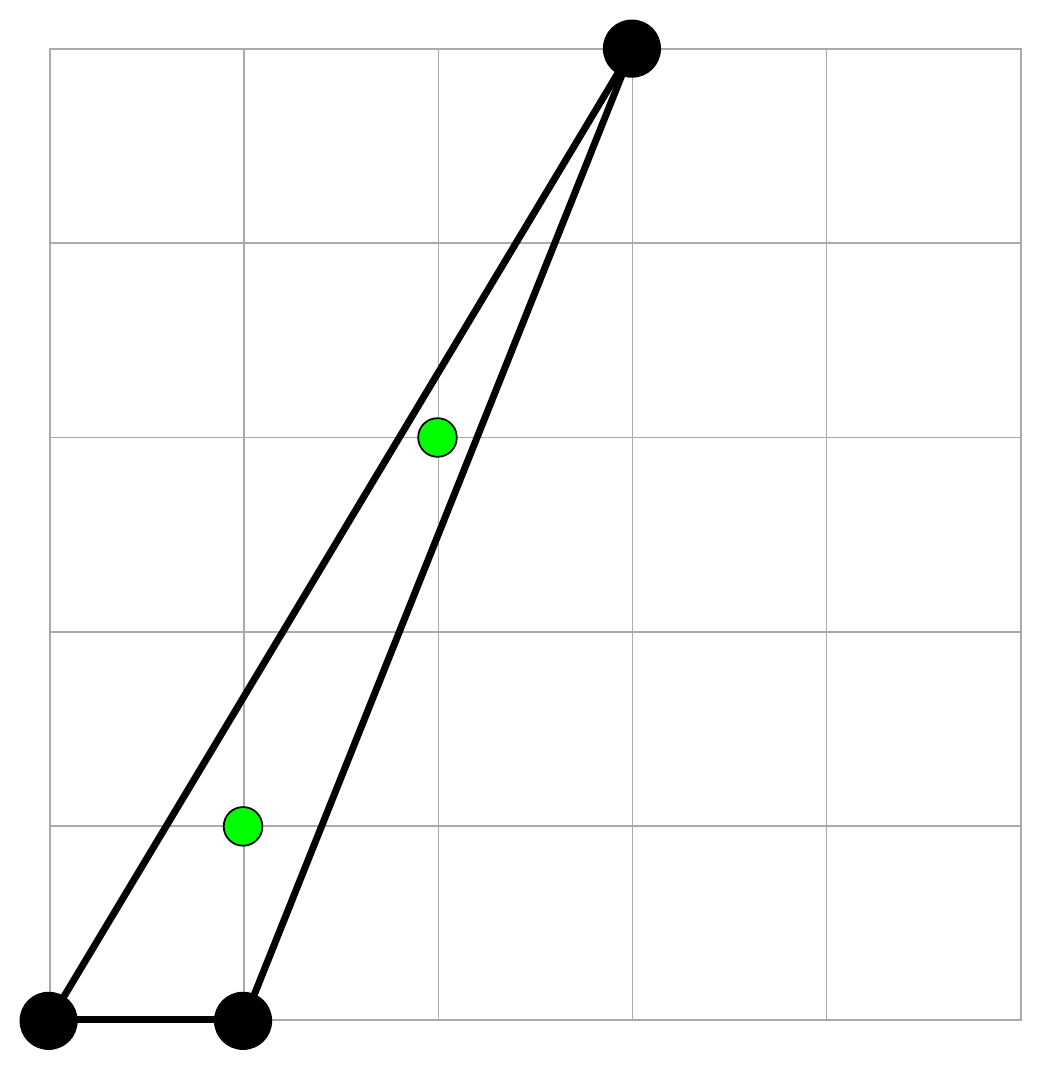}
\caption{Two toric diagrams corresponding to the orbifold $\mathbb{C}^{3}/\mathbb{Z}_{5}$ with area $N=5$. Internal points on the lattice are colored green.}
  \label{f5}
 \end{center}
\end{figure}

\textbf{Example.} Let us consider an example of two toric triangles and their equivalence relation. Taking $N=5$, it is possible to show explicitly that the two toric diagram triangles shown in \fref{f5} and corresponding to the orbifold $\mathbb{C}^{3}/\mathbb{Z}_{5}$ are equivalent.\\

First, we calculate the barycentric coordinates of the two toric diagram triangles to obtain the two topological characters,
\beal{e65b}
\tau_1~&=&~ 
\left\{
\left(1,0,0\right)~,~
\left(0,1,0\right)~,~
\left(0,0,1\right)~,~
\left(\frac{2}{5},\frac{1}{5},\frac{2}{5}\right)~,~
\left(\frac{1}{5},\frac{3}{5},\frac{1}{5}\right)
\right\}
\nn\\
\tau_2~&=&~
\left\{
\left(1,0,0\right)~,~
\left(0,1,0\right)~,~
\left(0,0,1\right)~,~
\left(\frac{1}{5},\frac{2}{5},\frac{2}{5}\right)~,~
\left(\frac{3}{5},\frac{1}{5},\frac{1}{5}\right)
\right\}~~.
\eea
Then, it can be seen that up to a permutation of the barycentric coordinate axes, $\{\hat{v}_1,\hat{v}_2,\hat{v}_3\}$, such that $(\hat{v}_1,\hat{v}_2,\hat{v}_3)\rightarrow(\hat{v}_2,\hat{v}_1,\hat{v}_3)$, the two characters in \eref{e65b} are equivalent, $\tau_1\sim\tau_2$. Hence, the corresponding toric diagram triangles and the dual orbifold actions of $\mathbb{C}^{3}/\mathbb{Z}_{5}$ are equivalent. \\

\textbf{Domain.} It is now possible to count the number of inequivalent toric diagram triangles of a given area $N$, and match the count with the number of inequivalent orbifold actions of a given order $N$ of the orbifold $\mathbb{C}^{3}/\Gamma_{N}$. In order to begin with the counting, we like to define a set of all toric diagram triangles of area $N$ where two triangles in the set are not related by a trivial symmetry transformation on the lattice $\mathbb{Z}^{2}$.\\

For the purpose of eliminating redundancies, we define the \textit{domain} $D(N)$ as the set of all lattice triangles with area $N$ and a corner point as the origin. The two other corner points are in the positive quadrant of $\mathbb{Z}^{2}$. These conditions ensure that no two lattice triangles in $D(N)$ are related by a rotation around the origin, or a translation of the lattice $\mathbb{Z}^{2}$.\\

Let a corner point $v_i$ of a toric diagram lattice triangle be given in Cartesian coordinates $(x_1,x_2)\in\mathbb{Z}^{2}$. The domain $D(N)$ can be generated using all possible transformations of the unit area lattice triangle into the positive quadrant of $\mathbb{Z}^{2}$ such that the transformed lattice triangle has area $N$ and has one corner point on the origin.\\

%It is now possible to count the number of inequivalent toric diagram triangles of a given area $N$, and match the count with the number of inequivalent orbifold actions of a given order $N$ of the orbifold $\mathbb{C}^{3}/\Gamma_{N}$. For the purpose of eliminating redundancies, we define the \textit{domain} $D(N)$ as a set of all lattice triangles with area $N$ where two triangles in the set are not related to each other by a lattice translation or a rotation around the $\mathbb{Z}^{2}$ origin.\\

%Let the corner points $v_i$ of a toric diagram lattice triangle be given in Cartesian coordinates $(x_1,x_2)\in\mathbb{Z}^{2}$. The domain $D(N)$ can be generated by transforming the unit area lattice triangle to lattice triangles with area $N$ such that the first corner point remains at the origin, the second corner point remains on the $x_2=0$ origin, and the third corner point is determined by the remaining translational freedom. All corner points remain in the positive quadrant of $\mathbb{Z}^{2}$.\\

The unit area triangle has corner points
\beql{e67}
v_{1}=(0,0)~~~,~~~v_{2}=(1,0)~~~,~~~v_{3}=(0,1)~~~,
\eeq
in Cartesian coordinates. The transformation matrices satisfying the domain conditions are of the form,
\beql{e68}
M~~=~~\left(\ba{cc} m_{11} & m_{12} \\ 0 & m_{22} \ea\right)~~,
\eeq
where $\det{M}=m_{11}m_{22}=N$ and $0\leq m_{12} < m_{22}$ with $m_{ij}\in\mathbb{N}_{0}$. These matrices are called the \textit{Hermite Normal Form}. Accordingly, the domain can be given as
\beal{e69}
D(N)~~=~~\Big\{&&
\left\{v_1,v_2,v_3\right\}=\left\{(0,0),(m_{11},0),(m_{12},m_{22})\right\}\nn\\
&~~|~~&
N=m_{11}m_{22}~,~~0\leq m_{12} < m_{22}~,~~m_{ij}\in\mathbb{N}_{0}~~~
\Big\}~~.
\eea\\

By defining the equivalence class for a toric diagram triangle $\sigma$ as
\beql{e70}
[\sigma]~=~\left\{
\tilde{\sigma}~~|~~\tau(\sigma)\sim\tau(\tilde{\sigma})
\right\}~,
\eeq
with $\tau(\sigma)\sim\tau(\tilde{\sigma})$ up to a permutation of the barycentric coordinate axes $\{\hat{v}_1,\hat{v}_2,\hat{v}_3\}$, the domain $D(N)$ can be decomposed into equivalence classes such that
\beql{e71}
\bigcup_{i=1}^{N_{E}}{[\sigma_{i}]}~=~D(N)~~,
\eeq
in analogy to the decomposition of the set of all orbifold actions  $\mathcal{A}_{N}$ in \eref{e7} into equivalence classes of orbifold actions $[A_{k}]$, with $N_E$ being both the number of equivalence classes of orbifold actions and toric diagram triangles. The size of an equivalence class $[\sigma_{i}]\in D(N)$, $|[\sigma_{i}]|=\mu_{i}$, is called the \textit{multiplicity} of the corresponding toric diagram triangle such that
\beql{e71b}
\sum_{i=1}^{N_{E}}{\mu_{i}}~=~|D(N)|~~,
\eeq
where $|D(N)|$ is called the \textit{domain size}.

%%%%%%%%%%%%%%%%%%%%%%%%%%%%%%%%%%%%%%%%%%%%%%%%%%%%%%%%%%%%%%%%%%%%%%%%%%%%%%%%%%%%%%%%%%%%%%%%%%%%%%%%%%%%%%%%%%%%%%%%%%%%

\section{Orbifolds of $\mathbb{C}^{4}$ \label{sc4}}

Moving on to $\mathbb{C}^{4}$ orbifolds, we illustrate how orbifold actions of $\mathbb{C}^{4}/\mathbb{Z}_{N}$, $\mathbb{C}^{4}/\mathbb{Z}_{n_1}\times\mathbb{Z}_{n_2}$ with $N=n_1 n_2$, and $\mathbb{C}^{4}/\mathbb{Z}_{n_1}\times\mathbb{Z}_{n_2}\times\mathbb{Z}_{n_3}$ with $N=n_1 n_2 n_3$ can be equivalent. We demonstrate the generalization of the methods introduced in Section \sref{sc3} and show that the correspondence between orbifolds and toric geometry continues to exist for $\mathbb{C}^{4}$ orbifolds. The question of equivalence between orbifold actions of $\mathbb{C}^{4}$ is reformulated in terms of lattice $3$-simplices (tetrahedra) of volume $N$. The concept of `scaling' a toric diagram lattice simplex is introduced for the purpose of defining equivalence between lattice simplices.

%--------------------------------------------
\subsection{Orbifolds and Orbifold Actions \label{sc4p1}}

Let $\mathbb{C}^{4}$ be parameterized by $\{z_1,z_2,z_3,z_4\}$. The orbifold is of the form $\mathbb{C}^{4}/\Gamma_{N}$ with $\Gamma_{N}=\mathbb{Z}_{n_1}\times\mathbb{Z}_{n_2}\times\mathbb{Z}_{n_3}\subset SU(4)$ and the order given by $N=n_1 n_2 n_3$. As a convention, the product is sorted such that $n_1\geq n_2 \geq n_3$.\\

Let a representation of $\Gamma_{N}=\mathbb{Z}_{n_1}\times\mathbb{Z}_{n_2}\times\mathbb{Z}_{n_3}$ be called $R_{(n_1,n_2,n_3)}$ with elements $\omega^{(\{a_i\},\{b_i\},\{c_i\})}$ and $i=1,\dots,4$. The elements of the representation $\omega^{(\{a_i\},\{b_i\},\{c_i\})}\in R_{(n_1,n_2,n_3)}$ are of the form
\beql{e37b}
\omega^{(\{a_i\},\{b_i\},\{c_i\})}~~=~~
\diag\left(\ba{c} e^{\frac{i2\pi a_1}{n_1}} \\ e^{\frac{i2\pi a_2}{n_1}} \\ e^{\frac{i2\pi a_3}{n_1}} \\ e^{\frac{i2\pi a_4}{n_1}} \ea \right) 
\diag\left(\ba{c} e^{\frac{i2\pi b_1}{n_2}} \\ e^{\frac{i2\pi b_2}{n_2}} \\ e^{\frac{i2\pi b_3}{n_2}} \\ e^{\frac{i2\pi b_4}{n_2}} \ea \right) 
\diag\left(\ba{c} e^{\frac{i2\pi c_1}{n_3}} \\ e^{\frac{i2\pi c_2}{n_3}} \\ e^{\frac{i2\pi c_3}{n_3}} \\ e^{\frac{i2\pi c_4}{n_3}} \ea \right) 
= \diag\left(\ba{c} e^{i2\pi (\frac{a_1}{n_1}+\frac{b_1}{n_2}+\frac{c_1}{n_3})} \\ e^{i2\pi (\frac{a_2}{n_1}+\frac{b_2}{n_2}+\frac{c_2}{n_3})} \\ e^{i2\pi (\frac{a_3}{n_1}+\frac{b_3}{n_2}+\frac{c_3}{n_3})} \\ e^{i2\pi (\frac{a_4}{n_1}+\frac{b_4}{n_2}+\frac{c_4}{n_3})} \ea \right)~~,
\eeq
where $\left(\sum_{i=1}^{4}a_i\right)\bmod{n_1}=0$, $\left(\sum_{i=1}^{4}b_i\right)\bmod{n_2}=0$ and $\left(\sum_{i=1}^{4}c_i\right)\bmod{n_3}=0$. The zero sum conditions are due to the Calabi-Yau condition manifested in the orbifold and the $\det=1$ property of $SU(4)$. We introduce notation to rewrite \eref{e37b} as
\beql{e37b}
\omega^{(\{a_i\},\{b_i\},\{c_i\})}
~=~
\omega^{(a_1,a_2,a_3,a_4)}\omega^{(b_1,b_2,b_3,b_4)}\omega^{(c_1,c_2,c_3,c_4)}
~=~
\omega^{((a_1,a_2,a_3,a_4),(b_1,b_2,b_3,b_4),(c_1,c_2,c_3,c_4))}~~.
\eeq \\

For the element $\omega^{(\{a_i\},\{b_i\},\{c_i\})}\in R_{(n_1,n_2,n_3)}$ to be also a generator of the representation $R_{(n_1,n_2,n_3)}$, it has to fulfill $\gcd{(n_1,\{a_i\})}=1$, $\gcd{(n_2,\{b_i\})}=1$ and $\gcd{(n_3,\{c_i\})}=1$. In addition, the identity operator is given by the period $(\omega^{(\{a_i\},\{b_i\},\{c_i\})})^{N}=1$, and by the Calabi-Yau condition $\mbox{det}(\omega^{(\{a_i\},\{b_i\},\{c_i\})})=1$.\\

The coordinate action of the generator $\omega^{(\{a_i\},\{b_i\},\{c_i\})}$ of the representation $R_{(n_1,n_2,n_3)}$ is defined as
\beql{e38}
\omega^{(\{a_i\},\{b_i\},\{c_i\})}~~:~~\bvec z_1\\z_2\\z_3\\z_4 \evec \mapsto \omega^{(\{a_i\},\{b_i\},\{c_i\})} \bvec z_1\\z_2\\z_3\\z_4 \evec =
\bvec 
z_1 ~e^{i2\pi (\frac{a_1}{n_1}+\frac{b_1}{n_2}+\frac{c_1}{n_3})} \\ 
z_2 ~e^{i2\pi (\frac{a_2}{n_1}+\frac{b_2}{n_2}+\frac{c_2}{n_3})} \\
z_3 ~e^{i2\pi (\frac{a_3}{n_1}+\frac{b_3}{n_2}+\frac{c_3}{n_3})} \\
z_4 ~e^{i2\pi (\frac{a_4}{n_1}+\frac{b_4}{n_2}+\frac{c_4}{n_3})}
\evec~~.
\eeq\\

The dual to the generator $\omega^{(\{a_i\},\{b_i\},\{c_i\})}$ of the representation $R_{(n_1,n_2,n_3)}$ is the orbifold action, now being a $3\times 4$ matrix and generating the representation $\tilde{R}_{(n_1,n_2,n_3)}$ with $\gcd{(n_1,\{a_i\})}=1$, $\gcd{(n_2,\{b_i\})}=1$ and $\gcd{(n_3,\{c_i\})}=1$.\\

%There is an isomorphism
%\beal{e39}
%\mathcal{R}~~:~~ R_{(n_1,n_2,n_3)}~&\rightarrow&~\tilde{R}_{(n_1,n_2,n_3)} \nn\\
%\omega^{(\{a_i\},\{b_i\},\{c_i\})}~&\mapsto&~((a_1,a_2,a_3,a_4),(b_1,b_2,b_3,b_4),(c_1,c_2,c_3,c_4)) ~~.
%\eea
For the case when $\gcd{(n_1,\{a_i\})}\neq 1$, $\gcd{(n_2,\{b_i\})}\neq 1$ and/or $\gcd{(n_3,\{c_i\})}\neq 1$, $((a_1,a_2,a_3,a_4),(b_1,b_2,b_3,b_4),(c_1,c_2,c_3,c_4))$ is not a generator of the representation $\tilde{R}_{(n_1,n_2,n_3)}$ of $\Gamma_{N}=\mathbb{Z}_{n_1}\times\mathbb{Z}_{n_2}\times\mathbb{Z}_{n_3}$ and is therefore not an orbifold action of $\mathbb{C}^{4}/\mathbb{Z}_{n_1}\times\mathbb{Z}_{n_2}\times\mathbb{Z}_{n_3}$.\\

Let the set of all possible orbifold actions at orbifold group order $N$ be given by
\beal{e40}
\mathcal{A}_{N=n_1 n_2 n_3}=
\left\{
\left(
\ba{cccccc} 
(&a_1,&a_2,&a_3,&a_4&) \\ 
(&b_1,&b_2,&b_3,&b_4&) \\ 
(&c_1,&c_2,&c_3,&c_4&) 
\ea
\right)
~~\Bigg|~~
\ba{ll} 
\left(\sum_{i=1}^{4}a_i\right)\bmod{n_1}=0~,
&~~~~\gcd{(n_1,\{a_i\})}=1~,\\
\left(\sum_{i=1}^{4}b_i\right)\bmod{n_2}=0~,
&~~~~\gcd{(n_2,\{b_i\})}=1~,\\
\left(\sum_{i=1}^{4}c_i\right)\bmod{n_3}=0~,
&~~~~\gcd{(n_3,\{c_i\})}=1~~\\
\ea
\right\}~~.\nn\\
\eea
The set of all orbifold actions is divided into equivalence classes $[A_k]$ such that the union $\cup_{i=1}^{N_E}[A_i]=\mathcal{A}_{N}$,
where $N_E$ is the number of equivalence classes in $\mathcal{A}_{N}$. The orbifold actions $A_{l}\in[A_{k}]$ and $A_{m}\in[A_{k}]$, which are both in the same equivalence class $[A_{k}]$, are generators of equivalent representations $\tilde{R}_{(n_1,n_2,n_3)}(A_{l})$ and $\tilde{R}_{(\tilde{n}_1,\tilde{n}_2,\tilde{n}_3)}(A_{m})$ respectively with $N=n_1 n_2 n_3=\tilde{n}_1 \tilde{n}_2 \tilde{n}_3$. The representations $\tilde{R}_{(n_1,n_2,n_3)}(A_{l})$ and $\tilde{R}_{(\tilde{n}_1,\tilde{n}_2,\tilde{n}_3)}(A_{m})$ are equivalent up to a permutation of the complex coordinates of $\mathbb{C}^{4}$. \\

An orbifold action $A_{k}$ of the orbifold $\mathbb{C}^{4}/\Gamma_{N}$ has three components corresponding to the three rows of the $3\times 4$ orbifold action matrix. The three components are $A_{k}^{(n_1)}=(a_1,a_2,a_3,a_4)$, $A_{k}^{(n_2)}=(b_1,b_2,b_3,b_4)$ and $A_{k}^{(n_3)}=(c_1,c_2,c_3,c_4)$ for $A_{k}=(A_{k}^{(n_1)},A_{k}^{(n_2)},A_{k}^{(n_3)})$. Given $A_{k}$ is the generator of the representation $\tilde{R}_{(n_1,n_2,n_3)}(A_{k})$ of the orbifold group $\Gamma_{N}=\mathbb{Z}_{n_1}\times\mathbb{Z}_{n_2}\times\mathbb{Z}_{n_3}$, its components $A_{k}^{(n_1)}=(a_1,a_2,a_3,a_4)$, $A_{k}^{(n_2)}=(b_1,b_2,b_3,b_4)$ and $A_{k}^{(n_3)}=(c_1,c_2,c_3,c_4)$ are each generators of the representations $\tilde{R}_{n_1}(A_{k}^{(n_1)})$, $\tilde{R}_{n_2}(A_{k}^{(n_2)})$ and $\tilde{R}_{n_3}(A_{k}^{(n_3)})$ of the groups $\mathbb{Z}_{n_1}$, $\mathbb{Z}_{n_2}$ and $\mathbb{Z}_{n_3}$ respectively. For the case when $n_{i}=1$ with $i\in\{1,2,3\}$, then it is more practical to use the non trivial component of an orbifold action and its representation rather than the orbifold action itself.\\

%---------------------------
\subsection{Equivalence of Orbifold Actions \label{s4} \label{sc4p2}}

As for all lower dimensional spaces, two orbifold actions that are related by a permutation of the $\mathbb{C}^{4}$ coordinates $\{z_1,z_2,z_3,z_4\}$ are equivalent. Permutations in the coordinates correspond to permutations of columns in the orbifold action matrix.\\

For orbifolds where $n_1\neq 1$ and $n_2,n_3=1$, such that the orbifold is of the form $\mathbb{C}^{4}/\mathbb{Z}_{n_1}$ with $N=n_1$, the effective action is the component $A_{k}^{(n_1)}=(a_1,a_2,a_3,a_4)$. Accordingly, for $\mathbb{C}^{4}/\mathbb{Z}_{n_1}$ two actions are said to be equivalent if their non trivial component actions satisfy the relation
\beql{e42}
(a_1,a_2,a_3,a_4)=m(\tilde{a}_1,\tilde{a}_2,\tilde{a}_3,\tilde{a}_4)~\bmod{N}  ~~,
\eeq
with $m\in\mathbb{Z}$, $1<m<n_1=N$ and $\mbox{gcd}(m,N)=1$.\\

To cover all cases of equivalence, there is a need to generalize the discussion of equivalence. There are in fact $4$ categories of equivalence that have to be considered in $\mathbb{C}^{4}$: actions of $\mathbb{C}^{4}/\mathbb{Z}_{n_i}$ that are equivalent to actions of $\mathbb{C}^{4}/\mathbb{Z}_{n_j}$, actions of $\mathbb{C}^{4}/\mathbb{Z}_{n_i}\times\mathbb{Z}_{n_j}$ that are equivalent to actions of $\mathbb{C}^{4}/\mathbb{Z}_{n_k}$, actions of $\mathbb{C}^{4}/\mathbb{Z}_{n_i}\times\mathbb{Z}_{n_j}\times\mathbb{Z}_{n_k}$ that are equivalent to actions of $\mathbb{C}^{4}/\mathbb{Z}_{n_s}\times\mathbb{Z}_{n_t}$, or actions of $\mathbb{C}^{4}/\mathbb{Z}_{n_i}\times\mathbb{Z}_{n_j}\times\mathbb{Z}_{n_k}$ that are equivalent to actions of $\mathbb{C}^{4}/\mathbb{Z}_{n_s}$. All these possibilities can be covered by an extension of the tools that have been introduced for orbifolds of $\mathbb{C}^{3}$ in Section \sref{s1}.\\

Let the orbifold action $A_{k}$, which is a generator of a representation of the orbifold group $\Gamma_{N}$, be written in terms of its three components, $A_{k}=(A_{k}^{(n_1)},A_{k}^{(n_2)},A_{k}^{(n_3)})$. Each component $A_{k}^{(n_1)}$, $A_{k}^{(n_2)}$ and $A_{k}^{(n_3)}$ is a generator of a representation $\tilde{R}_{n_1}(A_{k}^{(n_1)})$, $\tilde{R}_{n_2}(A_{k}^{(n_2)})$ and $\tilde{R}_{n_3}(A_{k}^{(n_3)})$ corresponding to the groups $\mathbb{Z}_{n_1}$, $\mathbb{Z}_{n_2}$ and $\mathbb{Z}_{n_3}$ respectively. From the representations generated by the components of the orbifold action, we wish to obtain the representation of the orbifold group $\Gamma_{N}$ generated by the orbifold action $A_{k}$ itself. For this purpose we define the map
\beal{e43}
\rho_{abc}^{n_1,n_2,n_3}~~:~~
(\tilde{R}_{n_1},\tilde{R}_{n_2},\tilde{R}_{n_3}) 
& ~~\rightarrow~~ & \bar{R}_{(n_1,n_2,n_3)}\left(\rho_{111}^{n_1,n_2,n_3}(A_{k})\right)
\nn\\
(A^{(n_1)}_{k},A_{k}^{(n_2)},A_{k}^{(n_3)}) 
& ~~\mapsto~~ &
(a n_2 n_3 A^{(n_1)}_{k}+b n_1 n_3 A^{(n_2)}_{k}+c n_1 n_2 A^{(n_3)}_{k})\bmod{N}~~,\nn\\
\eea
where $N=n_1 n_2 n_3$, and $1\leq a \leq n_1$, $1\leq b \leq n_2$ and $1\leq c \leq n_3$. The map $\rho_{abc}^{n_1,n_2,n_3}$ maps all elements of the component representations $\tilde{R}_{n_1}(A_{k}^{(n_1)})$, $\tilde{R}_{n_2}(A_{k}^{(n_2)})$ and $\tilde{R}_{n_3}(A_{k}^{(n_3)})$ into single elements of the representation $\bar{R}_{(n_1,n_2,n_3)}$ of the orbifold group $\Gamma_{N}$. It is also important to note that given $A_{k}^{(n_1)}$, $A_{k}^{(n_2)}$ and $A_{k}^{(n_3)}$ are the generators of the component representations $\tilde{R}_{n_1}(A_{k}^{(n_1)})$, $\tilde{R}_{n_2}(A_{k}^{(n_2)})$ and $\tilde{R}_{n_3}(A_{k}^{(n_3)})$ respectively, the map $\rho_{111}^{n_1,n_2,n_3}$ maps the component generators into the generator $\rho_{111}^{n_1,n_2,n_3}(A_{k})$ of the orbifold group representation $\bar{R}_{(n_1,n_2,n_3)}$.\\

In terms of orbifold operators $\omega^{(\{a_i\},\{b_i\},\{c_i\})}$, the action of the map in \eref{e43} can be identified as
\beal{e44}
\rho_{abc}^{n_1,n_2,n_3}~~:~~ 
\bvec 
\omega^{(a_1,a_2,a_3,a_4)} \\ 
\omega^{(b_1,b_2,b_3,b_4)} \\ 
\omega^{(c_1,c_2,c_3,c_4)} \evec
~~&\mapsto&~~ 
\left(\omega^{(a_1,a_2,a_3,a_4)}\right)^{a}
\left(\omega^{(b_1,b_2,b_3,b_4)}\right)^{b}
\left(\omega^{(c_1,c_2,c_3,c_4)}\right)^{c} \nn\\
&& = \bvec 
e^{i2\pi (\frac{a n_2 n_3 a_1+b n_1 n_3 b_1+c n_1 n_2 c_1}{N})} \\ 
e^{i2\pi (\frac{a n_2 n_3 a_2+b n_1 n_3 b_2+c n_1 n_2 c_2}{N})} \\ 
e^{i2\pi (\frac{a n_2 n_3 a_3+b n_1 n_3 b_3+c n_1 n_2 c_3}{N})} \\
e^{i2\pi (\frac{a n_2 n_3 a_4+b n_1 n_3 b_4+c n_1 n_2 c_4}{N})} 
\evec~~~.
\eea\\

Accordingly, for given two orbifold actions $A_{1}$ and $A_{2}$, $A_{1}\sim A_{2}$ if and only if $\bar{R}_{(n_1,n_2,n_3)}(\rho_{111}^{n_1,n_2,n_3}(A_{1}))\sim\bar{R}_{(\tilde{n}_{1},\tilde{n}_{2},\tilde{n}_{3})}(\rho_{111}^{\tilde{n}_{1},\tilde{n}_{2},\tilde{n}_{3}}(A_{2}))$ where $N=n_1 n_2 n_3 = \tilde{n}_{1}\tilde{n}_{2}\tilde{n}_{3}$ up to permutations of the complex coordinates of $\mathbb{C}^{4}$. Consequently, the set of all orbifold actions $\mathcal{A}_{N}$ of the orbifold $\mathbb{C}^{4}/\Gamma_{N}$ can be separated into equivalence classes of the form
\beql{e47}
[A_k]~~=~~\left\{
\tilde{A}_{k} ~\Bigg|~ A_{l}\sim A_{k}~~\Leftrightarrow~~
\bar{R}_{(n_1,n_2,n_3)}(\rho_{111}^{n_1,n_2,n_3}(A_{k}))
~\sim~
\bar{R}_{(\tilde{n}_{1},\tilde{n}_{2},\tilde{n}_{3})}(\rho_{111}^{\tilde{n}_{1},\tilde{n}_{2},\tilde{n}_{3}}(A_{l}))
\right\}~~,
\eeq
where $N=n_1 n_2 n_3 = \tilde{n}_{1}\tilde{n}_{2}\tilde{n}_{3}$. The set of all orbifold actions of a given orbifold group of order $N$, $\mathcal{A}_{N}$, can be rewritten as $\bigcup_{i=1}^{N_E}{[A_i]}=\mathcal{A}_{N}$ with $N_E$ being the number of equivalence classes in $\mathcal{A}_{N}$.\\

An alternative map to \eref{e29b} can be defined as
\beal{e46}
\tilde{\rho}^{n_1,n_2,n_3}_{abc}~~:~~(\tilde{R}_{n_1}(A^{(n_1)}_{k}),\tilde{R}_{n_2}(A^{(n_2)}_{k}),\tilde{R}_{n_3}(A^{(n_3)}_{k})) ~~&\rightarrow&~~ \tilde{R}_{(n_1,n_2)}(A_k) \nn\\
(A_{k}^{(n_1)},A_{k}^{(n_2)},A_{k}^{(n_3)})
~&\mapsto&~~
\bvec
(a A_{k}^{(n_1)})\bmod{n_1}\\
(b A_{k}^{(n_2)})\bmod{n_2}\\
(c A_{k}^{(n_3)})\bmod{n_3}
\evec ~~,\nn\\
\eea
where the representation $\tilde{R}_{(n_1,n_2)}(A_k)$ of $\Gamma_{N}=\mathbb{Z}_{n_1}\times\mathbb{Z}_{n_2}\times\mathbb{Z}_{n_3}$ is generated by the orbifold action $A_{k}=(A_{k}^{(n_1)},A_{k}^{(n_2)},A_{k}^{(n_3)})$ and not by $\rho^{n_1,n_2,n_3}_{111}(A_{k})$ as it is the case for the dual representation $\bar{R}_{(n_1,n_2)}(\rho^{n_1,n_2,n_3}_{111}(A_{k}))$. Different to the representation $\bar{R}_{(n_1,n_2)}(\rho^{n_1,n_2,n_3}_{111}(A_{k}))$, the representation $\tilde{R}_{(n_1,n_2)}(A_k)$ is not used to test equivalence between orbifold actions.
\\

%---------------------------
\subsection{Equivalence of Toric Tetrahedra \label{s5} \label{sc4p3}}

The notion of lattice tetrahedra with volume $N$ as toric diagrams of orbifolds $\mathbb{C}^{4}/\Gamma_{N}$ was introduced in \cite{HananyZaff08}. It can be shown that the set of topologically inequivalent lattice tetrahedra of a given volume $N$ are isomomorphic to the set of inequivalent orbifold actions of $\mathbb{C}^{4}/\Gamma_{N}$ with order $N$.\\

\textbf{Topology.} A lattice tetrahedron $\sigma$ is defined by $D=4$ corner points $I_{0}=\{v_1,v_2,v_3,v_4\}$ where each lattice point is identified by Cartesian coordinates $v_i=(x^{i}_1,x^{i}_2,x^{i}_3)\in\mathbb{Z}^{3}$ with $i=1,\dots 4=D$. Lattice points enclosed by the boundary of the tetrahedron are internal and form the set $I_3$. Lattice points on the edges excluding the corner points $I_0=\{v_i\}$ are in $I_1$, and points on faces of the tetrahedron excluding points in $I_0$ and $I_1$ form the set $I_2$. Accordingly, the set of points $\{w_k\}=I_0 \cup I_1 \cup I_2 \cup I_3$ making up a toric diagram tetrahedron $\sigma$ with $k=1,\dots,|I_0 \cup I_1 \cup I_2 \cup I_3|$ can be used to identify the topological character of the lattice tetrahedron.\\

Every point $w_k\in I_0 \cup I_1 \cup I_2 \cup I_3$ divides the tetrahedron into $D=4$ subtetrahedra with volumes $\lambda_{k1}$, $\lambda_{k2}$, $\lambda_{k3}$ and $\lambda_{k4}$. The barycentric coordinates of a point $w_k$ are defined accordingly as,
\beql{e72}
w_k~=~\frac{1}{N}\left(\lambda_{k1},\lambda_{k2},\lambda_{k3},\lambda_{k4}\right)~~,
\eeq
where $N=\sum_{i=1}^{4}{\lambda_{ki}}$. From \eref{e72}, it is possible to define the topological character of the toric diagram tetrahedron $\sigma$ with corner points $\{v_i\}$ as the set of the barycentric coordinates of the lattice points $w_k\in I_0 \cup I_1 \cup I_2 \cup I_3$,
\beql{e73}
\tau~~=~~\left\{
\frac{1}{N}\left(\lambda_{k1},\lambda_{k2},\lambda_{k3},\lambda_{k4}\right) ~~\Big|~~
w_k\in I_0 \cup I_1 \cup I_2 \cup I_3
\right\}~~,
\eeq
in analogy to the definition of the topological character of a toric diagram triangle in \eref{e61}. The topological character $\tau$ can be considered as a $k\times 4$ matrix, where the $k$ rows are the barycentric coordinates of the points $w_k\in I_0 \cup I_1 \cup I_2 \cup I_3$, and the $4$ columns correspond to the $4$ barycentric coordinate axes $\{\hat{v}_1,\hat{v}_2,\hat{v}_3,\hat{v}_4\}$.\\

\textbf{Equivalence.} Two toric diagram tetrahedra $\sigma_1$ and $\sigma_2$ with equal volume $N$ are considered to be equivalent if their corresponding topological characters $\tau_1=\tau(\sigma_1)$ and $\tau_2=\tau(\sigma_2)$ are equivalent up to permutations of the barycentric coordinate axes. Permutations of the barycentric coordinates axes correspond to permutations of the columns of the topological character matrix as seen in \eref{e62} for $\mathbb{C}^{3}$ orbifolds with now $i=1,\dots 4$ and $N_{IB}=|I_0 \cup I_1 \cup I_2 \cup I_3|$.\\

As in the discussion on toric diagram triangle equivalence in Section \sref{s3} for $\mathbb{C}^{3}$ orbifolds, it is interesting to observe the effect on the corner points $I_0=\{v_i\}$ of the toric diagram tetrahedron due to permutation of the barycentric coordinate axes $\{\hat{v}_1,\hat{v}_2,\hat{v}_3,\hat{v}_4\}$ for $\mathbb{C}^{4}$ orbifolds. The barycentric coordinates of the $4$ corner points of the toric diagram tetrahedron are
\beql{e74}
v_1=(0,0,0,1)~~,~~v_2=(0,0,1,0)~~,~~v_3=(0,1,0,0)~~,~~v_4=(1,0,0,0)~~,
\eeq
independent of the toric diagram tetrahedron volume $N$. It can be seen that the permutation of the barycentric coordinates axes is a permutation of the corner points $I_0$ of the toric diagram tetrahedron. As seen for toric diagram triangles corresponding to orbifolds of $\mathbb{C}^{3}$, the corner points for the tetrahedron in barycentric coordinates in \eref{e74} can be considered as unit vectors along the corresponding barycentric coordinate axes $\{\hat{v}_1,\hat{v}_2,\hat{v}_3,\hat{v}_4\}$.\\

A link can be now drawn to the orbifold action discussion for $\mathbb{C}^{4}/\Gamma_{N}$ orbifolds in Section \sref{s4}. As stated in Section \sref{s4}, two orbifold actions are equivalent if they are related by a permutation of the coordinates $\{z_1,z_2,z_3,z_4\}$ of $\mathbb{C}^{4}$. The correspondence to the invariance of the set of corner points $I_{0}$ can be made explicit via the map,
\beql{e75}
T~~:~~\{z_1,z_2,z_3,z_4\}~~\rightarrow~~\{\hat{v}_1,\hat{v}_2,\hat{v}_3,\hat{v}_4\}~~,
\eeq
analogous to the $\mathbb{C}^{3}$ mapping in \eref{e64}.\\

\begin{figure}[ht!]
\begin{center}
\includegraphics[totalheight=6.5cm]{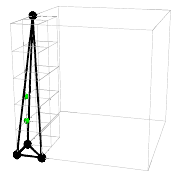}
\includegraphics[totalheight=6.5cm]{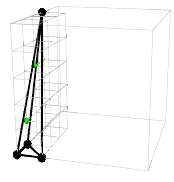}
\caption{Two toric diagrams corresponding to the orbifold $\mathbb{C}^{4}/\mathbb{Z}_{5}$ with volume $N=5$. Lattice points on the faces of the tetrahedra are colored green ($I_2$).}
  \label{f6}
 \end{center}
\end{figure}

\textbf{Example.} Let us consider an example of two toric diagram tetrahedra and their equivalence relation. We choose the volume for the toric diagram tetrahedra to be $N=5$ corresponding to the orbifold $\mathbb{C}^{4}/\mathbb{Z}_{5}$ and pick the two tetrahedra shown in \fref{f6}.

The first step is to calculate the barycentric coordinates of the points $w_k\in I_1 \cup I_2 \cup I_3$. The barycentric coordinates give the corresponding topological characters of the tetrahedra in \fref{f6},
\beal{e76}
\tau_1 ~~&=&~~ \left\{
(1,0,0,0)~,~
(0,1,0,0)~,~
(0,0,1,0)~,~
(0,0,0,1)~,~
\left(\frac{1}{5},\frac{3}{5},0,\frac{1}{5}\right)~,~
\left(\frac{2}{5},\frac{1}{5},0,\frac{2}{5}\right)
\right\}
\nn\\
\tau_2 ~~&=&~~ \left\{
(1,0,0,0)~,~
(0,1,0,0)~,~
(0,0,1,0)~,~
(0,0,0,1)~,~
\left(\frac{1}{5},\frac{2}{5},0,\frac{2}{5}\right)~,~
\left(\frac{3}{5},\frac{1}{5},0,\frac{1}{5}\right)
\right\}~~.\nn\\
\eea
It can be seen now that by a permutation of the barycentric coordinate axes, $(\hat{v}_1,\hat{v}_2,\hat{v}_3,\hat{v}_4)\mapsto(\hat{v}_2,\hat{v}_1,\hat{v}_3,\hat{v}_4)$, the two topological characters in \eref{e76} are equivalent, $\tau_1 \sim\tau_2$.\\

\begin{figure}[ht!]
\begin{center}
\includegraphics[totalheight=6.5cm]{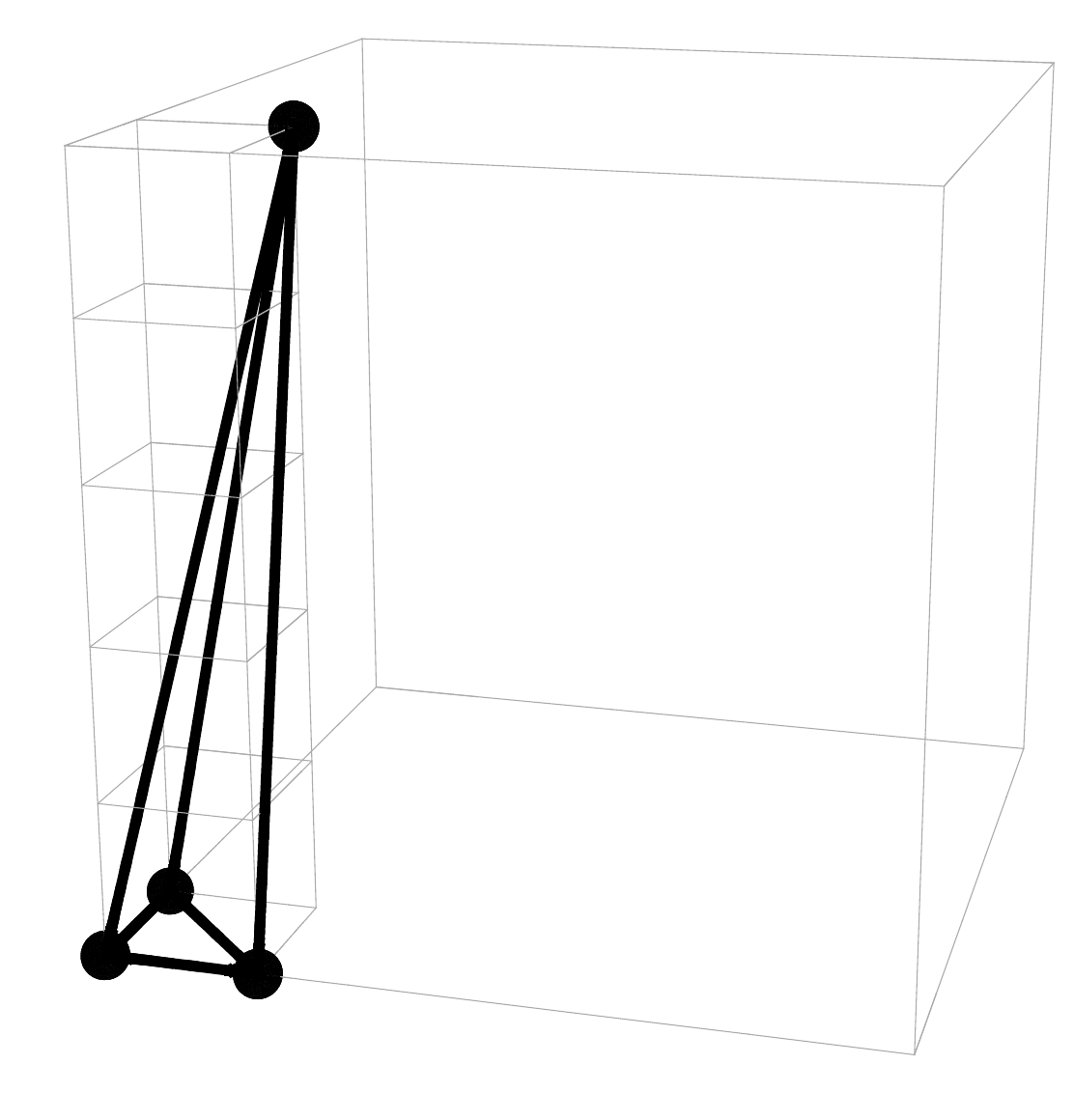}
\includegraphics[totalheight=6.5cm]{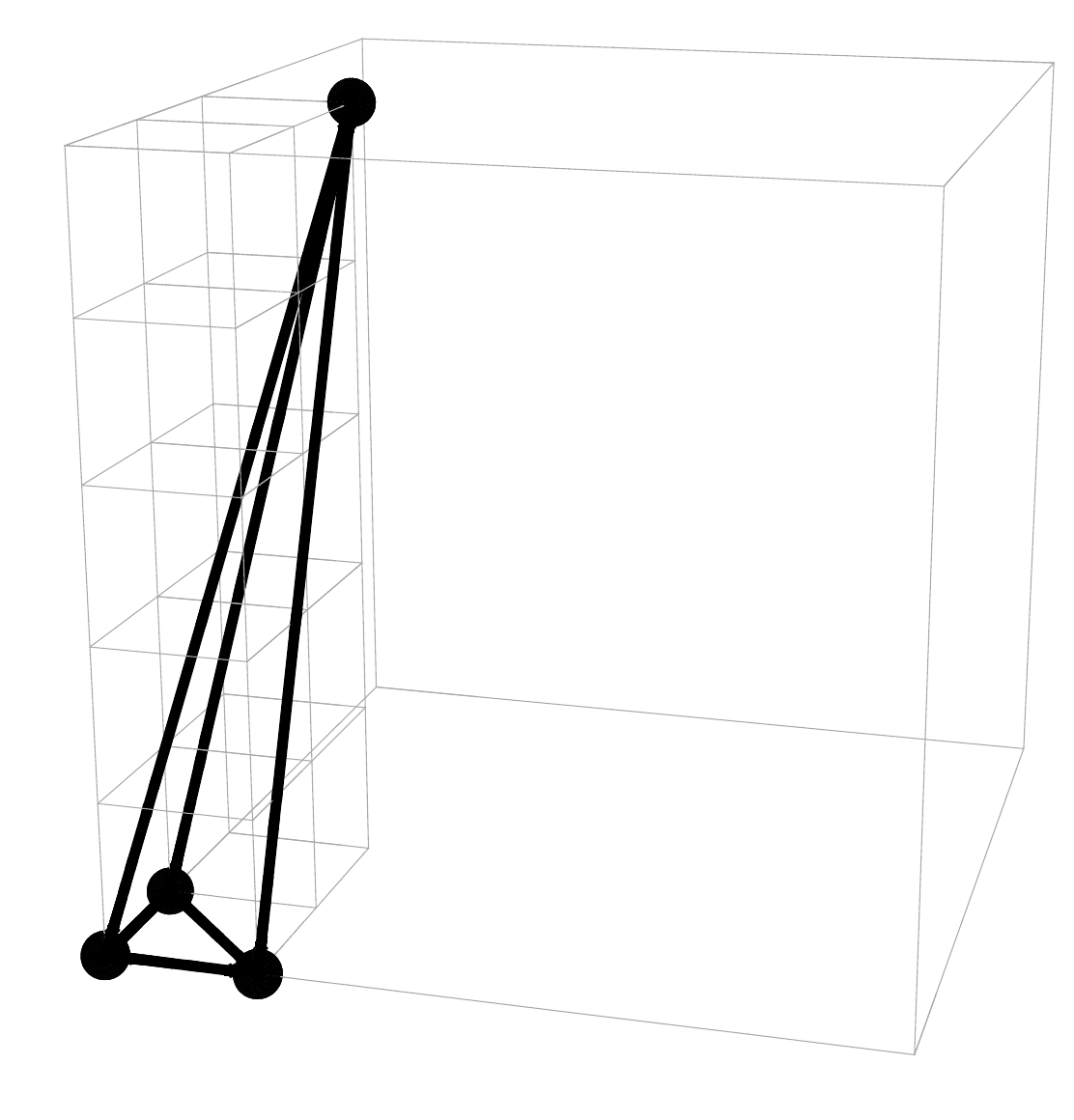}
\caption{Two toric diagrams corresponding to the orbifold $\mathbb{C}^{4}/\mathbb{Z}_{5}$ with volume $N=5$. These two tetrahedra have only four corner points, and no points which are internal or on faces or edges.}
  \label{f7}
 \end{center}
\end{figure}

\textbf{Scaling.} It is of interest to consider two more toric diagram tetrahedra of the same volume, $N=5$, as shown in \fref{f7}. Again these tetrahedra correspond to the orbifold $\mathbb{C}^{4}/\mathbb{Z}_{5}$, but are dual to orbifold actions of $\mathbb{C}^{4}/\mathbb{Z}_{5}$ that are inequivalent to the action dual to the tetrahedra in \fref{f6} as seen clearly by the absence of face points ($I_2$) in \fref{f7}.
The two tetrahedra in \fref{f7} appear to be equivalent, consisting both only of $4$ corner points and no additional topological points that are internal ($I_3$) or on faces ($I_2$) or edges ($I_1$) of the tetrahedron. The $4$ corner points inevitably give topological characters which turn out to be the same,
\beql{e77}
\tau_1~=~\tau_2~=~\left\{
(1,0,0,0)~,~
(0,1,0,0)~,~
(0,0,1,0)~,~
(0,0,0,1)
\right\}~~.
\eeq
This leads to the na\"ive conclusion that the toric diagram tetrahedra in \fref{f7} are equivalent. In fact, it turns out that the two tetrahedra are inequivalent and correspond to two inequivalent orbifold actions of $\mathbb{C}^{4}/\mathbb{Z}_{5}$.\\

It appears here that there is a need to introduce a complementary method to measure the `hidden' topological character of an `empty' lattice tetrahedron with $I_1,I_2,I_3=\emptyset$. In fact, some topologically important points have been left out in the character in \eref{e77}. To make these points visible for testing equivalence, we use a refinement of the toric diagram tetrahedron which we call the \textit{scaling} of the tetrahedron or equivalently the scaling of the $\mathbb{Z}^{3}$ lattice in which the toric diagram tetrahedron is embedded.\\

\begin{figure}[ht!]
\begin{center}
\includegraphics[totalheight=7cm]{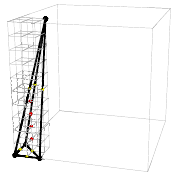}
\includegraphics[totalheight=7cm]{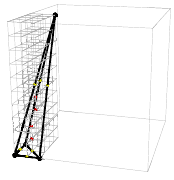}
\caption{Two toric diagrams corresponding to the orbifold $\mathbb{C}^{4}/\mathbb{Z}_{5}$ with volume $N=5$ and scaling $s=2$. These two tetrahedra are scaled from the tetrahedra in Figure 7 which has a unit scaling of $s=1$. Internal lattice points are colored red ($I_3$) while lattice points on edges are colored yellow ($I_1$).}
  \label{f8}
 \end{center}
\end{figure}

Given a toric diagram tetrahedron $\sigma$ with a corner point $(0,0,0)\in\{v_i\}$ in Cartesian coordinates, scaling can be considered as a map,
\beql{e78}
f_{s}~~:~~\{v_i\}~~\mapsto~~\{s v_i\}~=~\{(s x^{i}_1,s x^{i}_2,s x^{i}_3)\}~~,
\eeq
where $(x^{i}_1,x^{i}_2,x^{i}_3)$ are the Cartesian coordinates of a corner point $v_i\in\sigma$ and $s\in\mathbb{Z}^{+}$. Until now, scaling has been always unity, $s=1$, as it is the case for the tetrahedra in \fref{f7}. We can expect that for $s>1$, $I_i\neq\emptyset$ for some $i\in\{1,2,3\}$, as it is the case for a scaling of $s=2$ for the tetrahedra in \fref{f7} as shown in \fref{f8}.

As shown in \fref{f8}, the scaling by $s=2$ of the two tetrahedra in \fref{f7} has the result of additional points appearing on the edges and inside the tetrahedra with $|I_3|=4$ (shown red) and $|I_1|=6$ (shown yellow) for both tetrahedra. These points refine the initial topological characters of the two tetrahedra for a scaling of $s=1$ in \eref{e77} to the new scaled topological characters,
\beal{e79}
\tau_1~~=~~\Bigg\{&&
(1,0,0,0),~
(0,1,0,0),~
(0,0,1,0),~
(0,0,0,1),~\nn\\ &&
\left(0,0,\frac{1}{2},\frac{1}{2}\right),~
\left(0,\frac{1}{2},0,\frac{1}{2}\right),~
\left(0,\frac{1}{2},\frac{1}{2},0\right),~
\left(\frac{1}{2},0,0,\frac{1}{2}\right),~
\left(\frac{1}{2},0,\frac{1}{2},0\right),~
\left(\frac{1}{2},\frac{1}{2},0,0\right),~\nn\\ &&
\left(\frac{1}{10},\frac{2}{5},\frac{2}{5},\frac{1}{10}\right),~
\left(\frac{1}{5},\frac{3}{10},\frac{3}{10},\frac{1}{5}\right),~
\left(\frac{3}{10},\frac{1}{5},\frac{1}{5},\frac{3}{10}\right),~
\left(\frac{2}{5},\frac{1}{10},\frac{1}{10},\frac{2}{5}\right)~~
\Bigg\}~~~, \nn\\
\tau_2~~=~~\Bigg\{&&
(1,0,0,0)~,~
(0,1,0,0)~,~
(0,0,1,0)~,~
(0,0,0,1)~,~\nn\\ &&
\left(0,0,\frac{1}{2},\frac{1}{2}\right),~
\left(0,\frac{1}{2},0,\frac{1}{2}\right),~
\left(0,\frac{1}{2},\frac{1}{2},0\right),~
\left(\frac{1}{2},0,0,\frac{1}{2}\right),~
\left(\frac{1}{2},0,\frac{1}{2},0\right),~
\left(\frac{1}{2},\frac{1}{2},0,0\right),~\nn\\ &&
\left(\frac{1}{10},\frac{3}{10},\frac{2}{5},\frac{1}{5}\right),~
\left(\frac{1}{5},\frac{1}{10},\frac{3}{10},\frac{2}{5}\right),~
\left(\frac{3}{10},\frac{2}{5},\frac{1}{5},\frac{1}{10}\right),~
\left(\frac{2}{5},\frac{1}{5},\frac{1}{10},\frac{3}{10}\right)~~
\Bigg\}~~.
\eea
From the refined characters in \eref{e79}, one can see that the last $4$ character elements corresponding to the $4$ internal points in \fref{f8} are not equivalent, even after a permutation of the barycentric coordinate axes $\{\hat{v}_1,\hat{v}_2,\hat{v}_3,\hat{v}_4\}$. Accordingly, the topological characters of the toric diagram tetrahedra in \fref{f8} are not equivalent, and hence the toric diagram tetrahedra in \fref{f8} \textit{and} the unit scaled tetrahedra in \fref{f7} are not equivalent. As a result, the corresponding orbifold actions of $\mathbb{C}^{4}/\Gamma_5$ are expected to be not equivalent as well since scaling of toric diagram tetrahedra has no effect on the correspondence between orbifold actions and the toric lattice tetrahedra.\\

It is important to consider the optimal scaling $s$ of a given toric diagram tetrahedron for obtaining its corresponding topological character. For the example in \fref{f8}, we expect that the scaling to $s>2$ leads to the same conclusion that the two toric diagram tetrahedra are inequivalent. A greater scaling by $s>2$ would only increase the size of the inequivalent characters $\tau_1$ and $\tau_2$. Accordingly, $s=2$ can be considered as an optimal scaling for \fref{f7} to identify their inequivalence.\\

In general, let us consider the tetrahedron point sets $I_{1}(f_{s_1}(\sigma))$, $I_{2}(f_{s_2}(\sigma))$ and $I_{3}(f_{s_3}(\sigma))$ as functions of a scaled toric diagram tetrahedron $\sigma$ where $s_1,s_2,s_3\in\mathbb{Z}^{+}$ are the scaling parameters for $I_1$, $I_2$ and $I_3$ respectively. Then the optimal scaling coefficients $\{s_i\}$ are defined such that $I_{i}(f_{s_i\rightarrow\min{(s_i)}}(\sigma))\neq\emptyset$, where $\min{(s_i)}$ is the minimum value of $s_i$ with $i=1,\dots,3$. Accordingly, the topological character definition in \eref{e73} can be redefined as
\beql{e80}
\tau~~=~~\left\{
\frac{1}{N}(\lambda_{k1},\lambda_{k2},\lambda_{k3},\lambda_{k4})
~\Big|~
w_k\in I_{i}(f_{s}(\sigma))~,~i=1,2,3
\right\}~~,
\eeq 
where $s=\max{(\{s_i\})}$.
\\

\textbf{Domain.} The set of all toric diagram tetrahedra of volume $N$ with a corner point being the origin and the other corner points being in the positive sector of $\mathbb{Z}^{3}$ is called the domain $D(N)$. The conditions on the domain in $\mathbb{Z}^{3}$ ensure that no two lattice tetrahedra are trivially related under a rotation around the origin of $\mathbb{Z}^{3}$, or a translation.\\

Giving the corner points $v_i$ of the lattice tetrahedra in Cartesian coordinates $(x_1,x_2,x_3)\in\mathbb{Z}^{3}$, the domain can be generated using all transformations that map the unit volume tetrahedron into lattice tetrahedra of volume $N$ in the positive sector of $\mathbb{Z}^{3}$. The transformed tetrahedra require to have one corner point as the origin of $\mathbb{Z}^{3}$. \\

%We define the domain $D(N)$ as a set of all toric diagram tetrahedra of volume $N$ where two tetrahedra in the set are not related by a lattice translation or a rotation around the $\mathbb{Z}^{3}$ origin.\\

%Let the corner points $v_i$ of the lattice tetrahedra be expressed in Cartesian coordinates $(x_1,x_2,x_3)\in\mathbb{Z}^{3}$. Also, every tetrahedron in the domain has a corner point which is the origin of $\mathbb{Z}^{3}$. The domain can be generated by transformations of the unit volume tetrahedron into lattice tetrahedra of volume $N$ where the first corner point remains at the origin of $\mathbb{Z}^{3}$, the second corner point remains at the $x_2$-$x_3$-origin, the third corner point remains in the $x_3$-origin, and the fourth corner point is determined by the remaining translational freedom. All corner points of the tetrahedra are in the positive octant of $\mathbb{Z}^{3}$.\\

The Cartesian coordinates of the four corner points of the unit volume lattice tetrahedron are
\beql{e82}
v_{1}=(0,0,0)~,~v_{2}=(1,0,0)~,~v_{3}=(0,1,0)~,~v_{4}=(0,0,1)~~.
\eeq
The transformation matrices generating the domain have the form
\beql{e83}
M=\left(\ba{ccc}
m_{11} & m_{12} & m_{13} \\
0 & m_{22} & m_{23} \\
0 & 0 & m_{33}
\ea\right)~~,
\eeq
where $\det{M}=m_{11}m_{22}m_{33}=N$ and $0\leq m_{jk} < m_{jj}$ with $m_{jk}\in\mathbb{N}_{0}$. Accordingly, the domain can be given as
\beal{e84}
D(N)~~=~~\Big\{
&& \left\{v_1,v_2,v_3,v_4\right\}=\left\{(0,0,0),(m_{11},0,0),(m_{12},m_{22},0),(m_{13},m_{23},m_{33})\right\} \nn\\
&& ~~|~~
N=m_{11}m_{22}m_{33}~,~~0\leq m_{jk} < m_{jj}~,~~m_{jk}\in\mathbb{N}_{0}
\Big\}~~.\nn\\
\eea
With the definition of an equivalence class of a toric diagram tetrahedron $\sigma$, $[\sigma]$, in \eref{e70}, and using the modified definition of the topological character $\tau$ in \eref{e80}, it is possible to identify the equivalence classes of toric diagram tetrahedra of fixed volume $N$ in the domain $D(N)$ such that the domain can be rewritten as a union of equivalence classes as shown in \eref{e71} for $\mathbb{C}^{3}$ orbifolds. We expect that the count $N_{E}$ of equivalence classes of toric diagram tetrahedra of fixed volume $N$ exactly corresponds to the count of equivalence classes of orbifold actions of $\mathbb{C}^{4}/\Gamma_N$.\\

%%%%%%%%%%%%%%%%%%%%%%%%%%%%%%%%%%%%%%%%%%%%%%%%%%%%%%%%%%%%%%%%%%%%%%%%%%%%%%%%%%%%%%%%%%%%%%%%%%%%%%%%%%%%%%%%%%%%%%%%%%%%

\section{Orbifolds of $\mathbb{C}^{D}$ \label{scd}}

In this section, all lessons learned from the discussions on orbifolds of $\mathbb{C}^{2}$, $\mathbb{C}^{3}$ and $\mathbb{C}^{4}$ are generalized to $\mathbb{C}^{D}/\Gamma_{N}$ orbifolds for the purpose of stating a dimensionally independent definition of equivalence between orbifold actions and their dual lattice $(D-1)$-simplices of hypervolume $N$.
\\
%-----------------------------------------------------------
\subsection{Orbifolds and Orbifold Actions \label{sc5p1}}

$\mathbb{C}^{D}$ is a complex $D$-dimensional space parameterized by the coordinates $\{z_1,\dots,z_D\}$. The orbifold is obtained by taking the quotient with $\Gamma_{N}=\otimes_{j=1}^{D-1}{\mathbb{Z}_{n_j}}\subset SU(D)$ where the orbifold order is $N=\prod_{j=1}^{D-1}{n_j}$. Let us define the set 
\beql{e48}
\mathsf{G}_{N}=\left\{
(n_1,...n_{D-1}) ~\Big|~ N=\prod_{i=1}^{D-1}{n_i}~,~ n_1\leq n_2\leq\dots\leq n_{D-1}
\right\} ~,
\eeq
which is isomorphic to the set of all possible orbifold quotient groups, $\tilde{\mathsf{G}}_N=\{\mathbb{Z}_{n_1}\times\dots\times\mathbb{Z}_{n_{D-1}}\ ~|~ (n_1,\dots,n_{D-1})\in \mathsf{G}_N \}$. The orbifold is then defined as $\mathbb{C}^{D}/\Gamma_{(n_1,\dots,n_{D-1})}$ where $\Gamma_{(n_1,\dots,n_{D-1})}\in\tilde{\mathsf{G}}_N$.\\

Let a representation of the orbifold group $\Gamma_{(n_1,\dots,n_{D-1})}\in\tilde{\mathsf{G}}_N$ be called $R_{(n_1,\dots,n_{D-1})}$ with elements $\omega^{(\{a_{i}^{(n_j)}\})}$, $i=1,\dots,D$, $j=1,\dots,D-1$ and $|R_{(n_1,\dots,n_{D-1})}|=\prod_{j=1}^{D-1}{n_j}=N$. The elements $\omega^{(\{a_{i}^{(n_j)}\})}\in R_{(n_1,\dots,n_{D-1})}$ of the representation are of the form
\beql{e49}
\prod_{j=1}^{D-1}
{
\diag\bvec 
e^{\frac{i2\pi a^{(n_j)}_{1}}{n_j}} \\ 
\vdots \\ 
e^{\frac{i2\pi a^{(n_j)}_{D}}{n_{j}}}
\evec
}
~=~
\diag\bvec
e^{i2\pi \sum_{j=1}^{D-1}\frac{a^{(n_j)}_{1}}{n_j}} \\ 
\vdots \\ 
e^{i2\pi \sum_{j=1}^{D-1}\frac{a^{(n_j)}_{D}}{n_j}} 
\evec
\eeq
with $\left(\sum_{i=1}^{D} a^{(n_j)}_{i}\right)\bmod{n_j}=0$. The zero sum condition is a result of the Calabi-Yau condition on the orbifold singularity and the $\det=1$ property of $SU(D)$.\\

For the element $\omega^{(\{a_{i}^{(n_j)}\})}$ to be also a generator of the representation $R_{(n_1,\dots,n_{D-1})}$, it has to fulfill the conditions $\gcd{(n_j,\{a_{i}^{(n_j)}\})}=1$ for $i=1,\dots,D$ and $j=1,\dots,D-1$. In addition, the identity is defined as $\left(\omega^{(\{a_{i}^{(n_j)}\})}\right)^{N}=1$ and $\mbox{det}\left(\omega^{(\{a_{i}^{(n_j)}\})}\right)=1$ by the manifestation of the Calabi-Yau condition.\\

The orbifold operators $\omega^{(\{a_{i}^{(n_j)}\})}$ act on the coordinates of $\mathbb{C}^{D}$ as a map,
\beql{e50}
\omega^{\left(\{a_{i}^{(n_j)}\}\right)} ~:~
\left( \ba{c} z_1 \\ \vdots \\ z_D \ea \right) ~\mapsto~ \omega^{\left(\{a_{i}^{(n_j)}\}\right)} \left( \ba{c} z_1 \\ \vdots \\ z_D \ea \right)
~=~
\bvec
z_1 e^{i2\pi \sum_{j=1}^{D-1}\frac{a^{(n_j)}_{1}}{n_j}} \\ 
\vdots \\ 
z_D e^{i2\pi \sum_{j=1}^{D-1}\frac{a^{(n_j)}_{D}}{n_j}} 
\evec
~~,\eeq
with $i=1,\dots,D$ and $j=1,\dots,D-1$. This operation reduces to the form we have seen above for $\mathbb{C}^{2}$ in \eref{e2}, $\mathbb{C}^{3}$ in \eref{e15} and $\mathbb{C}^{4}$ in \eref{e38}. By the manifestation of the Calabi-Yau condition, it is possible to express in general $a^{(n_j)}_{D}=n_j-\sum_{i=1}^{D-1}a^{(n_j)}_{i}$ such that for example for $\mathbb{C}^3$, $a^{(n_j)}_3=n_j-a^{(n_j)}_1-a^{(n_j)}_2$.\\

The dual to the generator $\omega^{(\{a_{i}^{(n_j)}\})}$ of the representation $R_{(n_1,\dots,n_{D-1})}$ of $\Gamma_{N}$ is the $(D-1)\times D$ orbifold action  matrix 
\beql{e50b}
A_{k}~~=~~
\left(
\ba{cccccc} 
( & a_{1}^{(n_1)}, & a_{2}^{(n_1)}, & \dots, & a_{D}^{(n_1)} & ) \\ 
( & a_{1}^{(n_2)}, & a_{2}^{(n_2)}, & \dots, & a_{D}^{(n_2)} & ) \\
 & \vdots & \vdots & & \vdots & \\
( & a_{1}^{(n_{D-1})}, & a_{2}^{(n_{D-1})}, & \dots, & a_{D}^{(n_{D-1})} & ) 
\ea 
\right)
\eeq
generating the representation $\tilde{R}_{n_1,\dots,n_{D-1}}$ of $\Gamma_{N}$ with $\gcd{(n_j,\{a_{i}^{(n_j)}\})}=1$. 
%There is an isomorphism 
%\beal{e50c}
%\mathcal{R}~~:~~R_{(n_1,\dots,n_{D-1})}~~&\rightarrow&~~\tilde{R}_{(n_1,\dots,n_{D-1})} \nn\\
%\omega^{(\{a_{i}^{(n_j)}\})}~~&\mapsto&~~A_{k}~~.
%\eea
For the case when for some $j\in\{1,\dots,D-1\}$  $\gcd{(n_j,\{a_{i}^{(n_j)}\})}\neq 1$, then the convention is to call $A_{k}$ in \eref{e50b} not an orbifold action.
\\

Let the set of all orbifold actions of representations $\{\tilde{R}_{(n_1,\dots,n_{D-1})}\}$ of the orbifold group $\Gamma_{N}$ be called $\mathcal{A}_{N}=\{A_{k}\}$ with $k=1,\dots,|\mathcal{A}_{N}|$. This set in $D$ dimensions is defined as 
\beql{e50d}
\mathcal{A}_{N}~~=~~
\left\{
\left(
\ba{cccccc} 
( & a_{1}^{(n_1)}, & a_{2}^{(n_1)}, & \dots, & a_{D}^{(n_1)} & ) \\ 
( & a_{1}^{(n_2)}, & a_{2}^{(n_2)}, & \dots, & a_{D}^{(n_2)} & ) \\
 & \vdots & \vdots & & \vdots & \\
( & a_{1}^{(n_{D-1})}, & a_{2}^{(n_{D-1})}, & \dots, & a_{D}^{(n_{D-1})} & ) 
\ea 
\right)
~~\Bigg|~~
\ba{c}
\left(\sum_{i=1}^{D} a^{(n_j)}_{i}\right)\bmod{n_j}=0~~,
\\
\gcd{(n_j,\{a_{i}^{(n_j)}\})}\neq 1
\ea
\right\}~~,
\eeq
with $N=\prod_{j=1}^{D-1}{n_j}$. $\mathcal{A}_{N}$ decomposes into equivalence classes of orbifold actions, $[A_{k}]$, such that $\bigcup_{i=1}^{N_E}{[A_i]}=\mathcal{A}_{N}$ with $N_E$ being the number of equivalence classes in $\mathcal{A}_{N}$. If two orbifold actions $A_{l}\in[A_{k}]$ and $A_{m}\in[A_{k}]$ are of the same equivalence class $[A_{k}]$ and are both generators of the representations $\tilde{R}_{(n_1,\dots,n_{D-1})}(A_{l})$ and $\tilde{R}_{(\tilde{n}_1,\dots,\tilde{n}_{D-1})}(A_{m})$ respectively of the orbifold group $\Gamma_{N}$ with $N=\prod_{j=1}^{D-1}{n_j}=\prod_{j=1}^{D-1}{\tilde{n}_j}$, the the two representations $\tilde{R}_{(n_1,\dots,n_{D-1})}(A_{l})$ and $\tilde{R}_{(\tilde{n}_1,\dots,\tilde{n}_{D-1})}(A_{m})$ are equivalent up to permutations of the complex coordinates of $\mathbb{C}^{D}$.\\

An orbifold action $A_{k}$ in $D$ dimensions has $D-1$ components corresponding to the $D-1$ rows of the $(D-1)\times D$ orbifold action matrix. The $j$-th component has the form $A_{k}^{(n_j)}=(a_{1}^{(n_j)},a_{2}^{(n_j)},\dots,a_{D}^{(n_j)})$ and the overall orbifold action can be written in the compact form $A_{k}=(A_{k}^{(n_1)},A_{k}^{(n_2)},\dots,A_{k}^{(n_{D-1})})$. For $A_{k}$ being a generator of the representation $\tilde{R}_{(n_1,\dots,n_{D-1})}(A_{k})$ of the orbifold group $\Gamma_{N}=\otimes_{j=1}^{D-1}{\mathbb{Z}_{n_j}}$, the $j$-th component of the orbifold action, $A_{k}^{(n_j)}$, is the generator of the representation $\tilde{R}_{n_j}(A_{k}^{(n_j)})$ of the group $\mathbb{Z}_{n_j}$.
\\

%-----------------------------------------------------------
\subsection{Equivalence of Orbifold Actions \label{sc5p2}}

Two orbifold actions that are related by a permutation of the coordinates $\{z_1,\dots,z_D\}$ of $\mathbb{C}^{D}$ are equivalent. Permutations in the complex coordinates correspond to permutations of the $D$ columns in the $(D-1)\times D$ orbifold action matrix.\\

For the generalized equivalence condition, let the orbifold action $A_{k}$ be written in terms of its $D-1$ components, $A_{k}=(A_{k}^{(n_1)},A_{k}^{(n_2)},\dots,A_{k}^{(n_{D-1})})$. The $j$-th component $A_{k}^{(n_j)}$ generates a representation $\tilde{R}_{n_j}(A_{k}^{(n_j)})$ of the group $\mathbb{Z}_{n_j}$. For the purpose of obtaining the representation of the orbifold group $\Gamma_{N}$ generated by the overall orbifold action $A_{k}$, we define a map $\rho^{\{n_i\}}_{\{m_i\}}$ that maps all elements of the $j$ component action representations into all elements of the overall action representation of $\Gamma_{N}$. This map is defined as
\beal{e54}
\rho^{\{n_j\}}_{\{m_j\}}~~:~~
\left(
\tilde{R}_{n_1},\tilde{R}_{n_2},\dots,\tilde{R}_{n_{D-1}}
\right)
~~&\rightarrow&~~
\bar{R}_{(n_1,\dots,n_{D-1})}\left(\rho^{\{n_i\}}_{1\dots 1}(A_{k})\right)
\nn\\
\left(
A^{(n_1)}_{k},A^{(n_2)}_{k},\dots,A^{(n_{D-1})}_{k}
\right)
~~&\mapsto&~~ 
\left(
\sum_{s=1}^{D-1}{ m_{s}~A^{(n_s)}_{k}
\left( \prod_{\substack{j=2\\j\neq s}}^{D-1}{n_j} \right) }
\right)\bmod{N}~~,
\eea
where $N=\prod_{j=1}^{D-1}{n_j}$, $1\leq m_j \leq n_j$ and $m_j \in \mathbb{Z}$. The representation $\bar{R}_{(n_1,\dots,n_{D-1})}$ is the representation of the orbifold group $\Gamma_{N}$ and is generated by the generator from the identity mapping of all component generators, $\rho^{\{n_i\}}_{1\dots 1}(A_{k})$.\\

In terms of orbifold operators $\omega^{(\{a_{i}^{(n_j)}\})}$, the map $\rho^{\{n_i\}}_{\{m_i\}}$ acts as a simple multiplication operator such that
\beql{e55}
\rho^{\{n_j\}}_{\{m_j\}}~~:~~
\left(
\omega^{\left(a_{1}^{(n_1)},\dots,a_{D}^{(n_1)}\right)},
\dots,
\omega^{\left(a_{1}^{(n_{D-1})},\dots,a_{D}^{(n_{D-1})}\right)}
\right)
\mapsto
\prod_{j=1}^{D-1}{
\left(
\omega^{\left(a_{1}^{(n_{j})},\dots,a_{D}^{(n_{j})}\right)}
\right)^{m_j}
}~~.
\eeq\\

Accordingly, for given two orbifold actions $A_{1}$ and $A_{2}$, the orbifold actions are equivalent, $A_{1}\sim A_{2}$, if and only if their corresponding representations of orbifold groups of the same order $N$, $\bar{R}_{(n_1,\dots,n_{D-1})}\left(\rho^{\{n_i\}}_{1\dots 1}(A_{1})\right)$ and $\bar{R}_{(\tilde{n}_1,\dots,\tilde{n}_{D-1})}\left(\rho^{\{\tilde{n}_i\}}_{1\dots 1}(A_{2})\right)$ with $N=\prod_{j=1}^{D-1}{n_j}=\prod_{j=1}^{D-1}{\tilde{n}_j}$, are the same up to a permutation of the complex coordinates of $\mathbb{C}^{D}$. Consequently, the equivalence class of an orbifold action $A_{k}$ can be defined as
\beql{e58}
[A_{k}]~~=~~
\left\{ 
A_{l}
~\Bigg|~ 
A_{l}\sim A_{k}
~~\Leftrightarrow~~
\bar{R}_{(n_1,\dots,n_{D-1})}\left(\rho^{\{n_i\}}_{1\dots 1}(A_{k})\right)
~\sim~
\bar{R}_{(\tilde{n}_1,\dots,\tilde{n}_{D-1})}\left(\rho^{\{\tilde{n}_i\}}_{1\dots 1}(A_{l})\right)
\right\}~~,
\eeq
where $i=1,\dots,D$ and $N=\prod_{j=1}^{D-1}{n_j}=\prod_{j=1}^{D-1}{\tilde{n}_j}$. The set of all possible orbifold actions of a given orbifold group order $N$, $\mathcal{A}_{N}$, can be expressed in terms of the union of all equivalence classes in $\mathcal{A}_{N}$, $\bigcup_{i=1}^{N_E}{[A_i]}=\mathcal{A}_{N}$ with $N_E$ being the number of equivalence classes in $\mathcal{A}_{N}$.\\

The alternative mapping to $\rho^{\{n_j\}}_{\{m_j\}}$ in \eref{e54} is 
\beal{e58b}
\tilde{\rho}^{\{n_j\}}_{\{m_j\}}~~:~~
\left(
\tilde{R}_{n_1},\tilde{R}_{n_2},\dots,\tilde{R}_{n_{D-1}}
\right)
~~&\rightarrow&~~
\tilde{R}_{(n_1,\dots,n_{D-1})}\left(A_{k})\right)
\nn\\
\left(
A^{(n_1)}_{k},A^{(n_2)}_{k},\dots,A^{(n_{D-1})}_{k}
\right)
~~&\mapsto&~~ 
\bvec
(m_1 A_{k}^{(n_1)})\bmod{n_1} \\
(m_2 A_{k}^{(n_2)})\bmod{n_2} \\
\vdots \\
(m_{D-1} A_{k}^{(n_{D-1})})\bmod{n_{D-1}}
\evec
~~,
\eea
where the representation $\tilde{R}_{(n_1,\dots,n_{D-1})}(A_{k})$ of $\Gamma_{N}=\otimes_{j=1}^{D-1}{\mathbb{Z}_{n_j}}$ is generated by the orbifold action $A_{k}=(A^{(n_1)}_{k},\dots,A^{(n_{D-1})}_{k})$ and not by $\rho^{\{n_i\}}_{1\dots 1}(A_{k})$. The representation $\bar{R}_{(n_1,\dots,n_{D-1})}\left(\rho^{\{n_i\}}_{1\dots 1}(A_{k})\right)$ is used instead of the representation $\tilde{R}_{(n_1,\dots,n_{D-1})}(A_{k})$ for testing equivalence between orbifold actions.\\

%-----------------------------------------------------------
\subsection{Equivalence of $(D-1)$-simplices \label{sc5p3}}

We expect that the correspondence between inequivalent orbifold actions of $\mathbb{C}^{D}/\Gamma_{N}$ and inequivalent lattice polyhedra with $D$ corner points to hold for any dimension $D$. Orbifold actions of $\mathbb{C}^{2}/\Gamma_{N}$ correspond to lattice lines ($1$-simplices) of length $N$. Orbifold actions of $\mathbb{C}^{3}/\Gamma_N$ correspond to lattice triangles ($2$-simplices) of area $N$, while orbifold actions of $\mathbb{C}^{4}/\Gamma_N$ correspond to lattice tetrahedra ($3$-simplices) of volume $N$. By continuation, we expect that orbifold actions of $\mathbb{C}^{D}/\Gamma_{N}$ correspond to $(D-1)$-simplices of hypervolume $N$.\\

\textbf{Topology.} A lattice $(D-1)$-simplex has $D$ corner points which form the set $I_{0}=\{v_1,v_2,\dots,v_D\}$ where each corner point in Cartesian coordinates is expressed as $v_i=\{x_{1}^{i},x_{2}^{i},\dots,x_{D-1}^{i}\}\in\mathbb{Z}^{D-1}$ with $i=1,\dots,D$. It is important to note that a subset $J_{0}\subseteq I_0$ of size $d+1=|J_0|\leq D=|I_0|$ forms a $d$-dimensional hypersurface of a $d$-simplex with corner points $J_0$. The union of the lattice point sets enclosed by all possible $d$-simplices with fixed $|J_0|=d+1$ and $J_0\subseteq I_0$ as the corner points is defined as,
\beql{e85}
I_d~~=~~\Big\{
w_k=(x^{k}_1,\dots,x^{k}_{D-1}) \in \sigma^{d}[J_0]/\partial\sigma^{d}[J_0]
~~\Big|~~
\forall J_0\subseteq I_0~,~~|J_0|=d+1~,~~w_k\in\mathbb{Z}^{D-1}
\Big\}~~,
\eeq
where $\sigma^{d}[J_0]$ is the lattice $d$-simplex with corner points $J_0$, $\partial\sigma^{d}[J_0]$ is the boundary of the $d$-simplex with corner points $J_0$, and $(x^{k}_1,\dots,x^{k}_{D-1})\in\mathbb{Z}^{D-1}$ are the Cartesian coordinates of a $\mathbb{Z}^{D-1}$ lattice point $w_k$ where $k=1,\dots,|I_d|$. We note that by definition $0 < d < D$ such that for $d\geq D$, $I_d=\emptyset$.\\

As seen in Section \sref{s5} for toric diagram tetrahedra, the generalization of the sets of topologically important points of a lattice $(D-1)$-simplex in \eref{e85} can be used to obtain the sets for internal points $I_3$, face internal points $I_2$, and edge internal points $I_1$ for toric diagram tetrahedra. The union of all topologically important lattice point sets for a lattice $(D-1)$-simplex is given by
\beql{e86}
\bigcup_{d=0}^{D-1}{I_d}~=~\sigma^{D-1}~~,
\eeq
where $\sigma^{D-1}$ is the set of all lattice points on and enclosed by the boundary of the lattice $(D-1)$-simplex with corner points $\{v_1,\dots,v_{D}\}$.\\

Every lattice point $w_{k}\in\sigma^{D-1}$ divides the $(D-1)$-simplex into $D$ sub-simplices of dimension $D-1$ or less. Saying that these subsimplices have $(D-1)$-dimensional hypervolumes with values $\lambda_{k1},\lambda_{k2},\dots,\lambda_{kD}$ respectively, the lattice point $w_{k}\in\sigma^{D-1}$ can be given in terms of barycentric coordinates of the form
\beql{e87}
w_{k}~~=~~\frac{1}{N}(\lambda_{k1},\lambda_{k2},\dots,\lambda_{kD})~~,
\eeq
where $N$ is the $(D-1)$-dimensional hypervolume of the $(D-1)$-simplex $\sigma^{D-1}$.\\

The topological character of the $(D-1)$-dimensional toric diagram simplex needs to be defined in terms of optimized scaling conditions on the sets of topologically important lattice points, $I_d$. Using the definition of scaling in \eref{e78} generalized to any dimension $D$ such that
\beql{e88}
f_{s}~~:~~\{v_i\}~~\mapsto~~\{s v_i\}~=~\{(s x^{i}_1,s x^{i}_2,\dots,s x^{i}_{D-1})\}~~,
\eeq
where $(x^{i}_1,x^{i}_2,\dots,x^{i}_{D-1})$ are the Cartesian coordinates of the corner point $v_i\in\sigma^{D-1}$, one can re-define the set of topologically important points $I_d\subseteq\sigma^{D-1}$ as a function of a scaled $(D-1)$-simplex such that $I_d\neq\emptyset$. We re-define $I_d$ as
\beal{e89}
I_{d}(f_{s_{d}\rightarrow\min{(s_{d})}}(\sigma^{D-1}))~~=~~\Big\{&&
w_k=(x^{k}_1,\dots,x^{k}_{D-1}) \in \sigma^{d}[J_0]/\partial\sigma^{d}[J_0] \nn\\
&&~~\Big|~~
\forall J_0\subseteq I_{0}(f_{s_{d}\rightarrow\min{(s_{d})}}(\sigma^{D-1})),~~|J_0|=d+1~,~~w_k\in\mathbb{Z}^{D-1}
\Big\}~~,\nn\\
\eea
where $f_{s_{d}\rightarrow\min{(s_{d})}}(\sigma^{D-1})$ is the scaled $(D-1)$-simplex $\sigma^{D-1}$ by a minimal scaling factor with the limit $s_{d}\rightarrow\min{(s_{d})}\in\mathbb{Z}$ such that $I_{d}(f_{s_{d}\rightarrow\min{(s_{d})}}(\sigma^{D-1}))\neq\emptyset$ for all $d\in\{1,\dots,D-1\}$. $I_{0}(f_{s_{0}\rightarrow\min{(s_{0})}}(\sigma^{D-1}))$ is the set of corner points of the scaled $(D-1)$-simplex $f_{s_{0}\rightarrow\min{(s_{0})}}(\sigma^{D-1})$. The set of all topologically important points for the characterization of the $(D-1)$-simplex is defined as
\beql{e89b}
I(\sigma^{D-1})~~=~~\bigcup_{d=0}^{D-1}{I_{d}(f_{s}(\sigma^{D-1}))}~~,
\eeq 
where $s=\max({\{s_1,\dots,s_{D-1}\}})$.
\\

Accordingly, using the definition in \eref{e89}, we can define the optimized topological character of a $(D-1)$-simplex $\sigma^{D-1}$ with hypervolume $N$ as the set
\beql{e90}
\tau~~=~~\left\{
\frac{1}{N}(\lambda_{k1},\lambda_{k2},\dots,\lambda_{kD})~~\Big|~~
w_k\in I_{d}(f_{s}(\sigma^{D-1}))~,~~d=1,\dots,D-1
\right\}~~,
\eeq 
where the overall optimal scaling coefficient of the simplex $\sigma^{D-1}$ is $s=\max({\{s_1,s_2,\dots,s_{D-1}\}})$ such that $I_{d}(\sigma^{D-1})\neq\emptyset$ for all $d$. The topological character $\tau$ can be identified as the set of barycentric coordinates of the topologically important points $I(\sigma^{D-1})$ defined above in \eref{e89b}.\\

The topological character in \eref{e90} is a $k\times D$ matrix, where the $k$ rows are the barycentric coordinates of the topological relevant points $w_k\in I(\sigma^{D-1})$ of the lattice $(D-1)$-simplex, and the $D$ columns are the $D$ barycentric coordinates axes $\{\hat{v}_{1},\hat{v}_{2},\dots,\hat{v}_{D}\}$ corresponding to the $D$ corner points $w_k\in I_{0}$ of the $(D-1)$-simplex.\\

\textbf{Equivalence.} Two toric diagram lattice $(D-1)$-simplices $\sigma_{1}^{D-1}$ and $\sigma_{2}^{D-1}$ are equivalent if their corresponding topological characters  $\tau_1=\tau(\sigma_{1}^{D-1})$ and $\tau_2=\tau(\sigma_{2}^{D-1})$ are equivalent $\tau_1\sim\tau_2$ up to a permutation of the barycentric coordinates axes $\{\hat{v}_{1},\hat{v}_{2},\dots,\hat{v}_{D}\}$. We note that permutations of the barycentric coordinates correspond to permutations of the columns of the $k\times D$ topological character matrix of the $(D-1)$-simplex.\\

It is interesting to have a closer look on the barycentric coordinates of the corner points $\{v_i\}=I_{0}$ of the $(D-1)$-simplex with $i=1,\dots,D$. The $l$-th component of the barycentric coordinates of the corner point $v_i$ is
\beql{e91}
(v_i)_{l}~=~\delta_{il}~.
\eeq
Accordingly, we note that any permutation of the barycentric coordinates axes $\{\hat{v}_{1},\hat{v}_{2},\dots,\hat{v}_{D}\}$ leaves the set of corner points $\{v_i\}$ of a $(D-1)$-simplex invariant.\\

From this observation, a generalized link can be drawn to the coordinates on $\mathbb{C}^{D}$, $\{z_1,z_2,\dots,z_D\}$, which map under a permutation of coordinates an orbifold action generating a representation of an orbifold group $\Gamma_{N}$ to an equivalent orbifold action generating an equivalent representation of another orbifold group $\Gamma_{N}$ with the same order $N$. We can draw a correspondence between the coordinates of $\mathbb{C}^{D}$ and the barycentric coordinates axes of toric points as a map
\beql{e92}
\mathnormal{T}~~:~~\{z_1,z_2,\dots,z_{D}\}~\mapsto~\{\hat{v}_{1},\hat{v}_{2},\dots,\hat{v}_{D}\}~~,
\eeq
as seen above in \eref{e64} for $D=3$ and in \eref{e75} for $D=4$.\\

\textbf{Domain.} The domain is the set of all $(D-1)$-simplices in the positive sector of $\mathbb{Z}^{D-1}$ with one corner point of the lattice simplex being the origin of $\mathbb{Z}^{D-1}$. The domain conditions ensure that no two lattice simplices are related by a rotation around the origin, or a translation.\\

Given that the corner point $v_i$ of a lattice $(D-1)$-simplex is in Cartesian coordinates $(x_1,\dots,x_{D-1})\in\mathbb{Z}^{D-1}$, the domain $D(N)$ can be generated by all possible transformations of the unit hypervolume $(D-1)$-simplex into $(D-1)$-simplices of hypervolume $N$. The transformed $(D-1)$-simplices have one corner point as the origin of $\mathbb{Z}^{D-1}$ and are in the positive sector of $\mathbb{Z}^{D-1}$ to fulfill the domain conditions.\\
%The domain $D(N)$ is a set of all $(D-1)$-dimensional lattice simplices of hypervolume $N$ where two simplices in the set are not related by a lattice translation or a rotation around the $\mathbb{Z}^{D-1}$ origin.\\

%Let a corner point of every domain simplex be the origin of $\mathbb{Z}^{D-1}$ and let us express the corner points in Cartesian coordinates $(x_1,\dots,x_{D-1})\in\mathbb{Z}^{D-1}$. The domain $D(N)$ can be generated by transforming the unit hypervolume simplex into simplices of hypervolume $N$ where the transformation matrices are Hermite Normal Form in $D-1$ dimensions.\\

The $j$-th component of the Cartesian coordinates of the corner point $v_i$ belonging to the unit hypervolume $(D-1)$-simplex is
\beql{e94}
v_{i}^{j}=\delta_{i(j+1)}~,
\eeq
where $i=1,\dots,D$. The transformations generating the domain $D(N)$ are of the form
\beql{e95}
M~=~\left(\ba{cccccc}
m_{11} & m_{12} & \dots & m_{1j}     & \dots & m_{1(D-1)} \\
0 & m_{22} & \dots      & m_{2j}     & \dots & m_{2(D-1)} \\
0 & 0 &                 & m_{3j}     &       & m_{3(D-1)}  \\
\vdots & \vdots &       & \vdots     &       & \vdots  \\
0 & 0  &                & m_{(j-1)j} &       & m_{(j-1)(D-1)} \\
0 & 0 &                 & m_{jj}     &       & m_{j(D-1)} \\
0 & 0  &                & 0          &       & m_{(j+1)(D-1)}\\
\vdots & \vdots  &      & \vdots     &       & \vdots \\
0 & 0 & \dots           & 0          & \dots & m_{(D-1)(D-1)}
\ea\right)~~,
\eeq
where $j=1,\dots,D-1$ such that $\det{M}=\prod_{j=1}^{D-1}{m_{jj}}=N$ and $0\leq m_{jk} < m_{jj}$ with $m_{jk}\in\mathbb{N}_{0}$. \\

Accordingly, the domain of $(D-1)$-simplices of hypervolume $N$ is given as
\beql{e96}
D(N)~=~\Bigg\{
\{v_1,v_2,\dots,v_D\}=M.\sigma_{\Delta}
~~|~~N=\prod_{j=1}^{D-1}{m_{jj}}~,~~0\leq m_{jk} < m_{jj}~,~~m_{jk}\in\mathbb{N}_{0}
\Bigg\}~~,
\eeq
where $\sigma_{\Delta}$ is the unit hypervolume simplex with corner points given in \eref{e94}.\\

As defined in the context of $\mathbb{C}^{3}$ and $\mathbb{C}^{4}$, it is now possible to define the equivalence class of a toric diagram $(D-1)$-simplex $\sigma^{D-1}$ as
\beql{e96b}
[\sigma^{D-1}]~=~\{\tilde{\sigma}^{D-1}~|~\tau(\sigma^{D-1})\sim\tau(\tilde{\sigma}^{D-1})\}~~,
\eeq
where the definition of the optimized topological character $\tau$ in \eref{e90} is used such that the toric diagram lattice points are $w_k\in I(\sigma^{D-1})=\cup_{d=0}^{D-1}{I_{d}(f_{s}(\sigma^{D-1}))}$. The domain can be expressed like for the $\mathbb{C}^{3}$ and $\mathbb{C}^{4}$ orbifolds as
\beql{eq96c}
\bigcup_{i=1}^{N_E}[\sigma_{i}^{D-1}]~=~D(N)~~.
\eeq
The size of the equivalence class $|[\sigma_{i}^{D-1}]|$ is the multiplicity $\mu_i$ of the corresponding simplex $\sigma_{i}^{D-1}$, such that $\sum_{i=1}^{N_E}\mu_i~=~|D(N)|$. $N_E$ is the number of equivalence classes of $(D-1)$-simplices of hypervolume $N$ or equivalently the number of equivalence classes of orbifold actions of $\mathbb{C}^{D}/\Gamma_N$. 
\\

%%%%%%%%%%%%%%%%%%%%%%%%%%%%%%%%%%%%%%%%%%%%%%%%%%%%%%%%%%%%%%%%%%%%%%%%%%%%%
%%%%%%%%%%%%%%%%%%%%%%%%%%%%%%%%%%%%%%%%%%%%%%%%%%%%%%%%%%%%%%%%%%%%%%%%%%%%%
\clearpage

\begin{table}[h!]
\begin{center}
\scriptsize
\begin{tabular}{p{5mm}p{5mm}|p{15mm}p{15mm}p{15mm}p{15mm}p{15mm}p{15mm}}
\hline
\hline
\multicolumn{8}{c}{$\mathbb{C}^{D}/\Gamma_{N}$}
\\
\hline
\hline
 & & \multicolumn{6}{c}{$D$}
\\
& & 2 & 3 & 4 & 5 & 6 & 7 \\
\hline
\hline
\multirow{50}{*}{$N$} 
& 1 	& 1 & 1		& 1			& 1			& 1			& 1	 \\
& 2		& 1	& 1		& 2			& 2			& 3			& 3	 \\
& 3		&	1	& 2		& 3			& 4			& 6			& 7	 \\
& 4		&	1	& 3		& 7			&	10		& 17		& 23 	 \\
& 5		& 1	& 2		& 5			& 8			& 13		& 19	 \\
& 6		& 1	& 3		& 10		& 19		& 40		& 	65 \\
& 7		& 1	& 3		& 7			& 13		& 27		& 	46 \\
& 8		& 1	& 5		& 20		& 45		& 106		& 	 \\
& 9		& 1	& 4		& 14		& 33		& 78		& 	 \\
& 10	& 1	& 4		& 18		& 47		& 127		& 	 \\
& 11	& 1	& 3		& 11		& 30		& 79		& 	 \\
& 12	& 1	& 8		& 41		& 129		& 391		& 	 \\
& 13	& 1	& 4		&	15	 	& 43		& 129		& 	 \\
& 14	& 1	& 5		& 28		& 96		& 321		& 	 \\
& 15	& 1	& 6		& 31		& 108		& 358		& 	 \\
& 16	& 1	& 9		& 58		& 226		& 			& 	 \\
& 17	& 1	& 4		& 21		& 78		& 			& 	 \\
& 18	& 1	& 8		& 60		& 264		& 			& 	 \\
& 19	& 1	& 5		& 25		& 102		& 			& 	 \\
& 20	& 1	& 10	& 77		& 357		& 			& 	 \\
& 21	& 1	& 8		& 49		& 226		&	 			&	 	 \\
& 22	& 1	& 7		& 54		& 277		& 			& 	 \\
& 23	& 1	& 5		& 33		& 163		& 			& 	 \\
& 24	& 1	& 15	& 144		& 813		& 			& 	 \\
& 25	& 1	& 7		& 50		& 260		& 			& 	 \\
& 26	& 1	& 8		& 72		& 425		& 			& 	 \\
& 27	& 1	& 9		& 75		& 436		& 			& 	 \\
& 28	& 1	& 13	& 123		& 780		& 			& 	 \\
& 29	& 1	& 6		& 49		& 297		& 			& 	 \\
& 30	& 1	& 14	& 158		& 1092	& 			& 	 \\
& 31	& 1	& 7		& 55		& 			& 			& 	 \\
& 32	& 1	& 15	& 177		& 			& 			& 	 \\
& 33	& 1	& 10	& 97		& 			& 			& 	 \\
& 34	& 1	& 10	& 112		& 			& 			& 	 \\
& 35	& 1	& 10	& 99		& 			& 			& 	 \\
& 36	& 1	& 20	& 268		& 			& 			& 	 \\
& 37	& 1	& 8		& 75		& 			& 			& 	 \\
& 38	& 1	& 11	& 136		& 			& 			& 	 \\
& 39	& 1	& 12	& 129		& 			& 			& 	 \\
& 40	& 1	& 20	& 286		& 			& 			& 	 \\
& 41	& 1	& 8		& 89		& 			& 			& 	 \\
& 42	& 1	& 18	& 268		& 			& 			& 	 \\
& 43	& 1	& 9		& 97		& 			& 			& 	 \\
& 44	& 1	& 17	& 249		& 			& 			& 	 \\
& 45	& 1	& 16	& 218		& 			& 			& 	 \\
& 46	& 1	& 13	& 190 	& 			& 			& 	 \\
& 47	& 1	& 9		& 113 	& 			& 			& 	 \\
& 48	& 1	& 28	& 496		& 			& 			& 	 \\
& 49	& 1	& 12	& 146		& 			& 			& 	 \\
& 50	& 1	& 17	& 280		& 			& 			& 	 \\
\hline
\hline
\end{tabular}
\normalsize
\caption{Generated counting of the number of inequivalent orbifold actions ($N_E$) of $\mathbb{C}^{D}/\Gamma_{N}$ orbifolds and dual $(D-1)$-dimensional lattice simplices of hypervolume $N$.\label{tmain}}
\end{center}
\end{table}

\clearpage
%%%%%%%%%%%%%%%%%%%%%%%%%%%%%%%%%%%%%%%%%%%%%%%%%%%%%%%%%%%%%%%%%%%%%%%%%%%%%%%%%%%%%%%%%%%%%%%%%%%%%%%%%%%%%%%%%%%%%%%%%%%%

\section{Closed Form Formula of Counting Series \label{scount}}

It is now of great interest to implement the counting algorithms presented above using both the generation of inequivalent orbifold actions and the generation of the dual inequivalent toric diagram simplices of orbifolds $\mathbb{C}^{D}/\Gamma_{N}$. The main counting results are presented in \tref{tmain}. Ideally, we wish to obtain a closed form formula for the counting of inequivalent orbifold actions of $\mathbb{C}^{D}/\Gamma_{N}$ dependent on the dimension of the orbifold $D$ and the discrete group order $N$ of $\Gamma_{N}=\otimes_{j=1}^{D-1}{\mathbb{Z}_{n_j}}$ with $N=\prod_{j=1}^{D-1}{n_j}$. We present in this work a recursive form of the counting formula for the size of the domain $D(N)$ dependent on $N$, and the closed form formula for inequivalent orbifold actions dependent on $N$ for orbifolds $\mathbb{C}^{3}/\Gamma_{N}$.\\

The general form of a counting series formula is
\beql{e97}
g_{D}(t)~=~\sum_{N=1}^{\infty}{c_{ND}~t^{N}}~~,
\eeq
where $t$ is an arbitrary counting variable and the coefficient $c_{ND}\in\mathbb{Z}^{+}$ of the order $\mathcal{O}(t^{N})$ term of $g_{D}(t)$ is the count of a specific property of the orbifold $\mathbb{C}^{D}/\Gamma_{N}$ of order $N$ and dimension $D$. The counting series $g_{D}(t)$ can be reformulated in terms of a generating functional $f_{D}(t)$ such that \eref{e97} becomes
\beql{e98}
g_{D}(t)~=~\sum_{k=1}^{\infty}{m_{k}f_{D}(t^{k})}~~,
\eeq
where $m_{k}$ is some coefficient.
\\

\textbf{Counting the Domain Size $|D(N)|$.} Let us first consider the series of the domain size $|D(N)|$ with the domain being the set of all lattice simplices of hypervolume $N$ excluding equivalence due to translational and rotational symmetry along the lattice origin as discussed for orbifolds $\mathbb{C}^{3}/\Gamma_{N}$ in \eref{e69}, $\mathbb{C}^{4}/\Gamma_{N}$ in \eref{e84} and the generalized orbifolds $\mathbb{C}^{D}/\Gamma_{N}$ in \eref{e96}. The counting of the domain sizes $|D(N)|$ for the orbifolds $\mathbb{C}^{3}/\Gamma_{N}$ to $\mathbb{C}^{7}/\Gamma_{N}$ are shown in \tref{tc3} to \tref{tc7} respectively.\\

The closed form of the series of the domain size $|D(N)|$ for the orbifolds $\mathbb{C}^{2}/\Gamma_{N}$ is
\beql{e99}
g_{2}(t)~=~\sum_{k=1}^{\infty}{t^k}~=~\frac{t}{1-t}~~.
\eeq
For the orbifolds $\mathbb{C}^{3}/\Gamma_{N}$,
\beql{e100}
g_{3}(t)~=~\sum_{k=1}^{\infty}{\frac{t^k}{(1-t^k)^{2}}}~=~\sum_{k=1}^{\infty}{k\frac{t^{k}}{(1-t^{k})}}~~.
\eeq
Setting $g_{1}(t)=t$, we find a recursive expression for the domain size series of orbifolds $\mathbb{C}^{D}/\Gamma_{N}$ for any dimension $D$ as
\beql{e101}
g_{D}(t)~=~\sum_{k=1}^{\infty}{k^{D-2}g_{D-1}(t^{k})}~~.
\eeq
A proof is given in the appendix of \cite{HananyOrlando10}.
\\

\textbf{Counting the number of Equivalence Classes $N_E$.} We have shown in this work two methods of counting equivalence classes of abelian orbifolds: counting equivalence classes of representations of orbifold actions and counting equivalence classes of dual toric diagram simplices. Both methods led to the same counting of equivalence classes, $N_E$, and counts are shown for orbifolds $\mathbb{C}^{3}/\Gamma_{N}$ to $\mathbb{C}^{7}/\Gamma_{N}$ in \tref{tc3} to \tref{tc7} respectively.\\

Using the technology developed under Plethystics \cite{Benvenuti:2006qr,Feng:2007ur} and the work by \cite{HananyOrlando10}, the $N_{E}$ counting series generating function for $D=3$ orbifolds $\mathbb{C}^{3}/\Gamma_{N}$ can be found as
\beql{e101}
f_{3}(t)~=~\frac{1}{(1-t)(1+t^2)(1-t^3)}-1
\eeq
such that the closed form counting series for $N_E$ of $\mathbb{C}^{3}/\Gamma_{N}$ has the compact form
\beql{e102}
g_{3}(t)~=~\sum_{k=1}^{\infty}{f_{3}(t^{k})}~=~t+t^2+2t^3+3t^4+2t^5+\dots~~
\eeq
corresponding to the result shown in \tref{tc3}. A closed form formula for $D=4$ has been presented in \cite{HananyOrlando10} and the counting results in this work confirm it.\\

%\clearpage
%%%%%%%%%%%%%%%%%%%%%%%%%%%%%%%%%%%%%%%%%%%%%
%%%%%%%%%%%%% COUNTING TABLES %%%%%%%%%%%%%%%
%%%%%%%%%%%%%%%%%%%%%%%%%%%%%%%%%%%%%%%%%%%%%

%%%%%%%%%%%%% C3 %%%%%%%%%%%%%%%%%%%%%%%%%%%%
\begin{table}[h!]
\begin{tabular}{c|p{5mm}p{5mm}p{5mm}p{5mm}p{5mm}p{5mm}p{5mm}p{5mm}p{5mm}p{5mm}p{5mm}p{5mm}p{5mm}p{5mm}p{5mm}}
\hline
\hline
\multicolumn{16}{c}{$\mathbb{C}^{3}/\Gamma_{N}$}
\\
\hline
\hline
$N$ & 
1 & 2 & 3 & 4 & 5 & 6 & 7 & 8 & 9 & 10 & 11 & 12 & 13 & 14 & 15
\\
\hline
$|D(N)|$ &
1 & 3 & 4 & 7 & 6 & 12 & 8 & 15 & 13 & 18 & 12 & 28 & 14 & 24 & 24
\\
$N_E$ &
1 & 1 & 2 & 3 & 2 & 3 & 3 & 5 & 4 & 4 & 3 & 8 & 4 & 5 & 6
\\
\hline
\hline
$N$ & 
16 & 17 & 18 & 19 & 20 & 21 & 22 & 23 & 24 & 25 & 26 & 27 & 28 & 29 & 30
\\
\hline
$|D(N)|$ &
31 & 18 & 39 & 20 & 42 & 32 & 36 & 24 & 60 & 31 & 42 & 40 & 56 & 30 & 72
\\
$N_E$&
9 & 4 & 8 & 5 & 10 & 8 & 7 & 5 & 15 & 7 & 8 & 9 & 13 & 6 & 14
\\
\hline
\hline
$N$ & 
31 & 32 & 33 & 34 & 35 & 36 & 37 & 38 & 39 & 40 & 41 & 42 & 43 & 44 & 45
\\
\hline
$|D(N)|$ &
32 & 63 & 48 & 54 & 48 & 91 & 38 & 60 & 56 & 90 & 42 & 96 & 44 & 84 & 78
\\
$N_E$ &
7 & 15 & 10 & 10 & 10 & 20 & 8 & 11 & 12 & 20 & 8 & 18 & 9 & 17 & 16
\\
\hline
\hline
\end{tabular}
\caption{Series for the domain size $|D(N)|$ and the number of equivalence classes $N_E$ of $\mathbb{C}^{3}/\Gamma_{N}$ orbifold actions and dual lattice $2$-simplices. \label{tc3}}
\end{table}

%%%%%%%%%%%%% C4 %%%%%%%%%%%%%%%%%%%%%%%%%%%%
\begin{table}[h!]
\begin{tabular}{c|p{5mm}p{5mm}p{5mm}p{5mm}p{5mm}p{5mm}p{5mm}p{5mm}p{5mm}p{5mm}p{5mm}p{5mm}p{5mm}p{5mm}p{5mm}}
\hline
\hline
\multicolumn{16}{c}{$\mathbb{C}^{4}/\Gamma_{N}$}
\\
\hline
\hline
$N$ & 
1 & 2 & 3 & 4 & 5 & 6 & 7 & 8 & 9 & 10 & 11 & 12 & 13 & 14 & 15
\\
\hline
$|D(N)|$ &
1 & 7 & 13 & 35 & 31 & 91 & 57 & 155 & 130 & 217 & 133 & 455 & 183 & 399 & 403
\\
$N_E$ &
1 & 2 & 3 & 7 & 5 & 10 & 7 & 20 & 14 & 18 & 11 & 41 & 15 & 28 & 31
\\
\hline
\hline
$N$ & 
16 & 17 & 18 & 19 & 20 & 21 & 22 & 23 & 24 & 25 & 26 & 27 & 28 & 29 & 30
\\
\hline
$|D(N)|$ &
651 & 307 & 910 & 381 & 1085 & 741 & 931 & 553 & 2015 & 806 & 1281 & 1210 & 1995 & 871 & 2821
\\
$N_E$&
58 & 21 & 60 & 25 & 77 & 49 & 54 & 33 & 144 & 50 & 72 & 75 & 123 & 49 & 158
\\
\hline
\hline
$N$ & 
31 & 32 & 33 & 34 & 35 & 36 & 37 & 38 & 39 & 40 & 41 & 42 & 43 & 44 & 45
\\
\hline
$|D(N)|$ &
993 & 2667 & 1729 & 2149 & 1767 & 4550 & 1407 & 2667 & 2379 & 4805 & 1723 & 5187 & 1893 & 4655 & 4030
\\
$N_E$ &
55 & 177 & 97 & 112 & 99 & 268 & 75 & 136 & 129 & 286 & 89 & 268 & 97 & 249 & 218
\\
\hline
\hline
\end{tabular}
\caption{Series for the domain size $|D(N)|$ and the number of equivalence classes $N_E$ of $\mathbb{C}^{4}/\Gamma_{N}$ orbifold actions and dual lattice $3$-simplices.\label{tc4}}
\end{table}

%\clearpage
%%%%%%%%%%%%% C5 %%%%%%%%%%%%%%%%%%%%%%%%%%%%
\begin{table}[h!]
\begin{center}
\begin{tabular}{c|p{9mm}p{9mm}p{9mm}p{9mm}p{9mm}p{9mm}p{9mm}p{9mm}p{9mm}p{9mm}}
\hline
\hline
\multicolumn{11}{c}{$\mathbb{C}^{5}/\Gamma_{N}$}
\\
\hline
\hline
$N$ & 
1 & 2 & 3 & 4 & 5 & 6 & 7 & 8 & 9 & 10 
\\
\hline
$|D(N)|$ &
1 & 15 & 40 & 155 & 156 & 600 & 400 & 1395 & 1210 & 2340
\\
$N_E$ &
1 & 2 & 4 & 10 & 8 & 19 & 13 & 45 & 33 & 47  
\\
\hline
\hline
$N$ & 
11 & 12 & 13 & 14 & 15 & 16 & 17 & 18 & 19 & 20 
\\
\hline
$|D(N)|$ &
1464 & 6200 & 2380 & 6000 & 6240 & 11811 & 5220 & 18150 & 7240 & 24180 
\\
$N_E$&
30 & 129 & 43 & 96 & 108 & 226 & 78 & 264 & 102 & 357
\\
\hline
\hline
$N$ & 
21 & 22 & 23 & 24 & 25 & 26 & 27 & 28 & 29 & 30 
\\
\hline
$|D(N)|$ &
16000 & 21960 & 12720 & 55800 & 20306 & 35700 & 33880 & 62000 & 25260 & 93600 
\\
$N_E$&
226 & 277 & 163 & 813 & 260 & 425 & 436 & 780 & 297 & 1092
\\
\hline
\hline
\end{tabular}
\caption{Series for the domain size $|D(N)|$ and the number of equivalence classes $N_E$ of $\mathbb{C}^{5}/\Gamma_{N}$ orbifold actions and dual lattice $4$-simplices.\label{tc5}}
\end{center}
\end{table}

%%%%%%%%%%%%% C6 %%%%%%%%%%%%%%%%%%%%%%%%%%%%
\begin{table}[h!]
\begin{center}
\begin{tabular}{c|p{9mm}p{9mm}p{9mm}p{9mm}p{9mm}p{9mm}p{9mm}p{9mm}p{9mm}p{9mm}}
\hline
\hline
\multicolumn{11}{c}{$\mathbb{C}^{6}/\Gamma_{N}$}
\\
\hline
\hline
$N$ & 
1 & 2 & 3 & 4 & 5 & 6 & 7 & 8 & 9 & 10 
\\
\hline
$|D(N)|$ &
1 & 31 & 121 & 651 & 781 & 3751 & 2801 & 11811 & 11011 & 24211
\\
$N_E$ &
1 & 3 & 6 & 17 & 13 & 40 & 27 & 106 & 78 & 127
\\
\hline
\hline
$N$ & 
11 & 12 & 13 & 14 & 15 & 16 & 17 & 18 & 19 & 20 
\\
\hline
$|D(N)|$ &
16105 & 78711 & 30941 & 86831 & 94501 & 200787 & 88741 & 341341 & 137561 & 508431
\\
$N_E$&
79 & 391 & 129 & 321 & 358 & & & & &
\\
\hline
\hline
\end{tabular}
\caption{Series for the domain size $|D(N)|$ and the number of equivalence classes $N_E$ of $\mathbb{C}^{6}/\Gamma_{N}$ orbifold actions and dual lattice $5$-simplices.\label{tc6}}
\end{center}
\end{table}

%%%%%%%%%%%%% C7 %%%%%%%%%%%%%%%%%%%%%%%%%%%%
\begin{table}[h!]
\begin{center}
\begin{tabular}{c|p{9mm}p{9mm}p{9mm}p{9mm}p{9mm}p{9mm}p{9mm}p{9mm}p{9mm}p{9mm}}
\hline
\hline
\multicolumn{11}{c}{$\mathbb{C}^{7}/\Gamma_{N}$}
\\
\hline
\hline
$N$ & 
1 & 2 & 3 & 4 & 5 & 6 & 7 & 8 & 9 & 10 
\\
\hline
$|D(N)|$ &
1 & 63 & 364 & 2667 & 3906 & 22932 & 19608 & 97155 & 99463 & 246078
\\
$N_E$ &
1 & 3 & 7 & 23 & 19 & 65 & 46 & & &
\\
\hline
\hline
\end{tabular}
\caption{Series for the domain size $|D(N)|$ and the number of equivalence classes $N_E$ of $\mathbb{C}^{7}/\Gamma_{N}$ orbifold actions and dual lattice $6$-simplices.\label{tc7}}
\end{center}
\end{table}

%\clearpage

%%%%%%%%%%%%%%%%%%%%%%%%%%%%%%%%%%%%%%%%%%%%%%%%%%%%%%%%%%%%%%%%%%%%%%%%%%%%%%%%%%%%%%%%%%%%%%%%%%%%%%%%%%%%%%%%%%%%%%%%%%%%

\section{Conclusions}

We have presented methods which allow us to distinguish between inequivalent orbifold actions of any orbifold dimension and formulated closed form formulas for the counting series of the inequivalent orbifold actions on $\mathbb{C}^{3}$. The main counting results have been presented in \tref{tmain}. In this section, we summarize the observations made so far, draw additional applications of the counting methodology and suggest further directions of future study.\\

First, we notice that with the calculations presented in this work the counting of inequivalent lattice $(D-1)$-simplices of hypervolume $N$ is the same as the counting of inequivalent orbifold actions of $\mathbb{C}^{D}/\Gamma_{N}$ orbifolds. The method we have presented to distinguish inequivalent lattice simplices does not involve the explicit identification of $GL(D-1,\mathbb{Z})$ symmetry transformations that map equivalent lattice polyhedra in $D$-dimensional integer lattices into themselves. Instead, we have outlined an algorithm that characterizes the topology of a lattice simplex of hypervolume $N$, just like an orbifold action characterizing the orbifold singularity corresponding to the complex cone over the projective algebraic surface. The correspondence between orbifold actions and the toric diagram simplices are shown in the Appendix Section A for orbifolds of $\mathbb{C}^{3}$ with $N=1,\dots,10$ and orbifolds of $\mathbb{C}^{4}$ with $N=1,\dots,6$. We expect that this correspondence holds for any higher dimension $D$.\\

The task of counting lattice simplices has a novel application in condensed matter theory. Bernstein, Sloane and Wright \cite{Sloane97} discuss in their work a method of counting sublattices of a hexagonal lattice and obtained the same counting of inequivalent orbifold actions for $\mathbb{C}^{3}$ as we have. Their counting method corresponds exactly to the counting of inequivalent face labellings of brane tilings of the $\mathbb{C}^{3}/\Gamma_{N}$ orbifolds as discussed in Section \sref{s2}. In addition, the work by Hart and Forcade \cite{HartForcade08} presents an algorithm for counting $3$-dimensional inequivalent superlattices corresponding to atomic configurations in crystals. Their counting result matches exactly our counting of inequivalent orbifold actions of $\mathbb{C}^{4}/\Gamma_{N}$ orbifolds.\\

We have of course only touched with the closed form formula for $\mathbb{C}^{3}$ in \eref{e101} and \eref{e102} the tip of the problem of finding a generalized closed form formula for the series counting inequivalent orbifold actions for any higher dimension $D$ of the torus. It would be of great interest to find such a generalized closed form counting formula. Another interesting question to ask is whether the relationship between classical and stringy geometry is still meaningfully preserved for higher dimensional orbifolds beyond $\mathbb{C}^{4}/\Gamma_{N}$.

%%%%%%%%%%%%%%%%%%%%%%%%%%%%%%%%%%%%%%%%%%%%%%%%%%%%%%%%%%%%%%%%%%%%%%%%%%%%%%%%%%%%%%%%%%%%%%%%%%%%%%%%%%%%%%%%%%%%%%%%%%%%

\section*{Acknowledgments}

Discussions with Yang-Hui He, Noppadol Mekareeya and Guiseppe Torri are greatfully acknowledged. A. H. would like to thank Domenico Orlando and Susanne Reffert for collaboration on the topics discussed in \cite{HananyOrlando10} and the Simons Center for Geometry and Physics for their kind hospitality during the completion of this paper. R.-K. S. is very grateful for support and advice from Dimitri D. Vvedensky. He is also very grateful for the continuous support and encouragement from his parents. J.D. would like to thank STFC for his studentship and the occupiers of Huxley 606 for tolerating recent discussions.

%%%%%%%%%%%%%%%%%%%%%%%%%%%%%%%%%%%%%%%%%%%%%%%%%%%%%%%%%%%%%%%%%%%%%%%%%%%%%%%
%%%%%%%%%%%%%%%%%%%%%%%           APPENDIX      %%%%%%%%%%%%%%%%%%%%%%%%%%%%%%%
%%%%%%%%%%%%%%%%%%%%%%%%%%%%%%%%%%%%%%%%%%%%%%%%%%%%%%%%%%%%%%%%%%%%%%%%%%%%%%%
\clearpage

\appendix\label{appendix}

%%%%%%%%%%%%%%%%%%%%%%%%%%%%%%%%%%%%%%%%%%%%%%%%%%%%%%%%%%%%%%%%%%%%%%%%%%%%%%%
%%%%%%%%%%%%%%%%%%%%%%%%%%%%%%%%%%%%%%%%%%%%%%%%%%%%%%%%%%%%%%%%%%%%%%%%%%%%%%%
\section{$\mathbb{C}^{3}$ Orbifold Index \label{sa1}}

In the toric diagram triangles, lattice points on the edges of the triangle are colored yellow and lattice points enclosed by the triangle boundary are colored green (Tables 8-14).\\

%%%%%%%%%%%%%%%%%%%%%%%%%%%%%%%%%%%%%%%%%%%
%%%%%%%%%%%%%%%%%%% N=1..3 ================
\begin{table}[ht]

\begin{center}
\begin{tabular}{m{1cm}|m{0.8cm}|m{2.2cm}|m{3cm}|m{4.8cm}|m{2cm}}
\hline \hline
\# & $N$ & Orbifold & Orbifold Action & Toric Diagram & Multiplicity\\
\hline \hline

%%%%%%%%%%%%%%%%%%% N=1 %%%%%%%%%%%%%%%%%%%
(1.1) & 1 & $\mathbb{C}^{3}/\mathbb{Z}_{1}$ & $\bvec (0,0,0)\\(0,0,0) \evec$ & \includegraphics*[height=3.5cm]{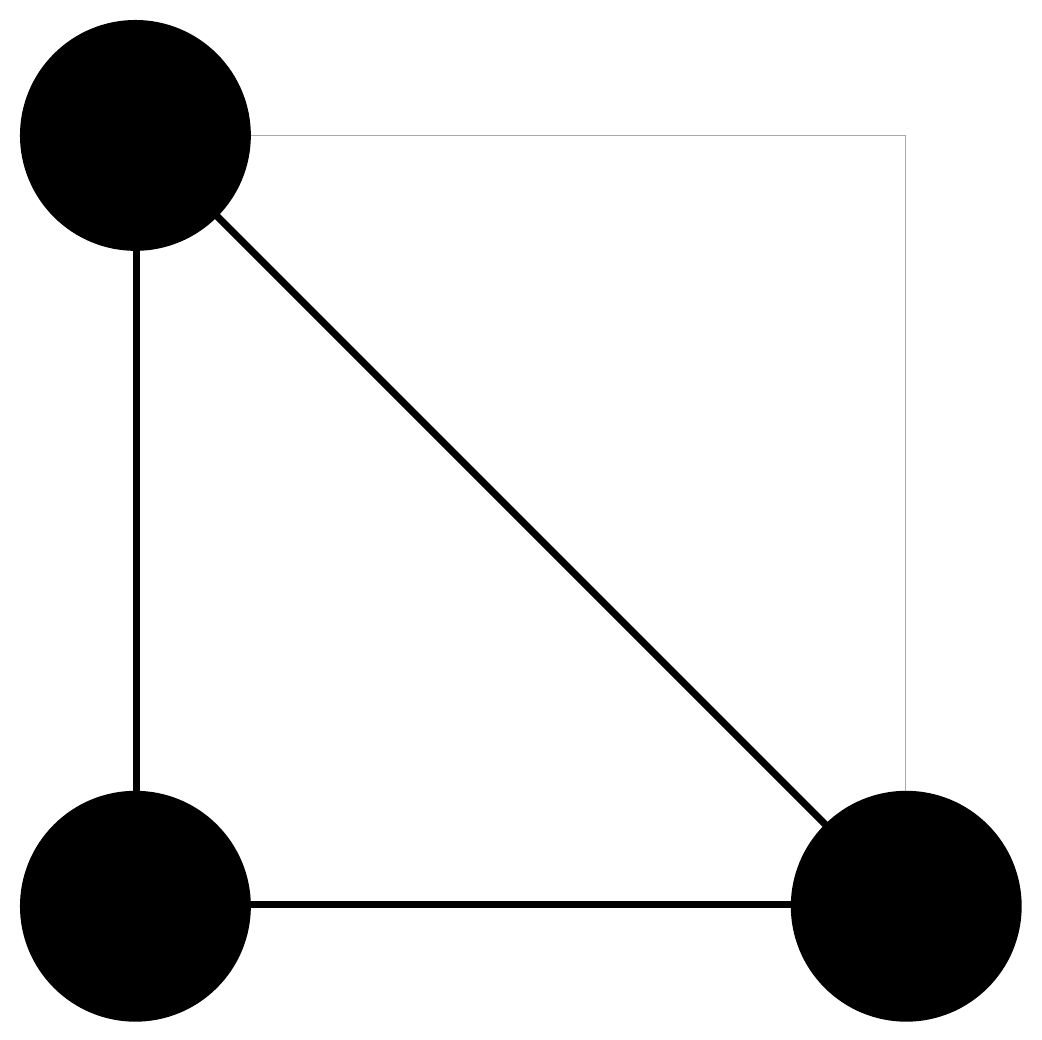} & 1
\\ 
\hline \hline

%%%%%%%%%%%%%%%%%%% N=2 %%%%%%%%%%%%%%%%%%%
(2.1) & 2 & $\mathbb{C}^{3}/\mathbb{Z}_{2}$ & $\bvec (0,1,1)\\(0,0,0) \evec$ &
\includegraphics*[height=3.5cm]{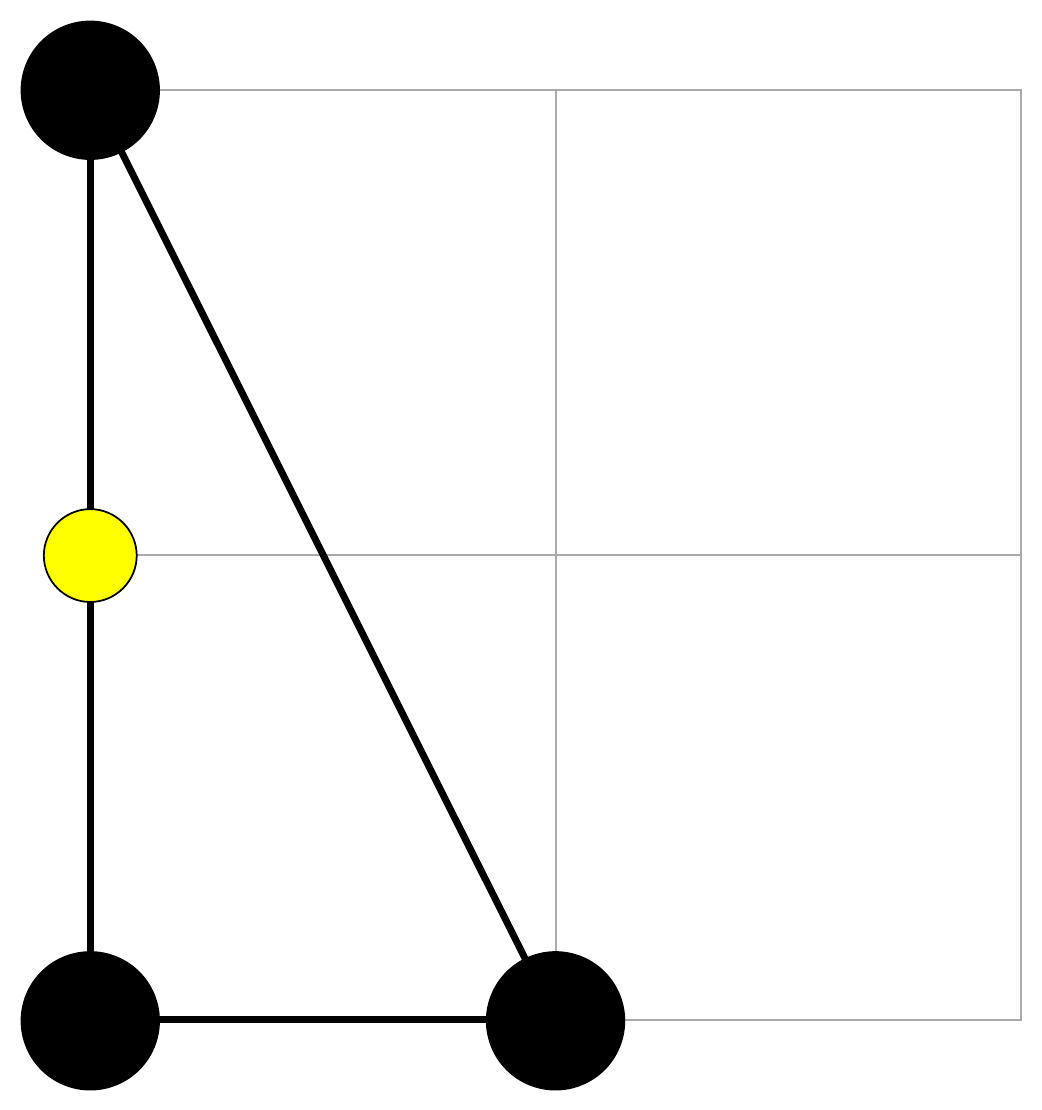} & 3
\\
\hline \hline

%%%%%%%%%%%%%%%%%%% N=3 %%%%%%%%%%%%%%%%%%%
(3.1) & 3 & $\mathbb{C}^{3}/\mathbb{Z}_{3}$ & $\bvec (0,1,2)\\(0,0,0) \evec$ &
\includegraphics*[height=3.5cm]{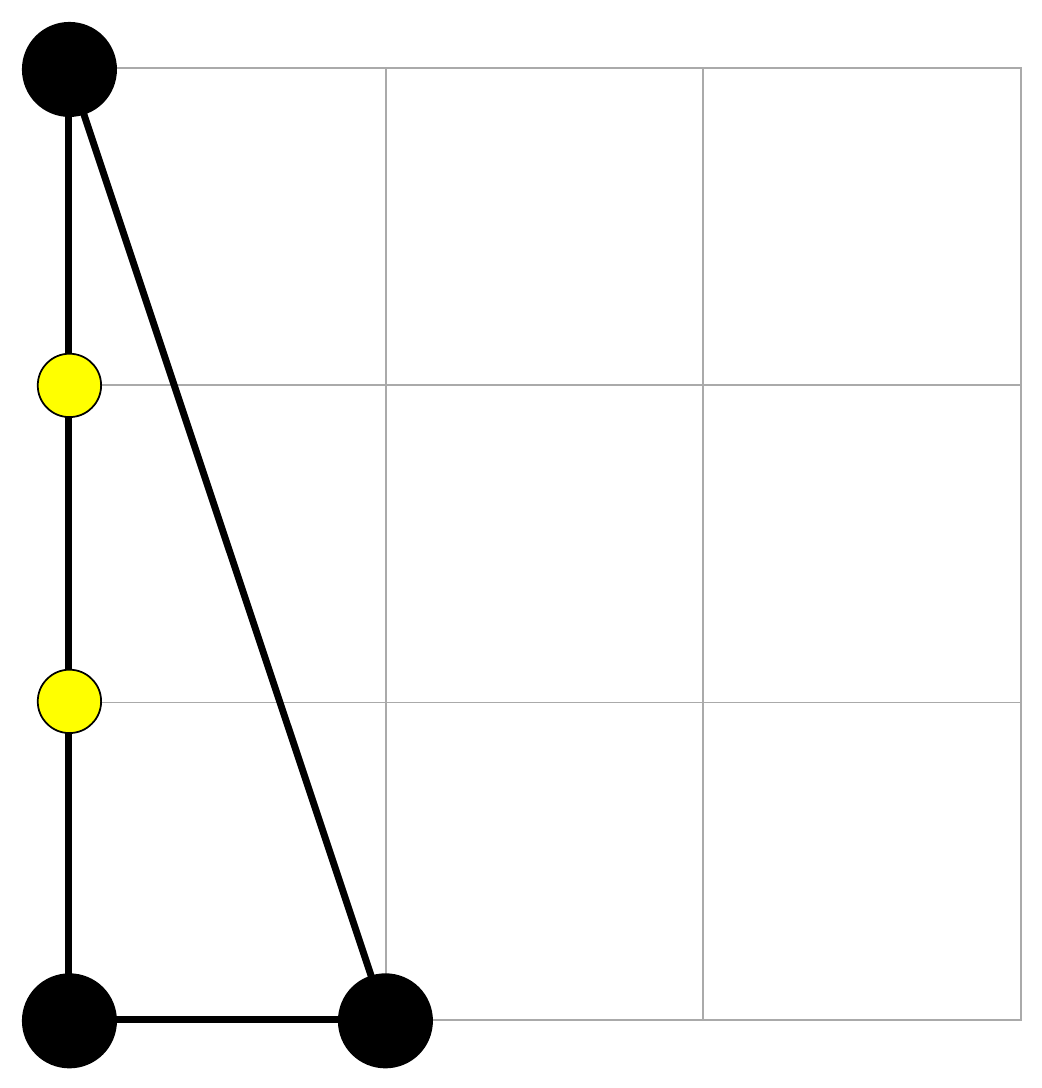} & 3
\\
\hline
(3.2) & 3 & $\mathbb{C}^{3}/\mathbb{Z}_{3}$ & $\bvec (1,1,1)\\(0,0,0) \evec$ &
\includegraphics*[height=3.5cm]{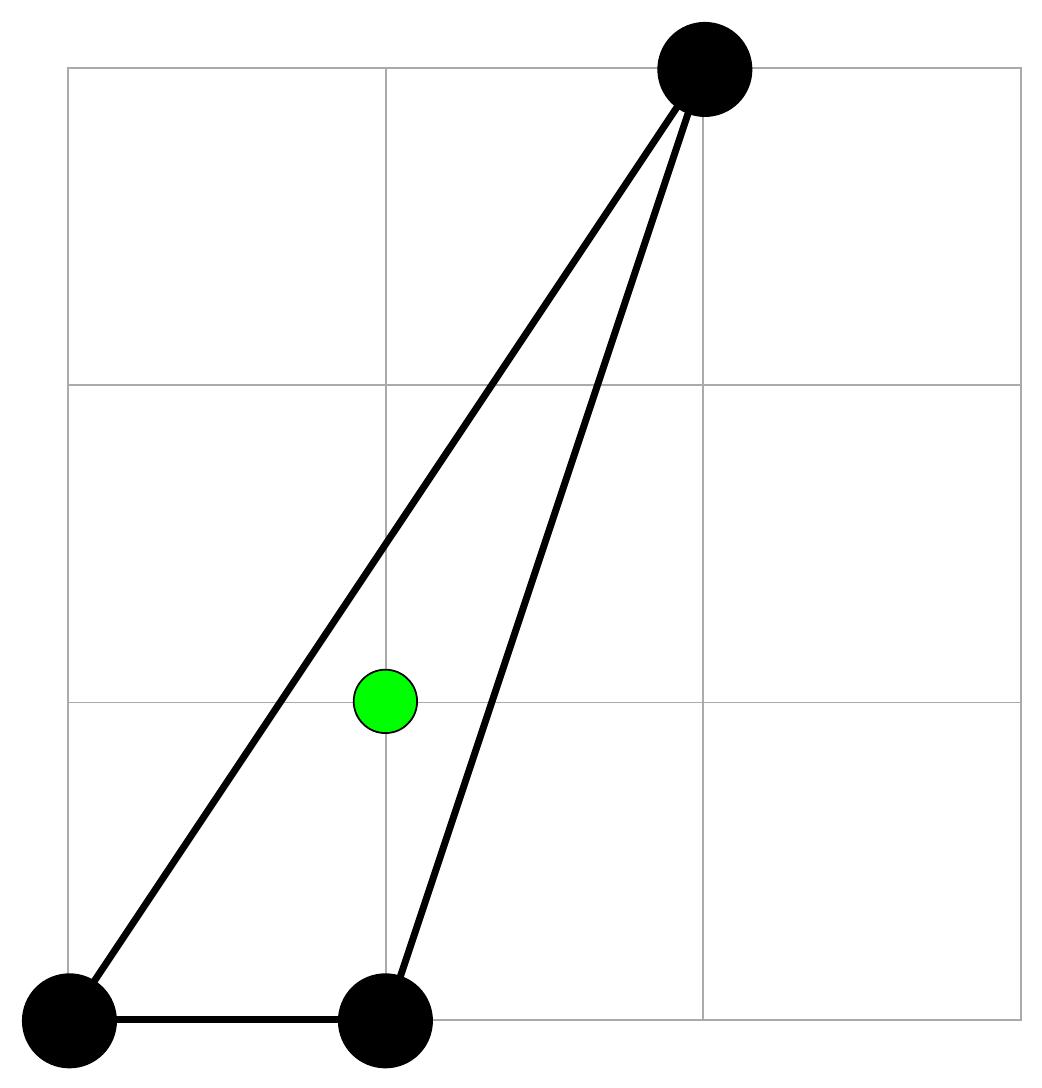} & 1
\\
\hline \hline

\end{tabular}
\end{center}

\caption{Orbifold Actions and corresponding Toric Diagrams for $\mathbb{C}^{3}/\Gamma_N$ orbifolds with order $N=1\dots 10$ \textbf{(Part 1/7)}.}
\label{t1}
\end{table}

\clearpage

%%%%%%%%%%%%%%%%%%%%%%%%%%%%%%%%%%%%%%%%%%%
%%%%%%%%%%%%%%%%%%% N=4..6 ================
\begin{table}[ht]

\begin{center}
\begin{tabular}{m{1cm}|m{0.8cm}|m{2.2cm}|m{3cm}|m{4.8cm}|m{2cm}}
\hline \hline
\# & $N$ & Orbifold & Orbifold Action & Toric Diagram & Multiplicity\\
\hline \hline

%%%%%%%%%%%%%%%%%%% N=4 %%%%%%%%%%%%%%%%%%%
(4.1) & 4 & $\mathbb{C}^{3}/\mathbb{Z}_{4}$ & $\bvec (0,1,3)\\(0,0,0) \evec$ & 
\includegraphics*[height=4.8cm]{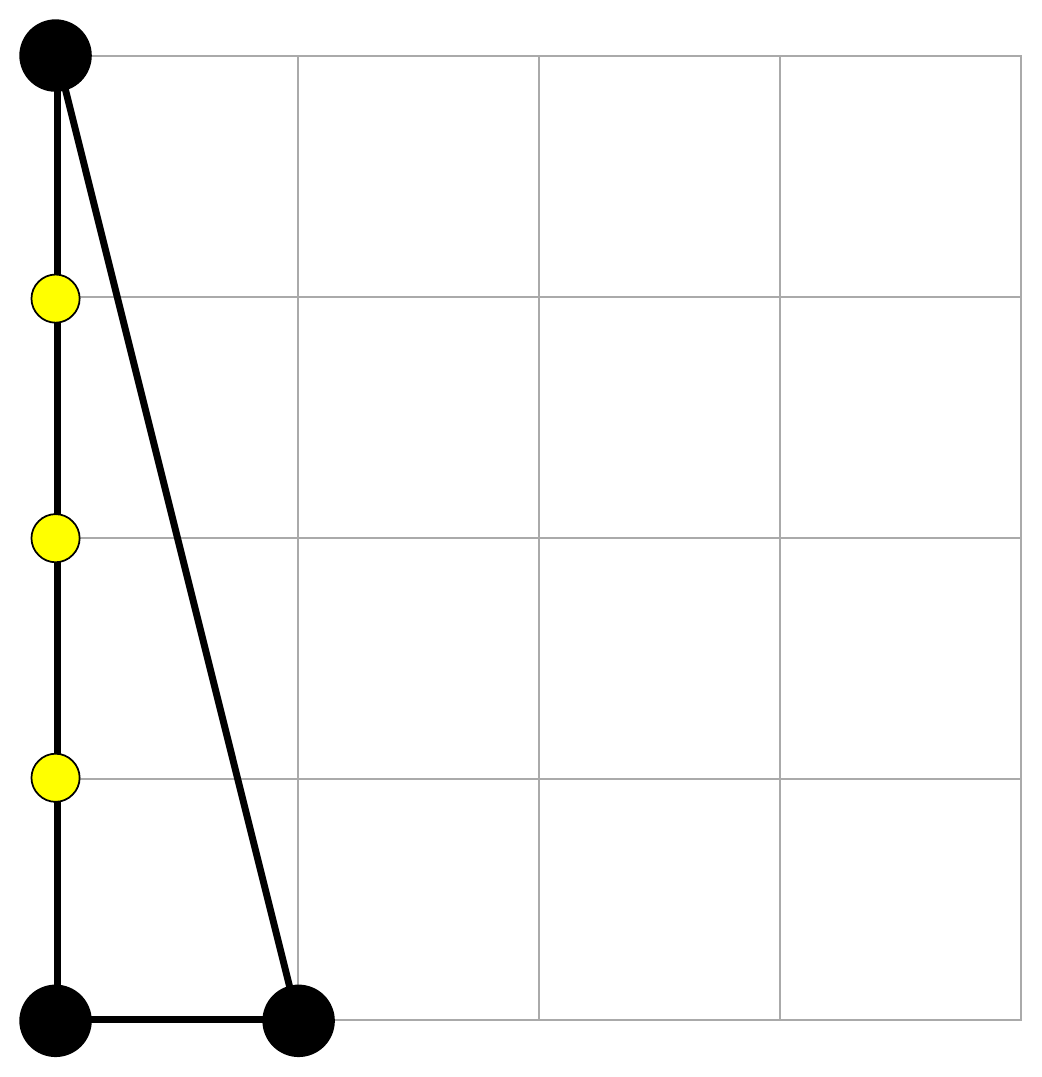} & 3
\\ 
\hline
(4.2) & 4 & $\mathbb{C}^{3}/\mathbb{Z}_{4}$ & $\bvec (1,1,2)\\(0,0,0) \evec$ & 
\includegraphics*[height=4.8cm]{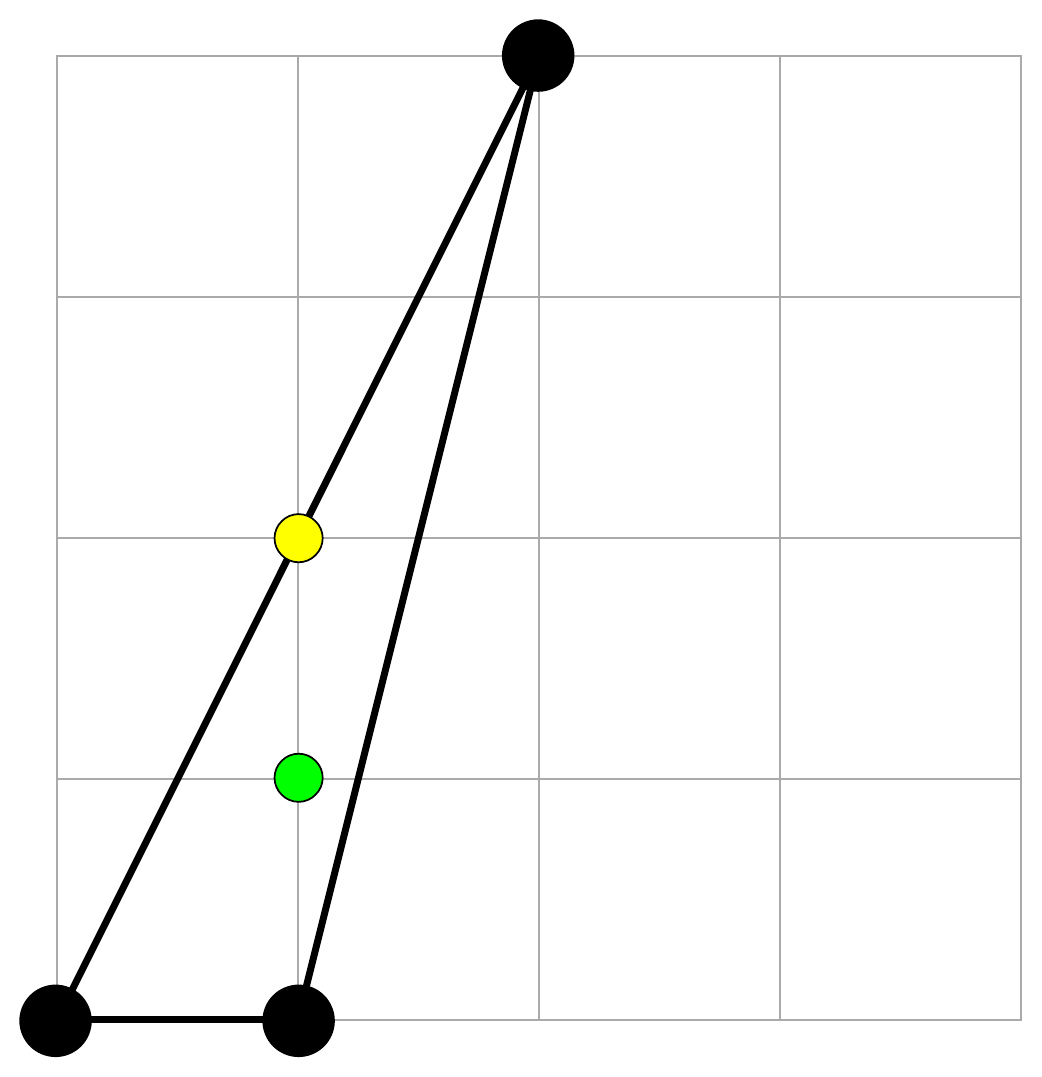} & 3
\\ 
\hline
(4.3) & 4 & $\mathbb{C}^{3}/\mathbb{Z}_{2}\times\mathbb{Z}_{2}$ & $\bvec (1,0,1)\\(0,1,1) \evec$ & 
\includegraphics*[height=4.8cm]{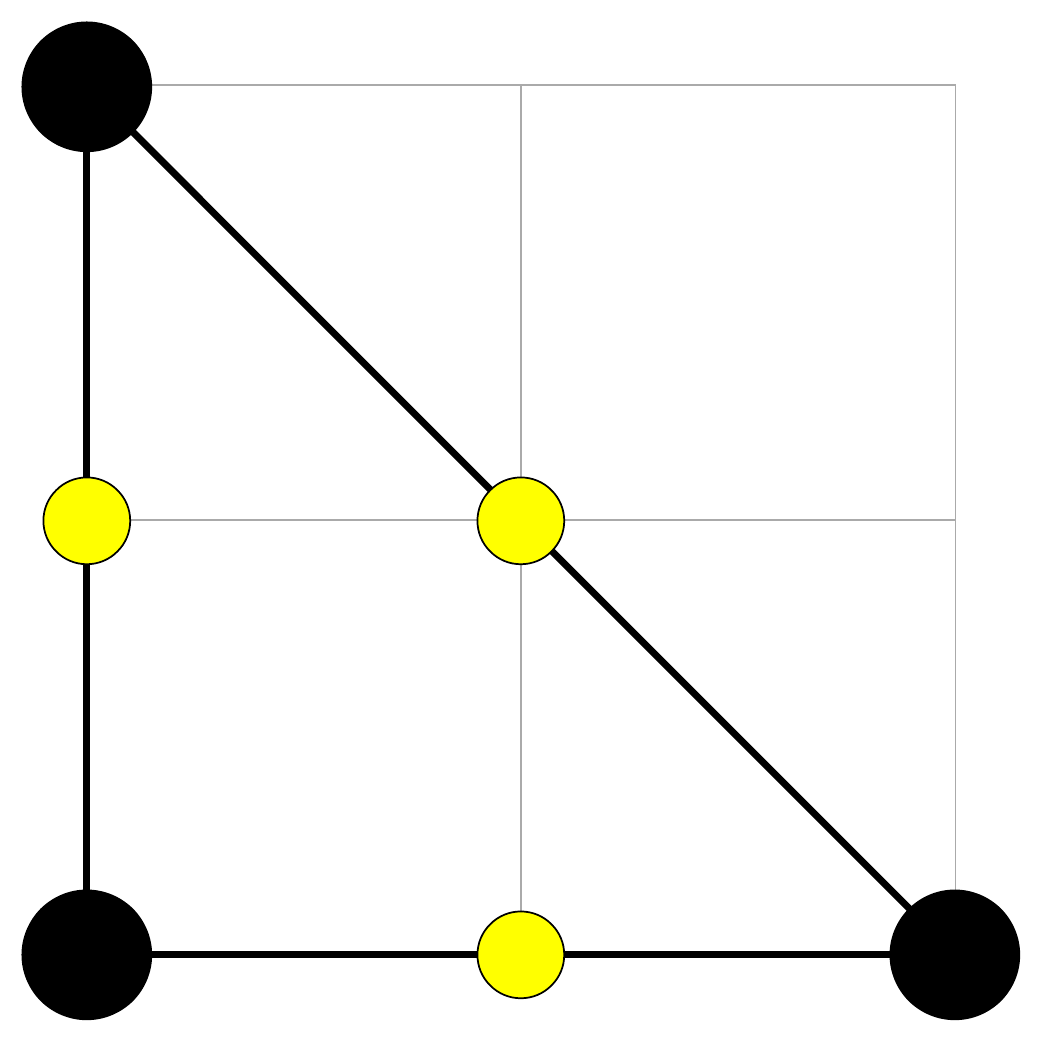} & 1
\\ 
\hline \hline

%%%%%%%%%%%%%%%%%%% N=5 %%%%%%%%%%%%%%%%%%%
(5.1) & 5 & $\mathbb{C}^{3}/\mathbb{Z}_{5}$ & $\bvec (0,1,4)\\(0,0,0) \evec$ &
\includegraphics*[height=4.8cm]{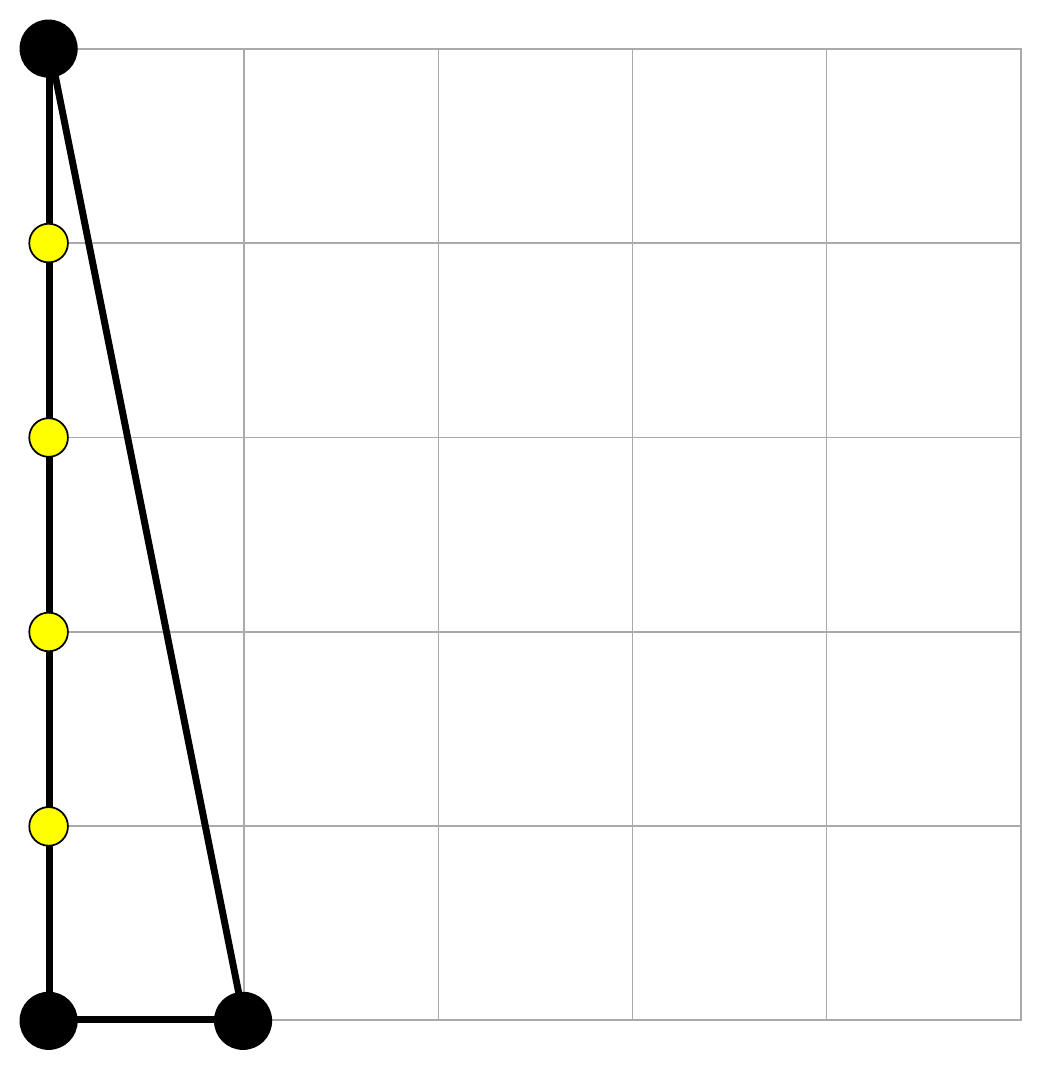} & 3
\\
\hline \hline

\end{tabular}
\end{center}

\caption{Orbifold Actions and corresponding Toric Diagrams for $\mathbb{C}^{3}/\Gamma_N$ orbifolds with order $N=1\dots 10$ \textbf{(Part 2/7)}.}
\label{t2}
\end{table}

\clearpage

%%%%%%%%%%%%%%%%%%%%%%%%%%%%%%%%%%%%%%%%%%%%%%%%
%%%%%%%%%%%%%%%%%%%%%%%%%%%%%%%%%%%%%%%%%%%%%%%%

\begin{table}[ht]

\begin{center}
\begin{tabular}{m{1cm}|m{0.8cm}|m{2.2cm}|m{3cm}|m{4.8cm}|m{2cm}}
\hline \hline
\# & $N$ & Orbifold & Orbifold Action & Toric Diagram & Multiplicity\\
\hline \hline
%%%%%%%%%%%%%%%%%%% N=5 %%%%%%%%%%%%%%%%%%%

(5.2) & 5 & $\mathbb{C}^{3}/\mathbb{Z}_{5}$ & $\bvec (1,1,3)\\(0,0,0) \evec$ &
\includegraphics*[height=4.8cm]{C3Tn5m2i1.pdf} & 3
\\
\hline \hline
%%%%%%%%%%%%%%%%%%% N=6 %%%%%%%%%%%%%%%%%%%
(6.1) & 6 & $\mathbb{C}^{3}/\mathbb{Z}_{6}$ & $\bvec (0,1,5)\\(0,0,0) \evec$ &
\includegraphics*[height=4.8cm]{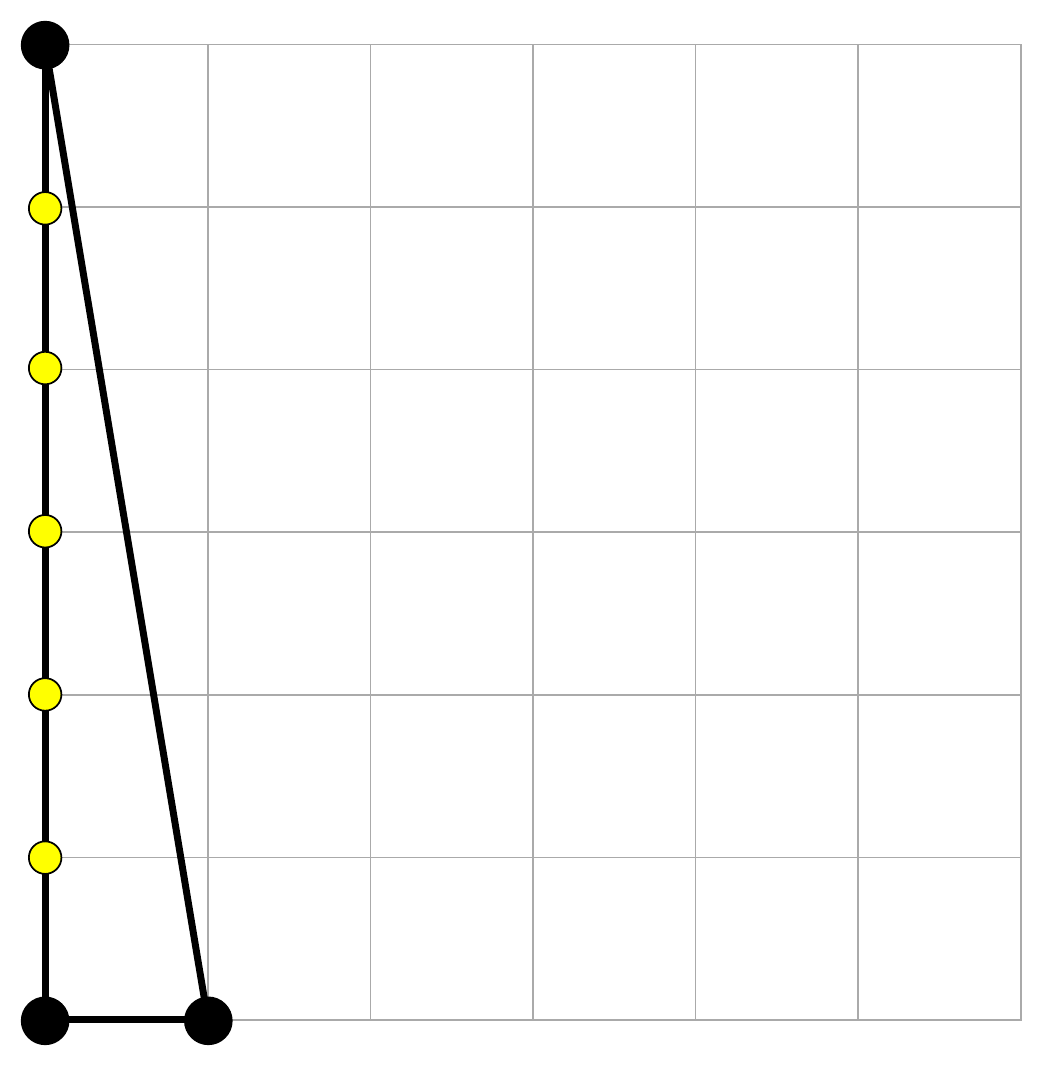} & 3
\\
\hline
(6.2) & 6 & $\mathbb{C}^{3}/\mathbb{Z}_{6}$ & $\bvec (1,1,4)\\(0,0,0) \evec$ &
\includegraphics*[height=4.8cm]{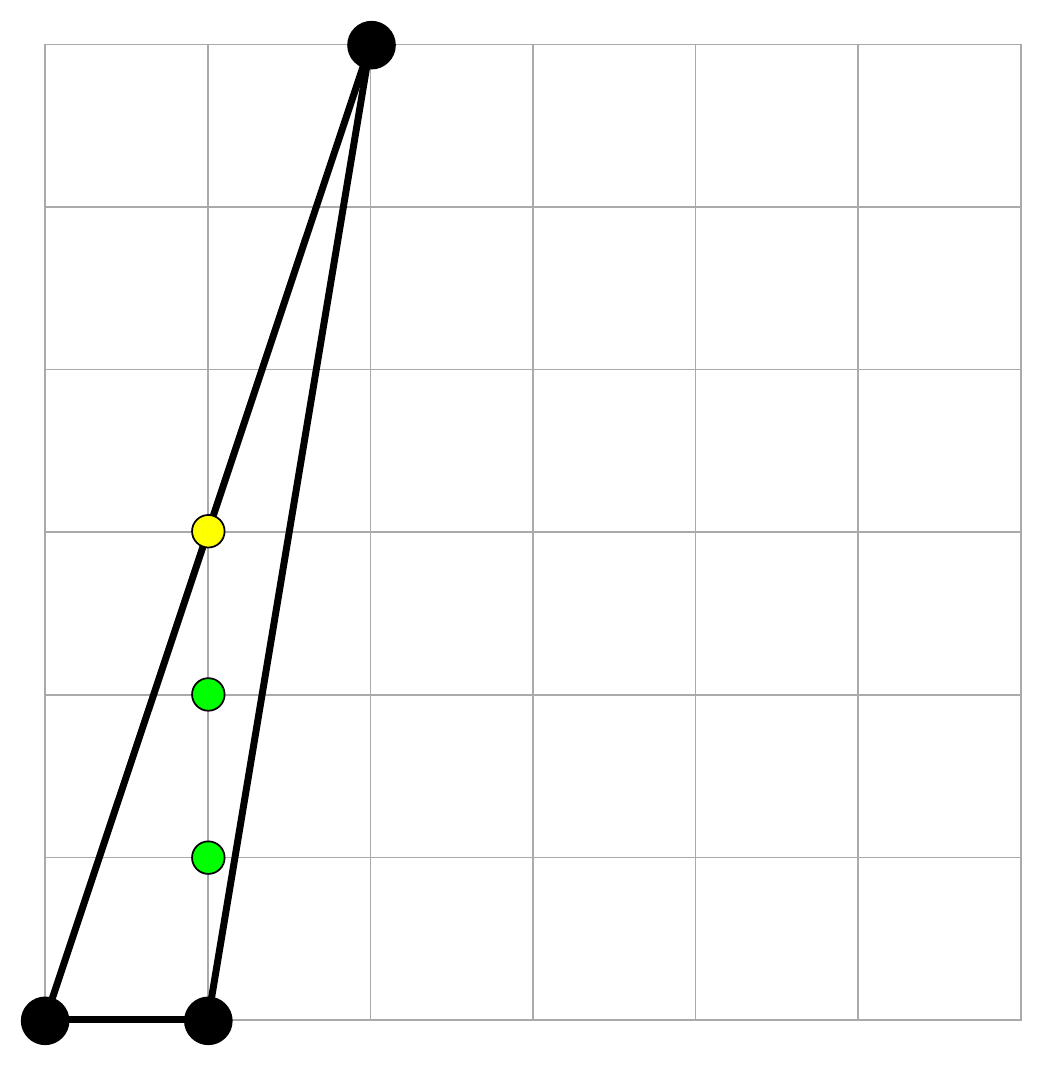} & 3
\\
\hline
(6.3) & 6 & $\mathbb{C}^{3}/\mathbb{Z}_{6}$ & $\bvec (1,2,3)\\(0,0,0) \evec$ &
\includegraphics*[height=4.8cm]{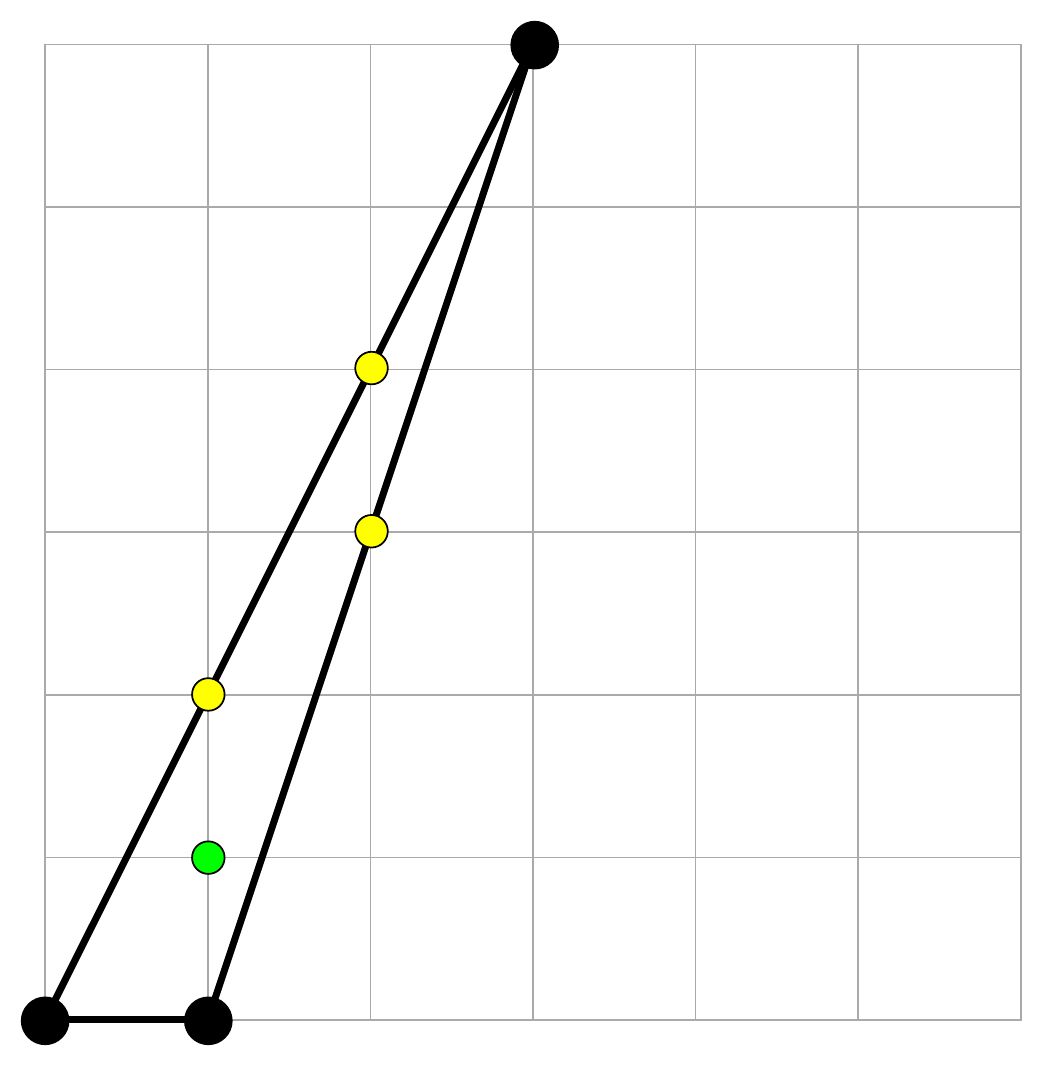} & 6
\\
\hline \hline

\end{tabular}
\end{center}

\caption{Orbifold Actions and corresponding Toric Diagrams for $\mathbb{C}^{3}/\Gamma_N$ orbifolds with order $N=1\dots 10$ \textbf{(Part 3/7)}.}
\label{t2b}
\end{table}

\clearpage

%%%%%%%%%%%%%%%%%%%%%%%%%%%%%%%%%%%%%%%%%%%%%%%%
%%%%%%%%%%%%%%%%%%%%%%%%%%%%%%%%%%%%%%%%%%%%%%%%

\begin{table}[ht]

\begin{center}
\begin{tabular}{m{1cm}|m{0.8cm}|m{2.2cm}|m{3cm}|m{4.8cm}|m{2cm}}
\hline \hline
\# & $N$ & Orbifold & Orbifold Action & Toric Diagram & Multiplicity\\
\hline \hline

%%%%%%%%%%%%%%%%%%% N=7 %%%%%%%%%%%%%%%%%%%
(7.1) & 7 & $\mathbb{C}^{3}/\mathbb{Z}_{7}$ & $\bvec (0,1,6)\\(0,0,0) \evec$ & 
\includegraphics*[height=4.8cm]{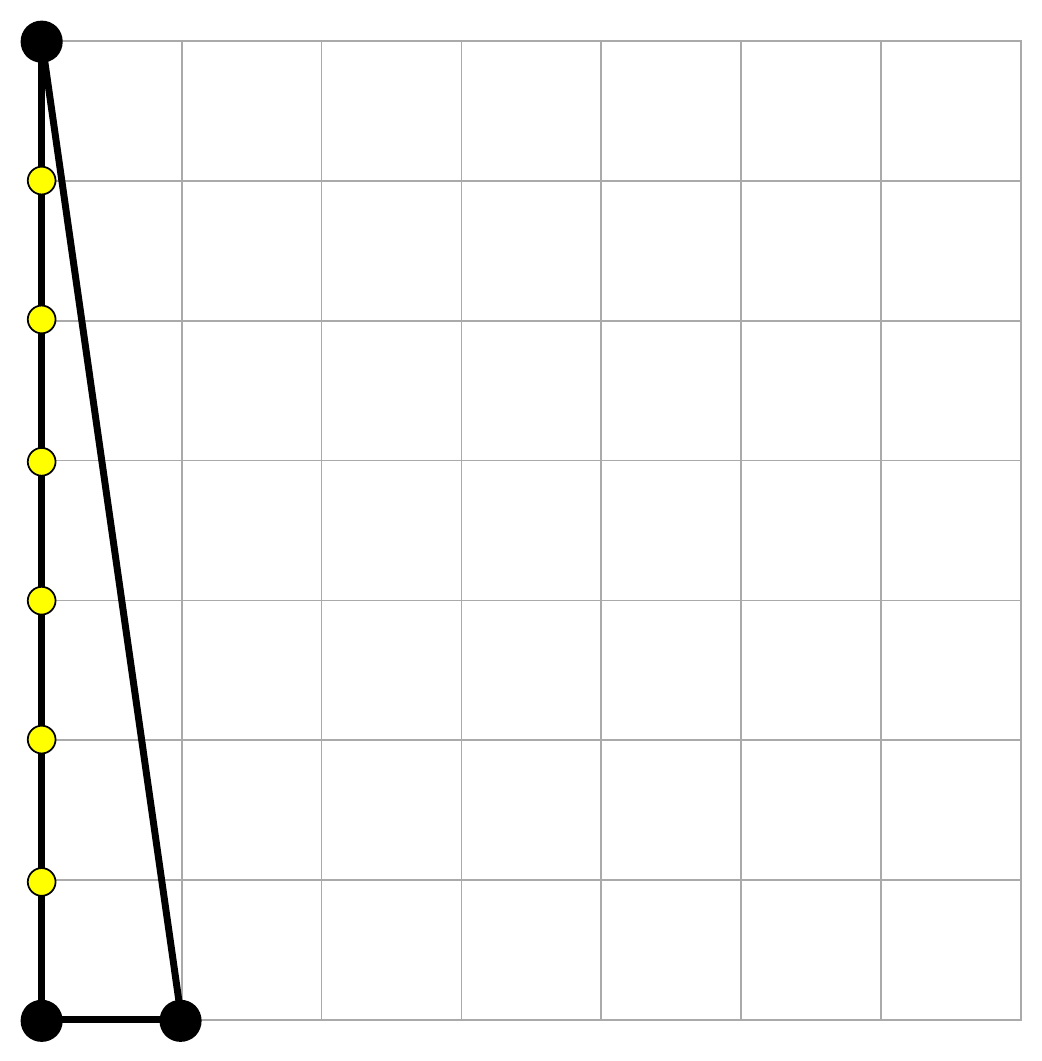} & 3
\\ 
\hline
(7.2) & 7 & $\mathbb{C}^{3}/\mathbb{Z}_{7}$ & $\bvec (1,1,5)\\(0,0,0) \evec$ & 
\includegraphics*[height=4.8cm]{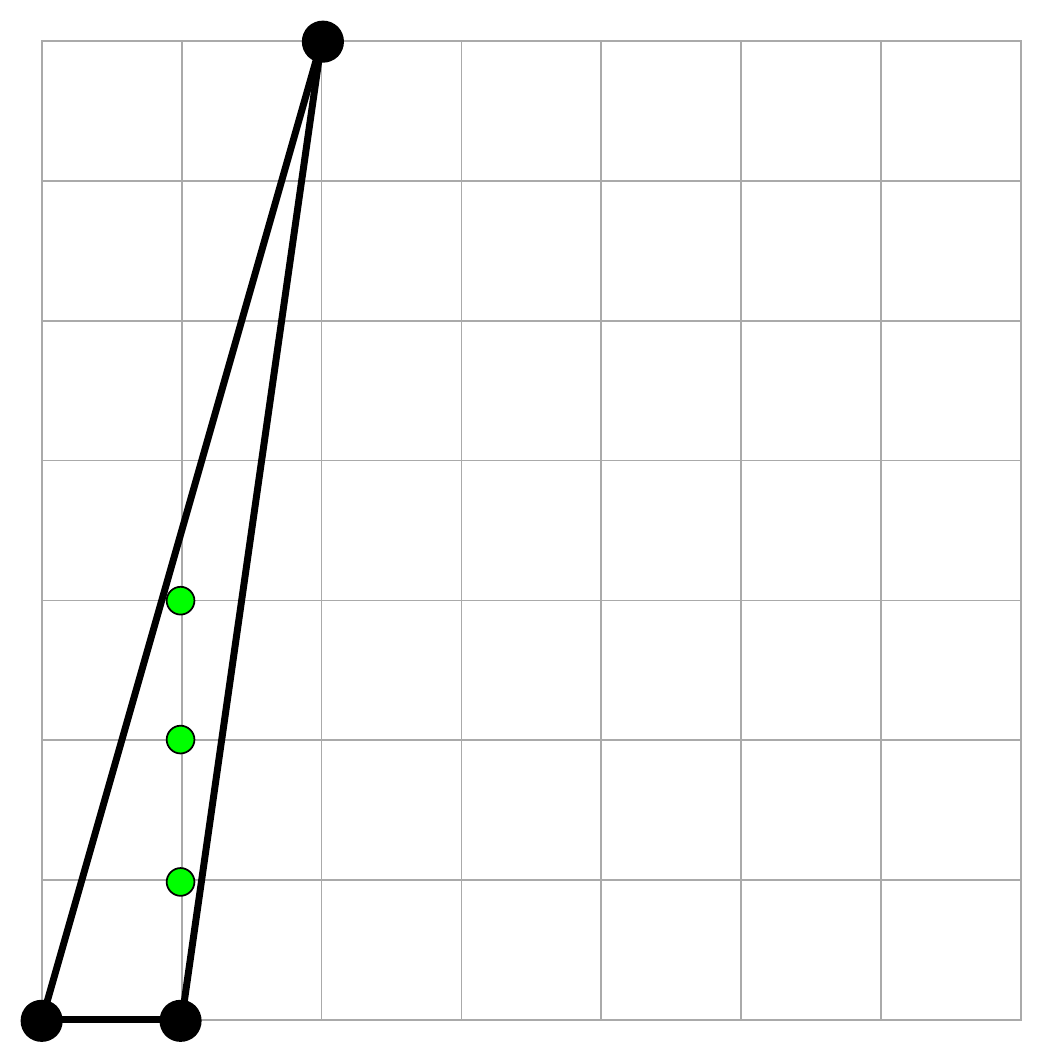} & 3
\\ 
\hline
(7.3) & 7 & $\mathbb{C}^{3}/\mathbb{Z}_{7}$ & $\bvec (1,2,4)\\(0,0,0) \evec$ & 
\includegraphics*[height=4.8cm]{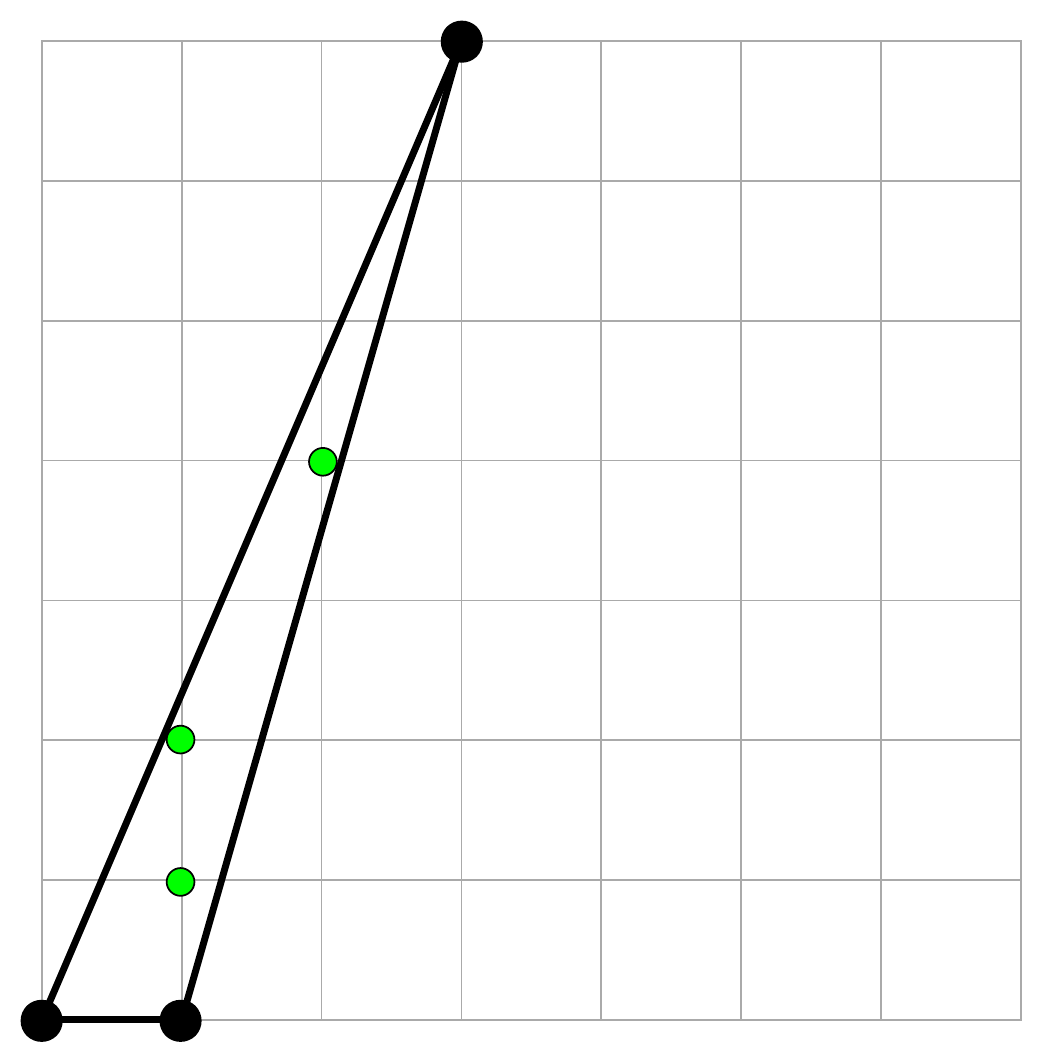} & 2
\\ 
\hline \hline

%%%%%%%%%%%%%%%%%%% N=8 %%%%%%%%%%%%%%%%%%%
(8.1) & 8 & $\mathbb{C}^{3}/\mathbb{Z}_{8}$ & $\bvec (0,1,7)\\(0,0,0) \evec$ &
\includegraphics*[height=4.8cm]{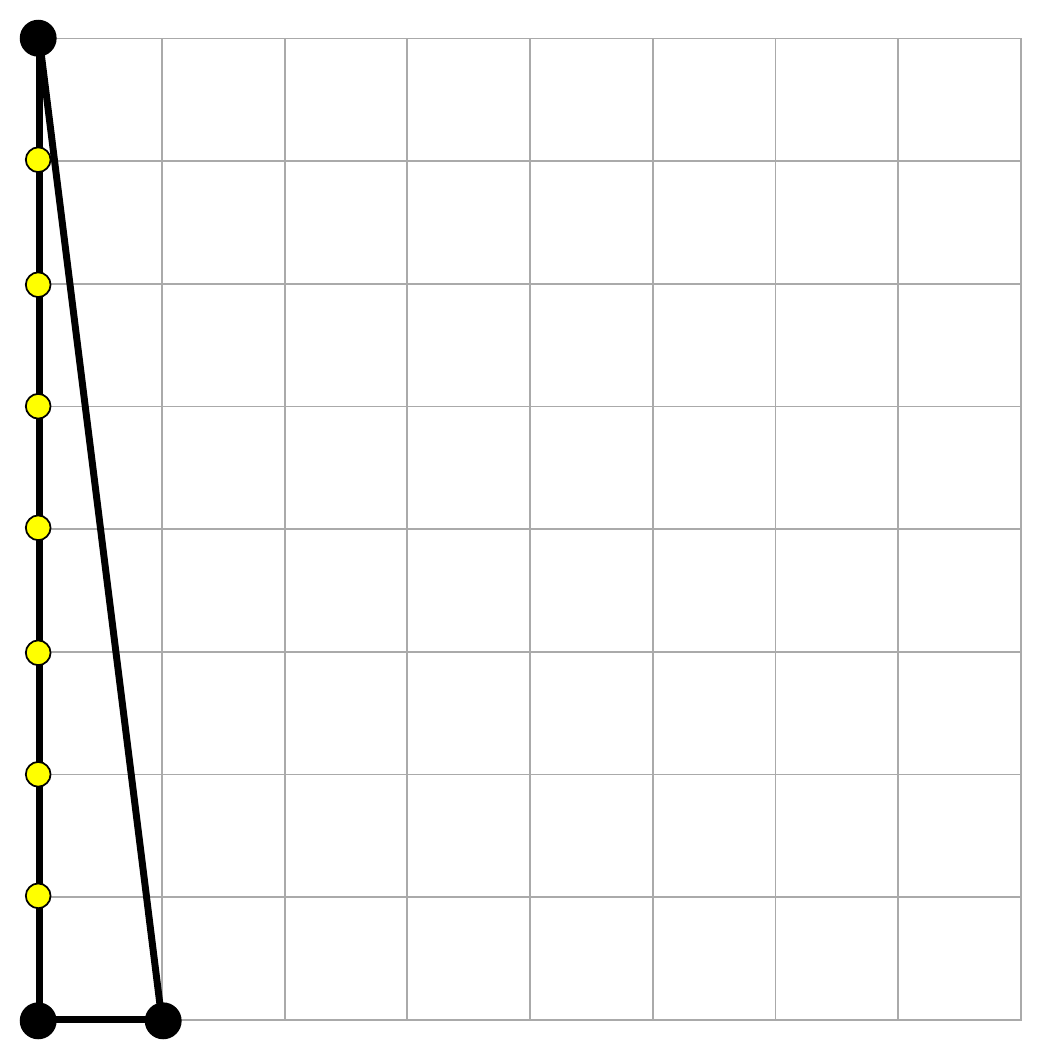} & 3
\\
\hline \hline

\end{tabular}
\end{center}

\caption{Orbifold Actions and corresponding Toric Diagrams for $\mathbb{C}^{3}/\Gamma_N$ orbifolds with order $N=1\dots 10$ \textbf{(Part 4/7)}.}
\label{t3}
\end{table}

\clearpage

%%%%%%%%%%%%%%%%%%%%%%%%%%%%%%%%%%%%%%%%%%%%%%%%
%%%%%%%%%%%%%%%%%%%%%%%%%%%%%%%%%%%%%%%%%%%%%%%%

\begin{table}[ht]

\begin{center}
\begin{tabular}{m{1cm}|m{0.8cm}|m{2.2cm}|m{3cm}|m{4.8cm}|m{2cm}}
\hline \hline
\# & $N$ & Orbifold & Orbifold Action & Toric Diagram & Multiplicity\\
\hline \hline
%%%%%%%%%%%%%%%%%%% N=8 %%%%%%%%%%%%%%%%%%%
(8.2) & 8 & $\mathbb{C}^{3}/\mathbb{Z}_{8}$ & $\bvec (1,1,6)\\(0,0,0) \evec$ &
\includegraphics*[height=4.8cm]{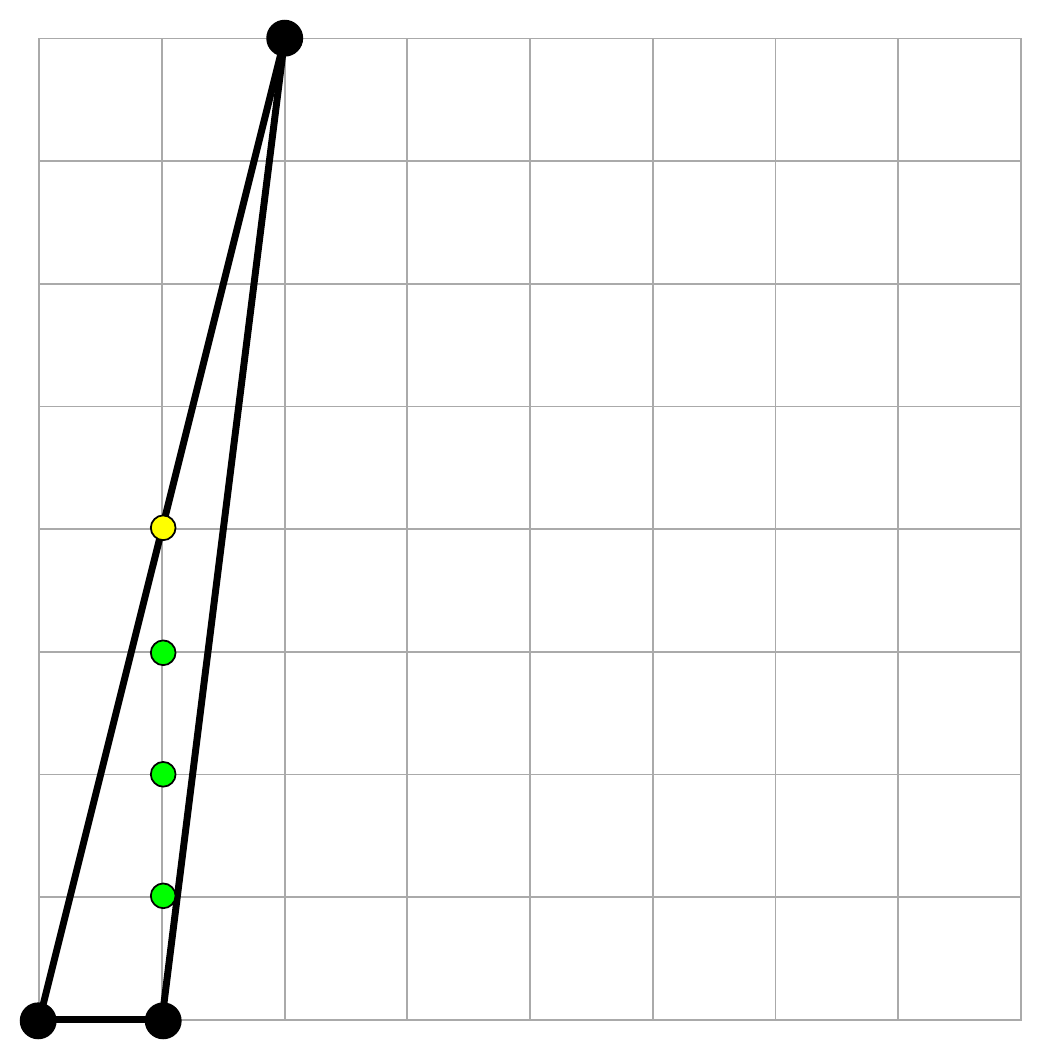} & 3
\\
\hline 
(8.3) & 8 & $\mathbb{C}^{3}/\mathbb{Z}_{8}$ & $\bvec (1,2,5)\\(0,0,0) \evec$ &
\includegraphics*[height=4.8cm]{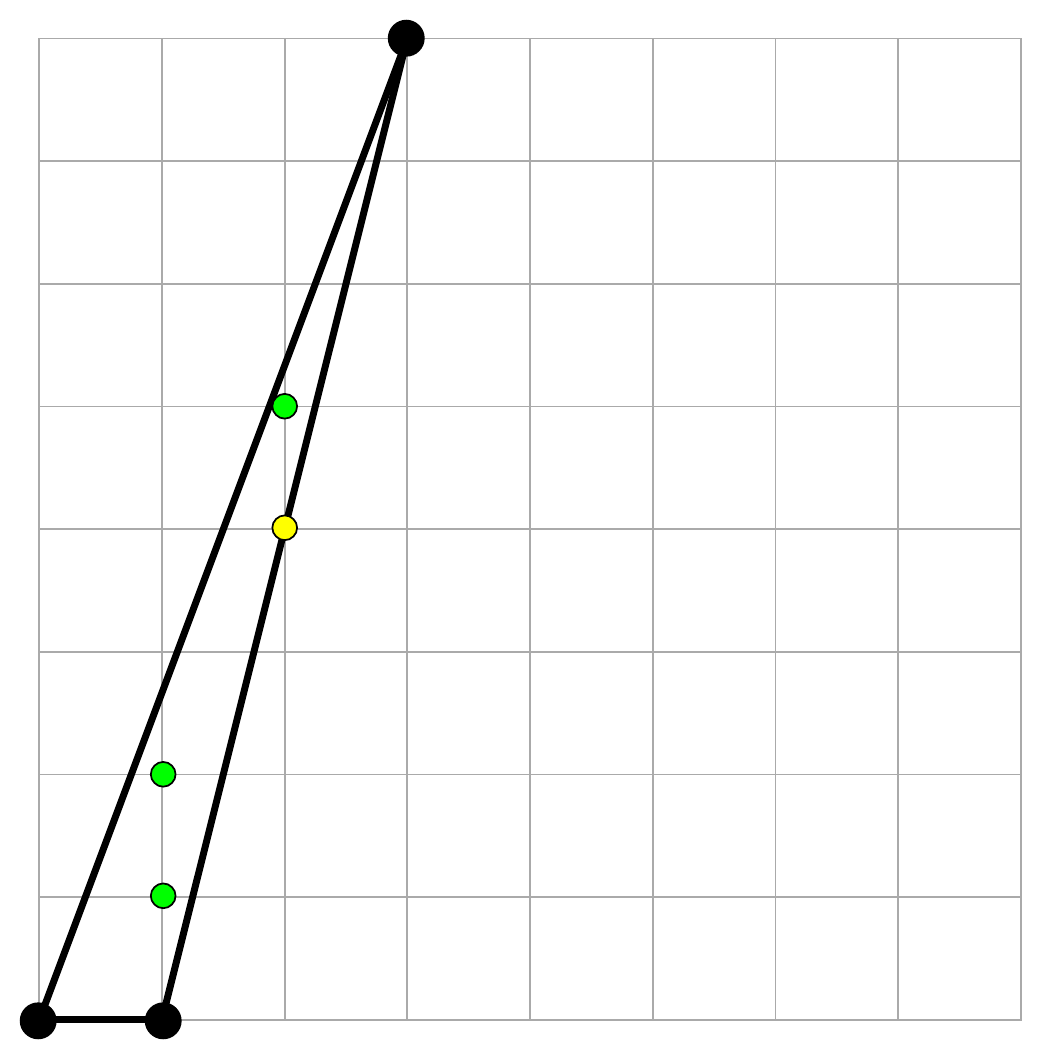} & 3
\\
\hline
(8.4) & 8 & $\mathbb{C}^{3}/\mathbb{Z}_{8}$ & $\bvec (1,3,4)\\(0,0,0) \evec$ &
\includegraphics*[height=4.8cm]{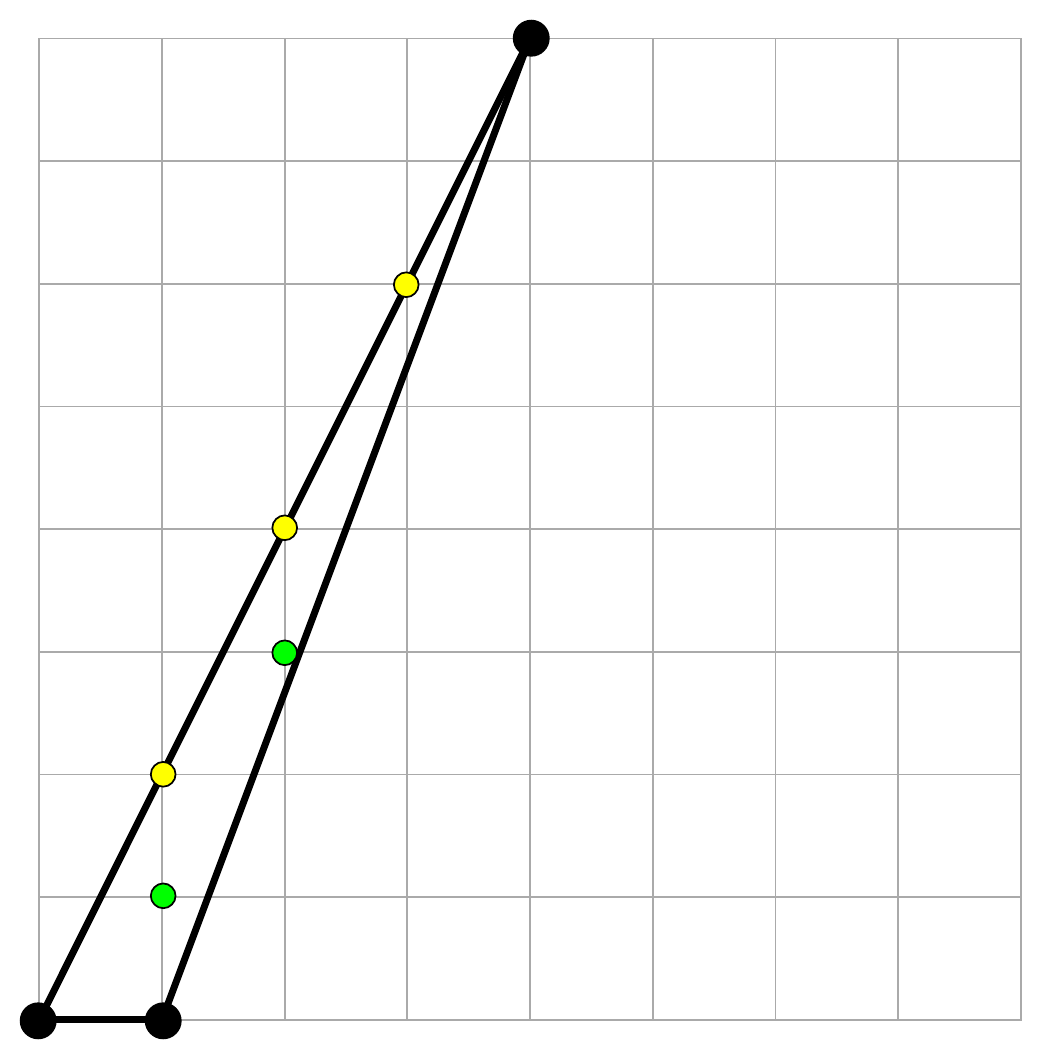} & 3
\\
\hline
(8.5) & 8 & $\mathbb{C}^{3}/\mathbb{Z}_{4}\times\mathbb{Z}_{2}$ & $\bvec (1,0,3)\\(0,1,1) \evec$ &
\includegraphics*[height=4.8cm]{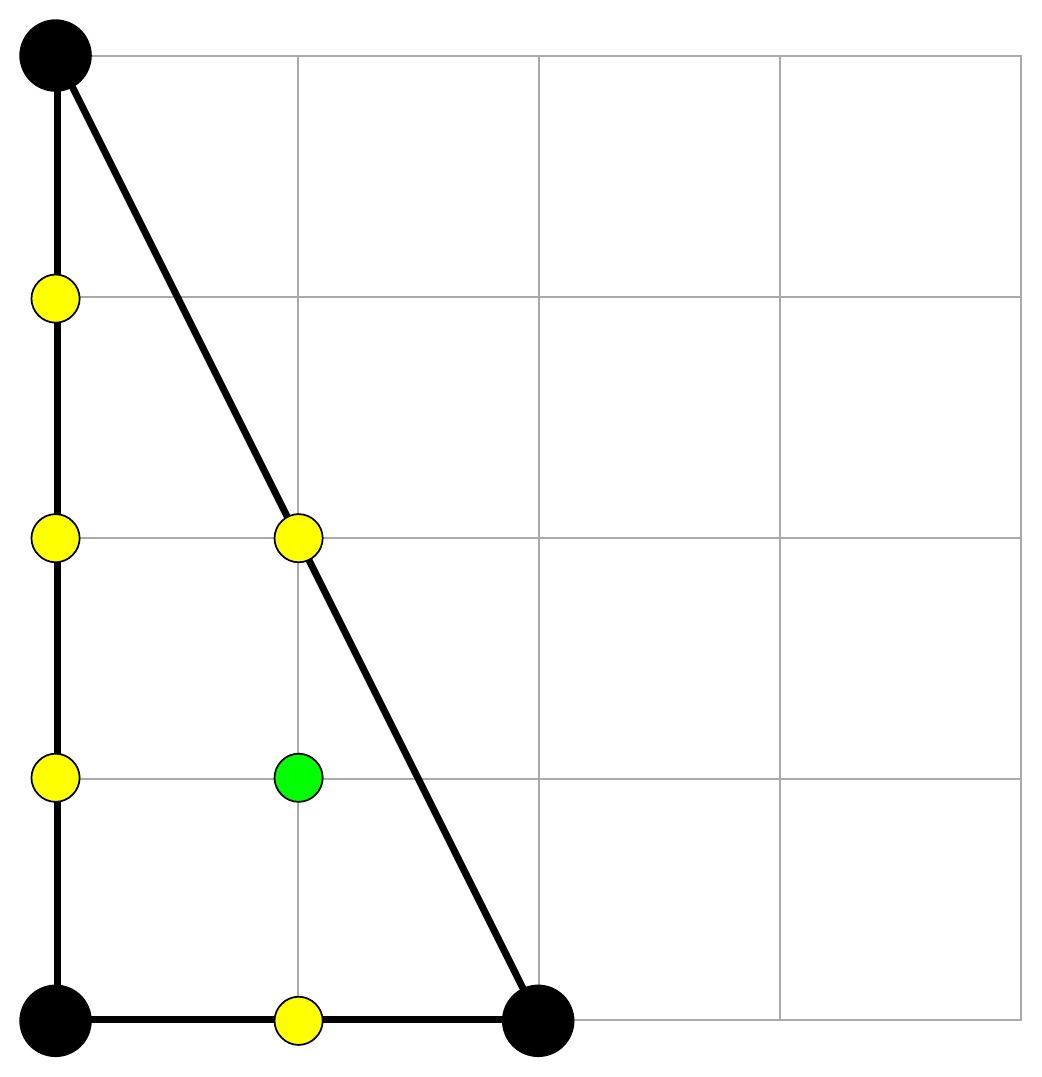} & 3
\\
\hline \hline

\end{tabular}
\end{center}

\caption{Orbifold Actions and corresponding Toric Diagrams for $\mathbb{C}^{3}/\Gamma_N$ orbifolds with order $N=1\dots 10$ \textbf{(Part 5/7)}.}
\label{t3b}
\end{table}

\clearpage

%%%%%%%%%%%%%%%%%%%%%%%%%%%%%%%%%%%%%%%%%%%%%%%%
%%%%%%%%%%%%%%%%%%%%%%%%%%%%%%%%%%%%%%%%%%%%%%%%

\begin{table}[ht]

\begin{center}
\begin{tabular}{m{1cm}|m{0.8cm}|m{2.2cm}|m{3cm}|m{4.8cm}|m{2cm}}
\hline \hline
\# & $N$ & Orbifold & Orbifold Action & Toric Diagram & Multiplicity\\
\hline \hline

%%%%%%%%%%%%%%%%%%% N=9 %%%%%%%%%%%%%%%%%%%
(9.1) & 9 & $\mathbb{C}^{3}/\mathbb{Z}_{9}$ & $\bvec (0,1,8)\\(0,0,0) \evec$ & 
\includegraphics[height=4.8cm]{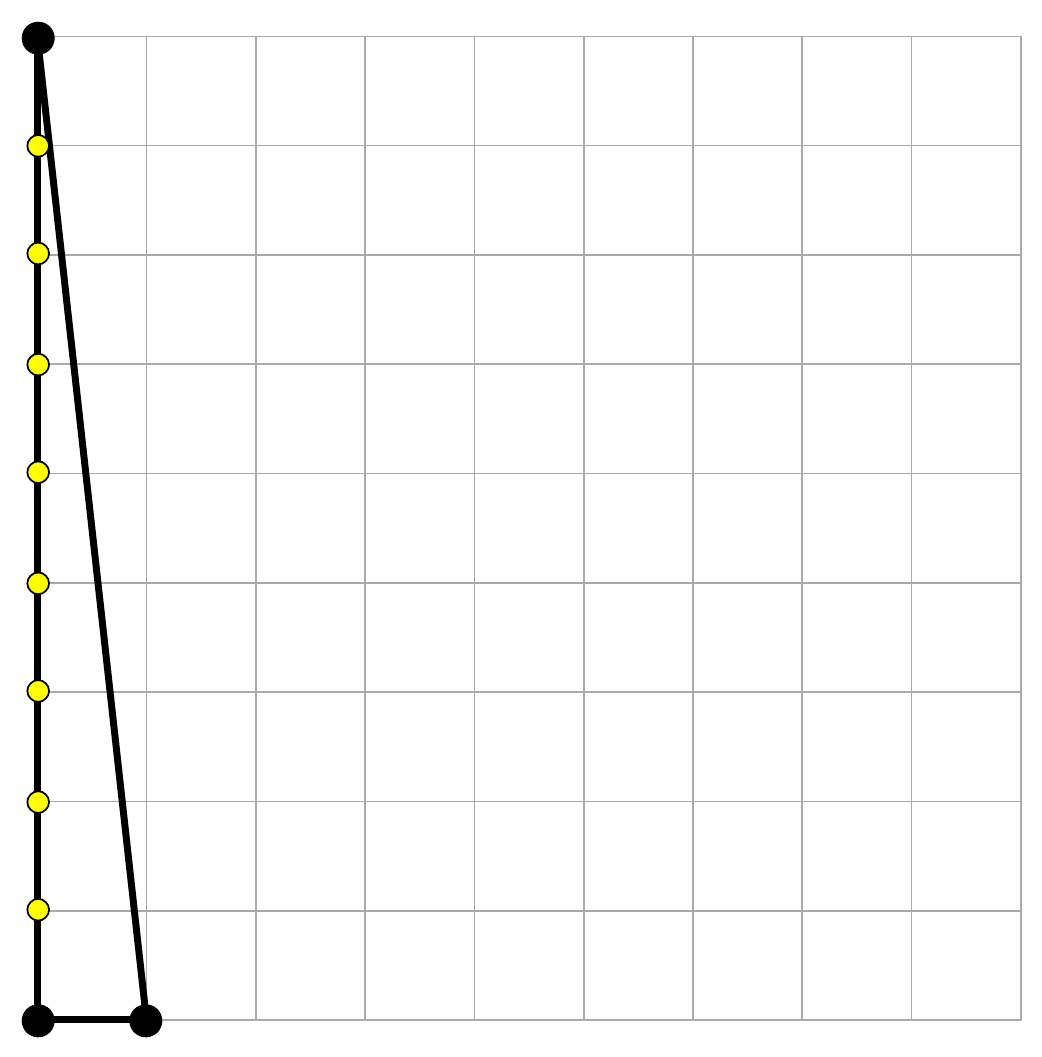}  & 3
\\ \hline
(9.2) & 9 & $\mathbb{C}^{3}/\mathbb{Z}_{9}$ & $\bvec (1,1,7)\\(0,0,0) \evec$ & 
\includegraphics[height=4.8cm]{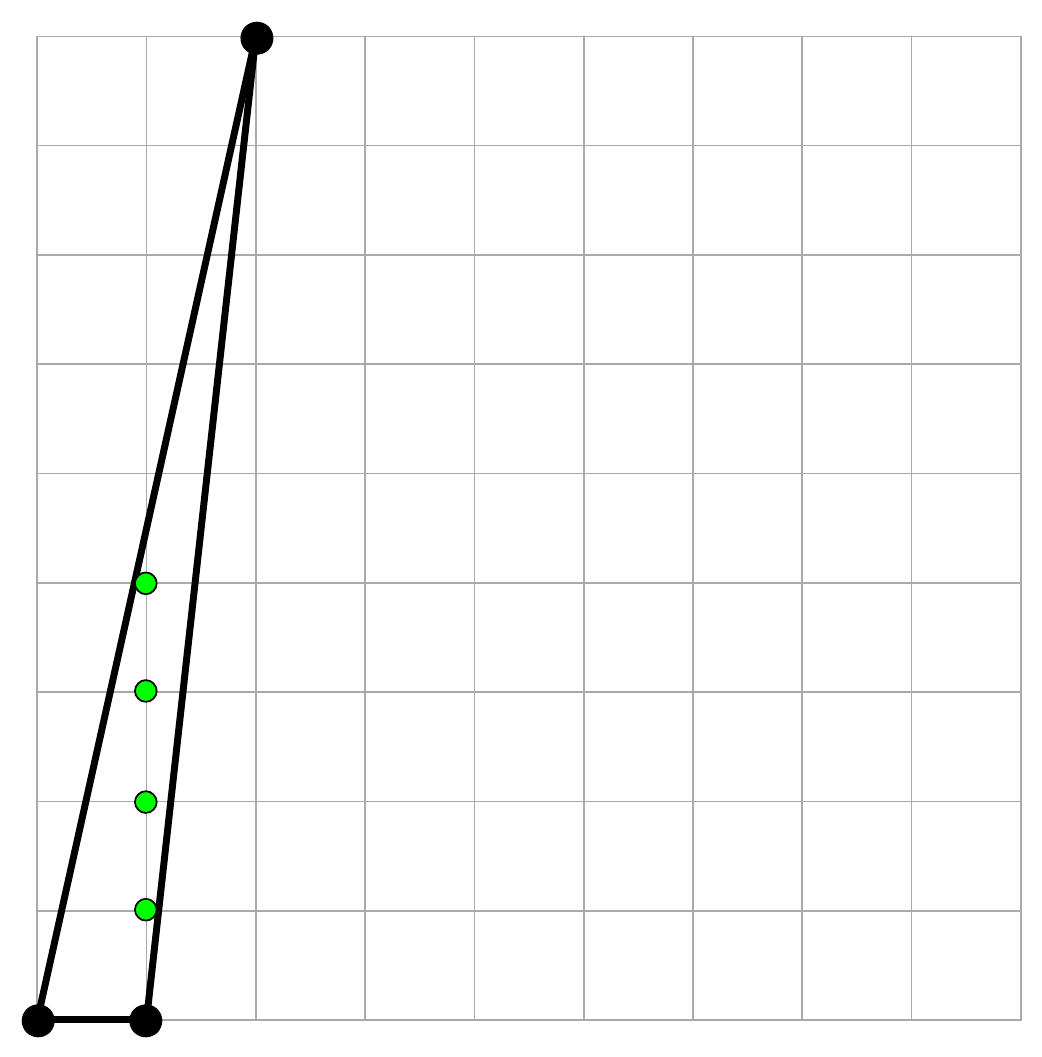} & 3
\\ \hline
(9.3) & 9 & $\mathbb{C}^{3}/\mathbb{Z}_{9}$ & $\bvec (1,2,6)\\(0,0,0) \evec$ & 
\includegraphics[height=4.8cm]{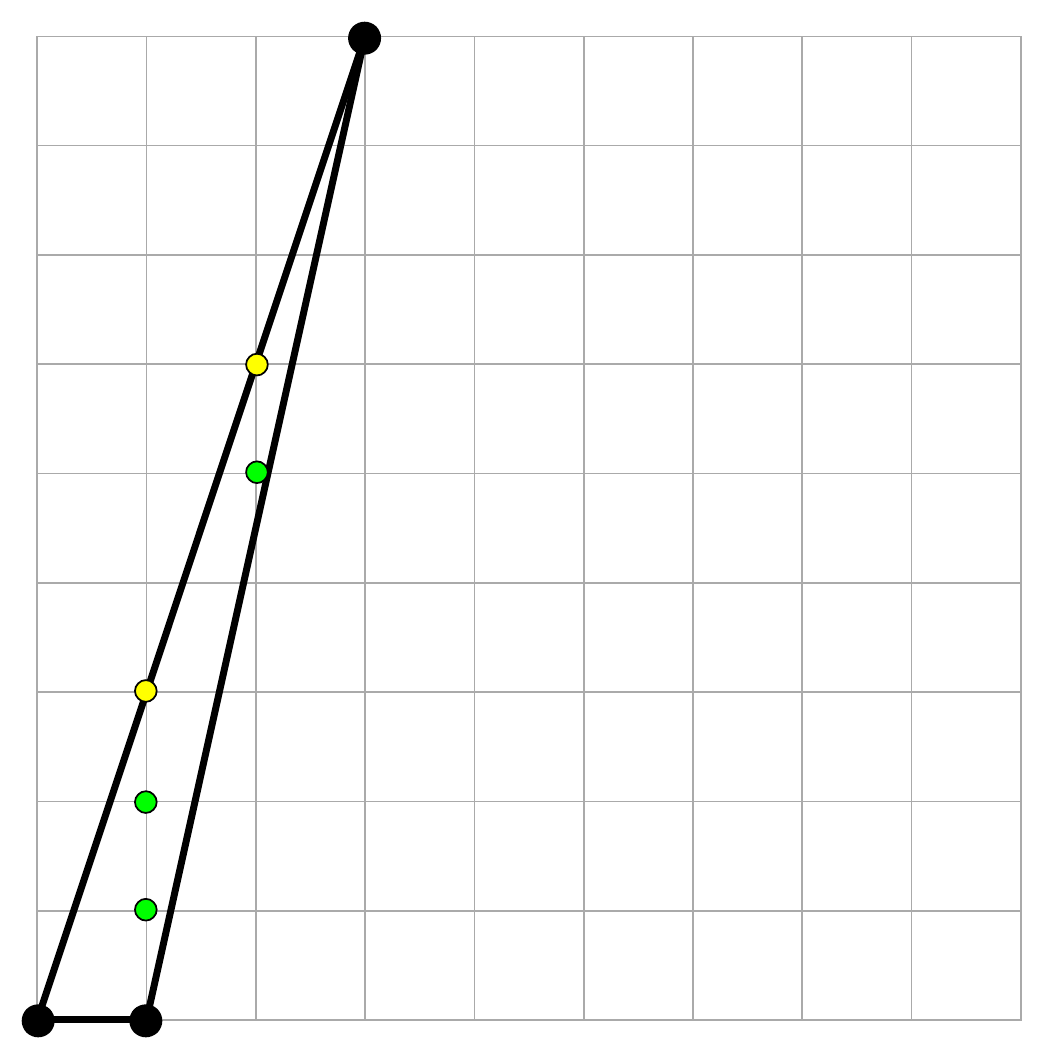} & 6
\\ \hline
(9.4) & 9 & $\mathbb{C}^{3}/\mathbb{Z}_{3}\times\mathbb{Z}_{3}$ & $\bvec (0,1,2)\\(1,0,2) \evec$ & 
\includegraphics[height=4.8cm]{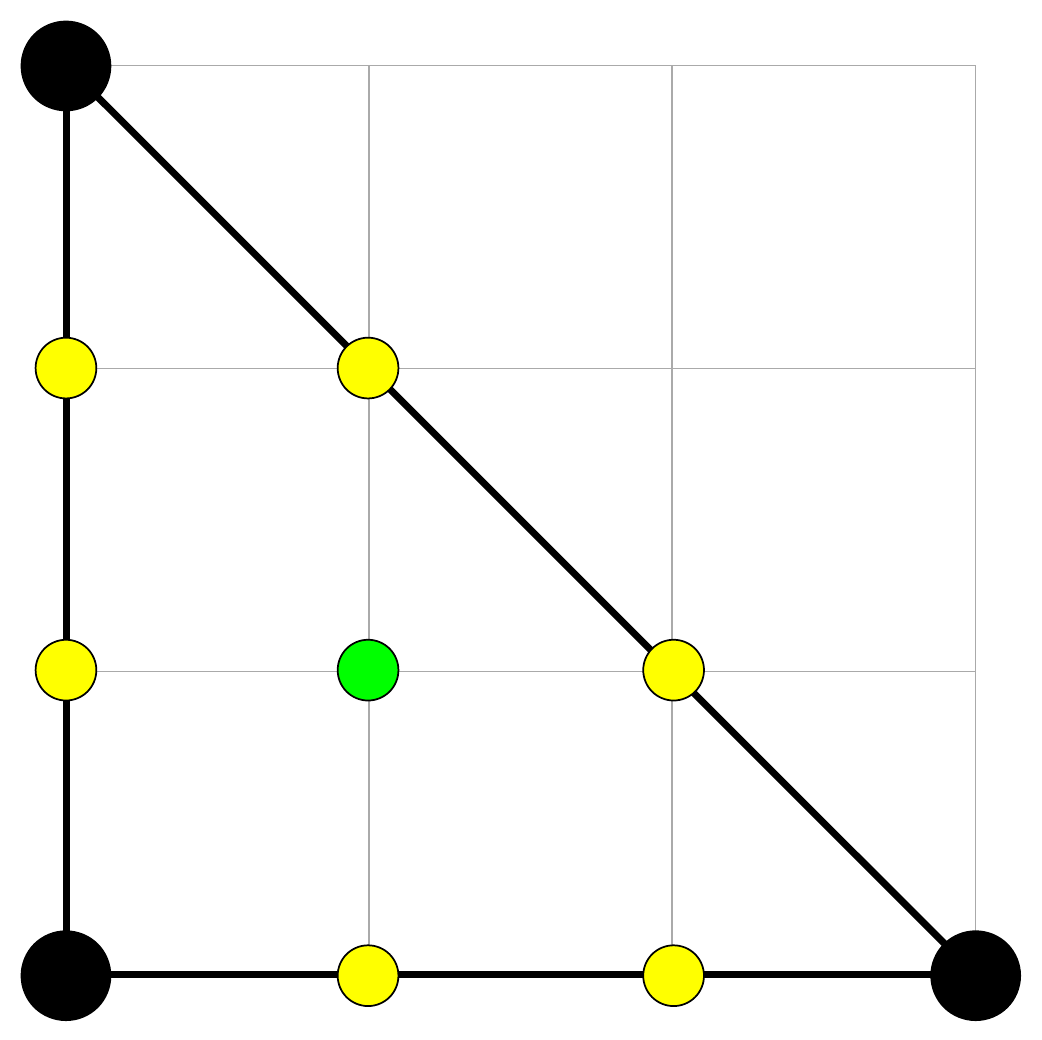} & 1
\\ \hline \hline

\end{tabular}
\end{center}

\caption{Orbifold Actions and corresponding Toric Diagrams for $\mathbb{C}^{3}/\Gamma_N$ orbifolds with order $N=1\dots 10$ \textbf{(Part 6/7)}.}
\label{t4}
\end{table}

\clearpage

%%%%%%%%%%%%%%%%%%%%%%%%%%%%%%%%%%%%%%%%%%%%%%%%
%%%%%%%%%%%%%%%%%%%%%%%%%%%%%%%%%%%%%%%%%%%%%%%%

\begin{table}[ht]

\begin{center}
\begin{tabular}{m{1cm}|m{0.8cm}|m{2.2cm}|m{3cm}|m{4.8cm}|m{2cm}}
\hline \hline
\# & $N$ & Orbifold & Orbifold Action & Toric Diagram & Multiplicity\\
\hline \hline

%%%%%%%%%%%%%%%%%%% N=10 %%%%%%%%%%%%%%%%%%%
(10.1) & 10 & $\mathbb{C}^{3}/\mathbb{Z}_{10}$ & $\bvec (0,1,9)\\(0,0,0) \evec$ &
\includegraphics*[height=4.8cm]{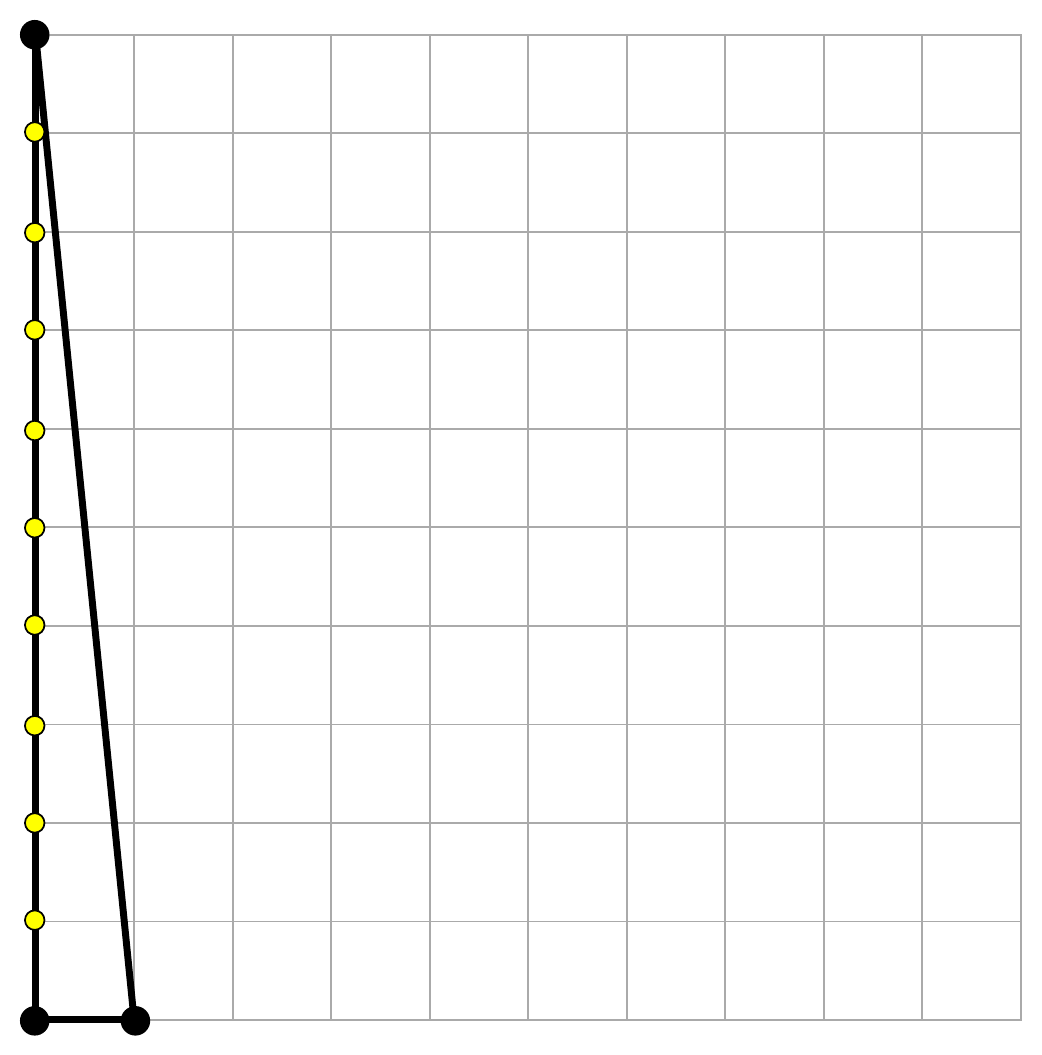} & 3
\\ \hline
(10.2) & 10 & $\mathbb{C}^{3}/\mathbb{Z}_{10}$ & $\bvec (1,1,8)\\(0,0,0) \evec$ &
\includegraphics[height=4.8cm]{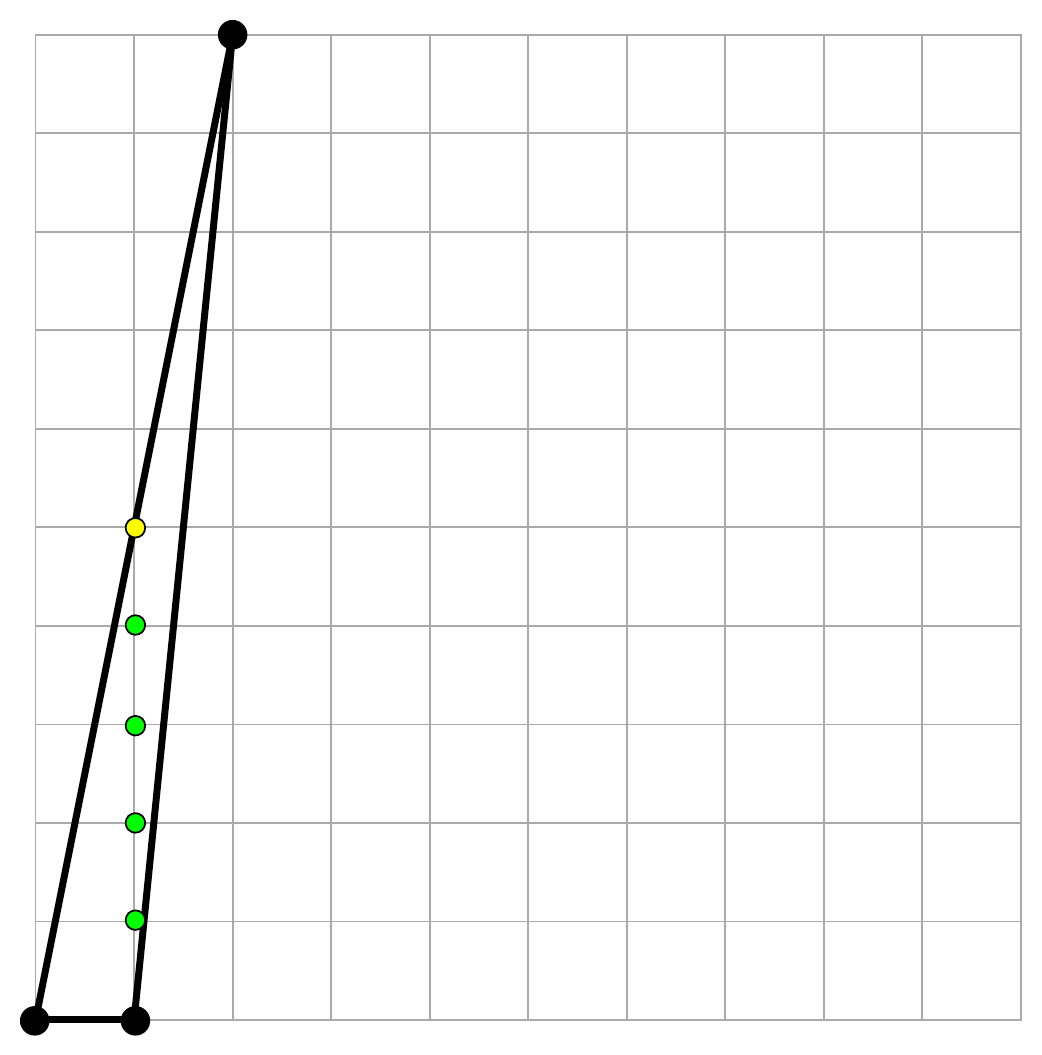} & 3
\\ \hline
(10.3) & 10 & $\mathbb{C}^{3}/\mathbb{Z}_{10}$ & $\bvec (1,2,7)\\(0,0,0) \evec$ &
\includegraphics[height=4.8cm]{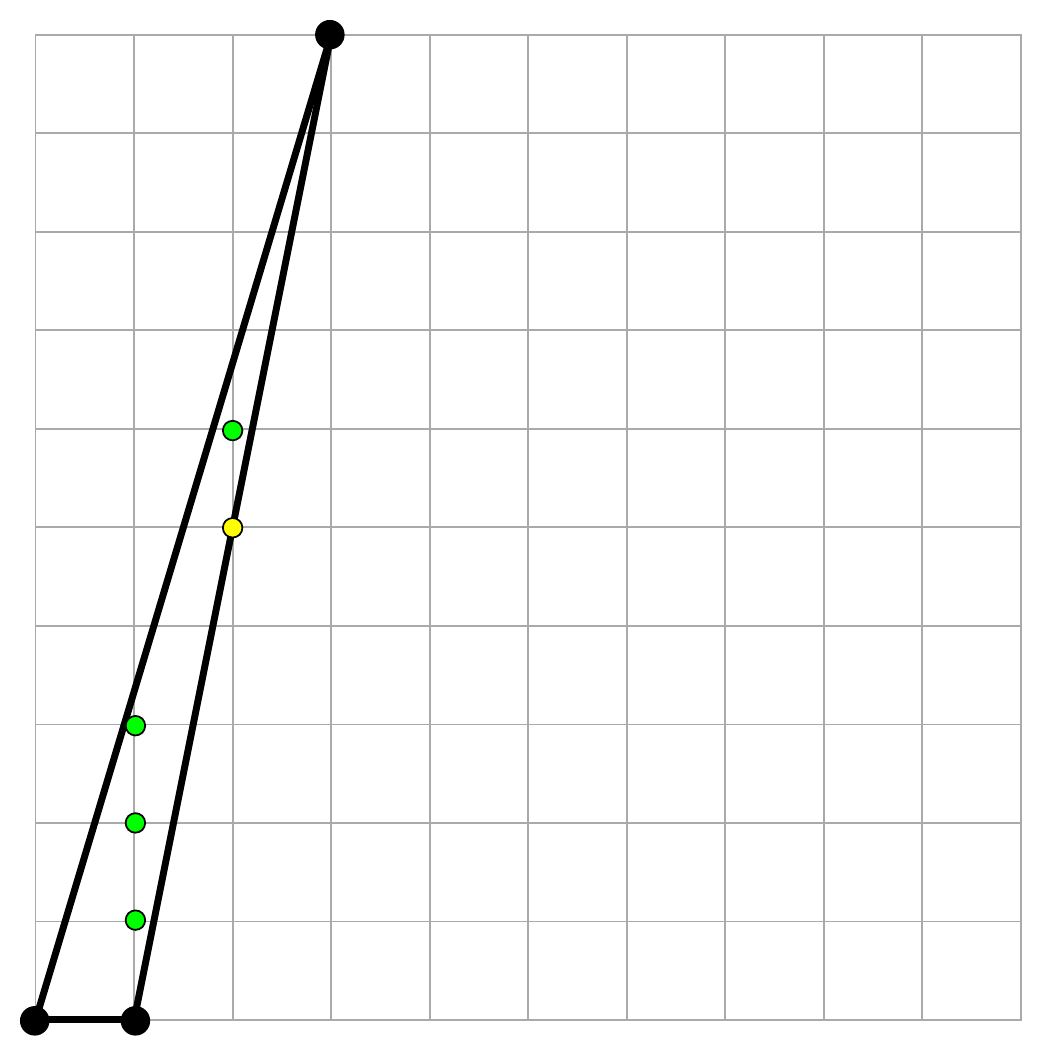} & 6
\\ \hline
(10.4) & 10 & $\mathbb{C}^{3}/\mathbb{Z}_{10}$ & $\bvec (1,4,5)\\(0,0,0) \evec$ &
\includegraphics[height=4.8cm]{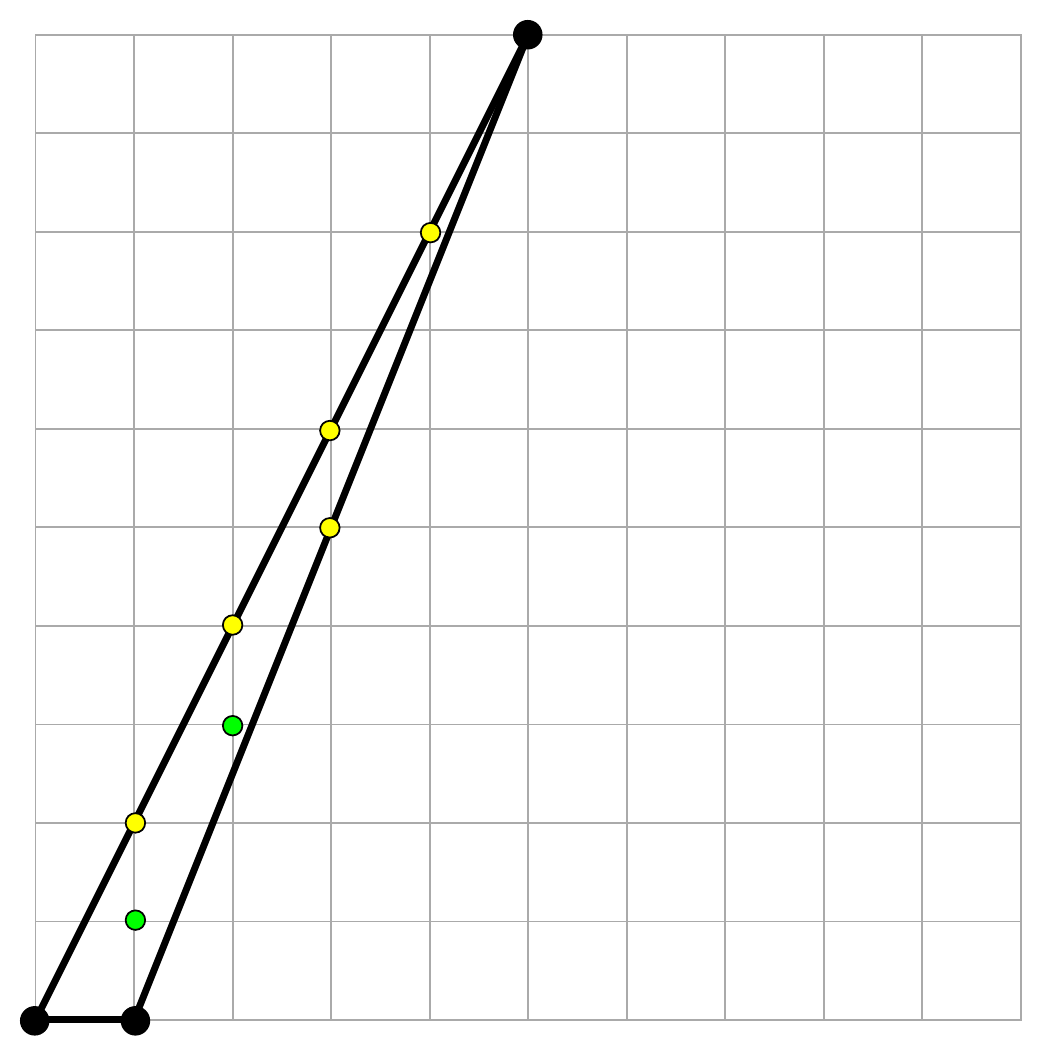} & 6
\\ \hline \hline

\end{tabular}
\end{center}

\caption{Orbifold Actions and corresponding Toric Diagrams for $\mathbb{C}^{3}/\Gamma_N$ orbifolds with order $N=1\dots 10$ \textbf{(Part 7/7)}.}
\label{t4b}
\end{table}

\clearpage

%%%%%%%%%%%%%%%%%%%%%%%%%%%%%%%%%%%%%%%%%%%%%%%%%%%%%%%%%%%%%%%%%%%%%%%%%%%%%%%
%%%%%%%%%%%%%%%%%%%%%%%%%%%%%%%%%%%%%%%%%%%%%%%%%%%%%%%%%%%%%%%%%%%%%%%%%%%%%%%
\section{$\mathbb{C}^{4}$ Orbifold Index \label{sa2}}

In the toric diagram tetrahedra, internal lattice points ($I_3$) are colored red, lattice points on the faces are colored green ($I_2$) and lattice points on edges are colored yellow ($I_1$) (Tables 15-24).

%%%%%%%%%%%%%%%%%%%%%%%%%%%%%%%%%%%%%%%%%%%%%%%
%%%%%%%%%%%%%%%%%%%%%%%%%%%%%%%%%%%%%%%%%%%%%%%
\begin{table}[ht]

\begin{center}
\begin{tabular}{m{1cm}|m{0.8cm}|m{2.2cm}|m{3cm}|m{6cm}|m{2cm}}
\hline \hline
\# & $N$ & Orbifold & Orbifold Action & Toric Diagram & Multiplicity\\
\hline \hline

%%%%%%%%%%%%%%%%%%% N=1 %%%%%%%%%%%%%%%%%%%
(1.1) & 1 & $\mathbb{C}^{4}/\mathbb{Z}_{1}$ & $\bvec (0,0,0,0)\\(0,0,0,0)\\(0,0,0,0) \evec$ & 
\includegraphics*[height=4.8cm]{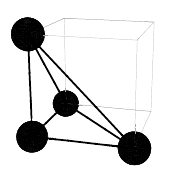} & 1
\\ 
\hline \hline

%%%%%%%%%%%%%%%%%%% N=2 %%%%%%%%%%%%%%%%%%%
(2.1) & 2 & $\mathbb{C}^{4}/\mathbb{Z}_{2}$ & $\bvec (0,0,1,1)\\(0,0,0,0)\\(0,0,0,0) \evec$ &
\includegraphics*[height=4.8cm]{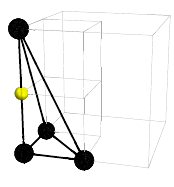} & 6
\\
\hline
(2.2) & 2 & $\mathbb{C}^{4}/\mathbb{Z}_{2}$ & $\bvec (1,1,1,1)\\(0,0,0,0)\\(0,0,0,0) \evec$ &
\includegraphics*[height=4.8cm]{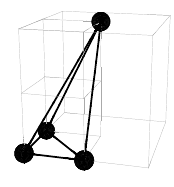} & 1
\\
\hline \hline
\end{tabular}
\end{center}

\caption{Orbifold Actions and corresponding Toric Diagrams for $\mathbb{C}^{4}/\Gamma_N$ orbifolds with order $N=1\dots 6$ \textbf{(Part 1/10)}.}
\label{t5}
\end{table}

%%%%%%%%%%%%%%%%%%%%%%%%%%%%%%%%%%%%%%%%%%%%%%%
%%%%%%%%%%%%%%%%%%%%%%%%%%%%%%%%%%%%%%%%%%%%%%%
\begin{table}[ht]

\begin{center}
\begin{tabular}{m{1cm}|m{0.8cm}|m{2.2cm}|m{3cm}|m{6cm}|m{2cm}}
\hline \hline
\# & $N$ & Orbifold & Orbifold Action & Toric Diagram & Multiplicity\\
\hline \hline

%%%%%%%%%%%%%%%%%%% N=3 %%%%%%%%%%%%%%%%%%%
(3.1) & 3 & $\mathbb{C}^{4}/\mathbb{Z}_{3}$ & $\bvec (0,0,1,2)\\(0,0,0,0)\\(0,0,0,0) \evec$ &
\includegraphics*[height=6cm]{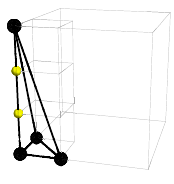} & 6
\\
\hline
(3.2) & 3 & $\mathbb{C}^{4}/\mathbb{Z}_{3}$ & $\bvec (0,1,1,1)\\(0,0,0,0)\\(0,0,0,0) \evec$ &
\includegraphics*[height=6cm]{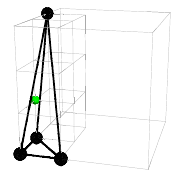} & 4
\\
\hline
(3.3) & 3 & $\mathbb{C}^{4}/\mathbb{Z}_{3}$ & $\bvec (1,1,2,2)\\(0,0,0,0)\\(0,0,0,0) \evec$ &
\includegraphics*[height=6cm]{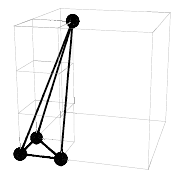} & 3
\\
\hline \hline
\end{tabular}
\end{center}

\caption{Orbifold Actions and corresponding Toric Diagrams for $\mathbb{C}^{4}/\Gamma_N$ orbifolds with order $N=1\dots 6$ \textbf{(Part 2/10)}.}
\label{t5b}
\end{table}

\clearpage

%%%%%%%%%%%%%%%%%%%%%%%%%%%%%%%%%%%%%%%%%%%
%%%%%%%%%%%%%%%%%%% N=4..4 ================
\begin{table}[ht]

\begin{center}
\begin{tabular}{m{1cm}|m{0.8cm}|m{2.2cm}|m{3cm}|m{6cm}|m{2cm}}
\hline \hline
\# & $N$ & Orbifold & Orbifold Action & Toric Diagram & Multiplicity\\
\hline \hline

%%%%%%%%%%%%%%%%%%% N=4 %%%%%%%%%%%%%%%%%%%
(4.1) & 4 & $\mathbb{C}^{4}/\mathbb{Z}_{4}$ & $\bvec (0,0,1,3)\\(0,0,0,0)\\(0,0,0,0) \evec$ & 
\includegraphics*[height=6cm]{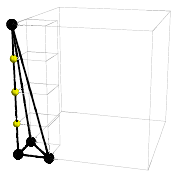} & 6
\\ 
\hline 
(4.2) & 4 & $\mathbb{C}^{4}/\mathbb{Z}_{4}$ & $\bvec (0,1,1,2)\\(0,0,0,0)\\(0,0,0,0) \evec$ &
\includegraphics*[height=6cm]{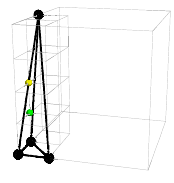} & 12
\\
\hline
(4.3) & 4 & $\mathbb{C}^{4}/\mathbb{Z}_{4}$ & $\bvec (1,1,3,3)\\(0,0,0,0)\\(0,0,0,0) \evec$ &
\includegraphics*[height=6cm]{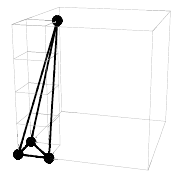} & 3
\\
\hline \hline
\end{tabular}
\end{center}

\caption{Orbifold Actions and corresponding Toric Diagrams for $\mathbb{C}^{4}/\Gamma_N$ orbifolds with order $N=1\dots 6$ \textbf{(Part 3/10)}.}
\label{t5c}
\end{table}

%%%%%%%%%%%%%%%%%%%%%%%%%%%%%%%%%%%%%%%%%%%%%%%
%%%%%%%%%%%%%%%%%%%%%%%%%%%%%%%%%%%%%%%%%%%%%%%

\begin{table}[ht]

\begin{center}
\begin{tabular}{m{1cm}|m{0.8cm}|m{2.2cm}|m{3cm}|m{6cm}|m{2cm}}
\hline \hline
\# & $N$ & Orbifold & Orbifold Action & Toric Diagram & Multiplicity\\
\hline \hline
(4.4) & 4 & $\mathbb{C}^{4}/\mathbb{Z}_{4}$ & $\bvec (1,2,2,3)\\(0,0,0,0)\\(0,0,0,0) \evec$ &
\includegraphics*[height=6cm]{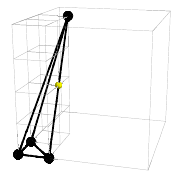} & 6
\\
\hline
(4.5) & 4 & $\mathbb{C}^{4}/\mathbb{Z}_{4}$ & $\bvec (1,1,1,1)\\(0,0,0,0)\\(0,0,0,0) \evec$ &
\includegraphics*[height=6cm]{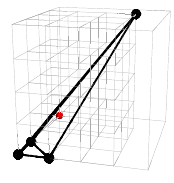} & 1
\\
\hline
(4.6) & 4 & $\mathbb{C}^{4}/\mathbb{Z}_{2}\times\mathbb{Z}_{2}$ & $\bvec (0,1,0,1)\\(0,0,1,1)\\(0,0,0,0) \evec$ &
\includegraphics*[height=6cm]{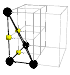} & 4
\\
\hline \hline 

\end{tabular}
\end{center}

\caption{Orbifold Actions and corresponding Toric Diagrams for $\mathbb{C}^{4}/\Gamma_N$ orbifolds with order $N=1\dots 6$ \textbf{(Part 4/10)}.}
\label{t5d}
\end{table}

\clearpage

%%%%%%%%%%%%%%%%%%%%%%%%%%%%%%%%%%%%%%%%%%%
%%%%%%%%%%%%%%%%%%% N=5..5 ================
\begin{table}[ht]

\begin{center}
\begin{tabular}{m{1cm}|m{0.8cm}|m{2.2cm}|m{3cm}|m{6cm}|m{2cm}}
\hline \hline
\# & $N$ & Orbifold & Orbifold Action & Toric Diagram & Multiplicity\\
\hline \hline
%%%%%%%%%%%%%%%%%%% N=4 %%%%%%%%%%%%%%%%%%%
(4.7) & 4 & $\mathbb{C}^{4}/\mathbb{Z}_{2}\times\mathbb{Z}_{2}$ & $\bvec  (0,0,1,1)\\(1,1,1,1)\\(0,0,0,0) \evec$ &
\includegraphics*[height=6cm]{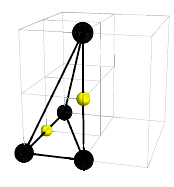} & 3
\\
\hline\hline
%%%%%%%%%%%%%%%%%%% N=5 %%%%%%%%%%%%%%%%%%%
(5.1) & 5 & $\mathbb{C}^{4}/\mathbb{Z}_{5}$ & $\bvec (0,0,1,4)\\(0,0,0,0)\\(0,0,0,0) \evec$ & 
\includegraphics*[height=6cm]{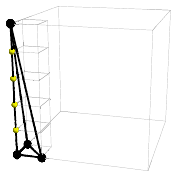} & 6
\\ 
\hline
(5.2) & 5 & $\mathbb{C}^{4}/\mathbb{Z}_{5}$ & $\bvec (0,1,1,3)\\(0,0,0,0)\\(0,0,0,0) \evec$ &
\includegraphics*[height=6cm]{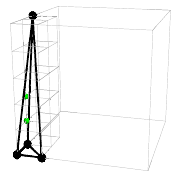} & 12
\\
\hline \hline

\end{tabular}
\end{center}

\caption{Orbifold Actions and corresponding Toric Diagrams for $\mathbb{C}^{4}/\Gamma_N$ orbifolds with order $N=1\dots 6$ \textbf{(Part 5/10)}.}
\label{t6}
\end{table}

%%%%%%%%%%%%%%%%%%%%%%%%%%%%%%%%%%%%%%%%%%%%%%%
%%%%%%%%%%%%%%%%%%%%%%%%%%%%%%%%%%%%%%%%%%%%%%%

\begin{table}[ht]

\begin{center}
\begin{tabular}{m{1cm}|m{0.8cm}|m{2.2cm}|m{3cm}|m{6cm}|m{2cm}}
\hline \hline
\# & $N$ & Orbifold & Orbifold Action & Toric Diagram & Multiplicity\\
\hline \hline
(5.3) & 5 & $\mathbb{C}^{4}/\mathbb{Z}_{5}$ & $\bvec (1,1,4,4)\\(0,0,0,0)\\(0,0,0,0) \evec$ &
\includegraphics*[height=6cm]{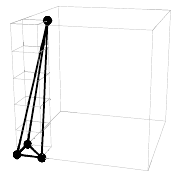} & 3
\\
\hline
(5.4) & 5 & $\mathbb{C}^{4}/\mathbb{Z}_{5}$ & $\bvec (1,2,3,4)\\(0,0,0,0)\\(0,0,0,0) \evec$ &
\includegraphics*[height=6cm]{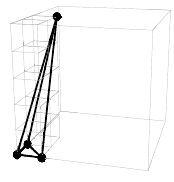} & 6
\\
\hline
(5.5) & 5 & $\mathbb{C}^{4}/\mathbb{Z}_{5}$ & $\bvec (1,1,1,2)\\(0,0,0,0)\\(0,0,0,0) \evec$ &
\includegraphics*[height=6cm]{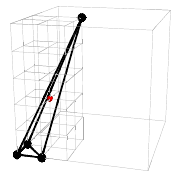} & 4
\\
\hline \hline

\end{tabular}
\end{center}

\caption{Orbifold Actions and corresponding Toric Diagrams for $\mathbb{C}^{4}/\Gamma_N$ orbifolds with order $N=1\dots 6$ \textbf{(Part 6/10)}.}
\label{t6b}
\end{table}

\clearpage

%%%%%%%%%%%%%%%%%%%%%%%%%%%%%%%%%%%%%%%%%%%
%%%%%%%%%%%%%%%%%%% N=6..6 PART I ================
\begin{table}[ht]

\begin{center}
\begin{tabular}{m{1cm}|m{0.8cm}|m{2.2cm}|m{3cm}|m{6cm}|m{2cm}}
\hline \hline
\# & $N$ & Orbifold & Orbifold Action & Toric Diagram & Multiplicity\\
\hline \hline

%%%%%%%%%%%%%%%%%%% N=6 %%%%%%%%%%%%%%%%%%%
(6.1) & 6 & $\mathbb{C}^{4}/\mathbb{Z}_{6}$ & $\bvec (0,0,1,5)\\(0,0,0,0)\\(0,0,0,0) \evec$ & 
\includegraphics*[height=6cm]{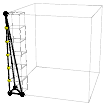} & 6
\\ 
\hline
(6.2) & 6 & $\mathbb{C}^{4}/\mathbb{Z}_{6}$ & $\bvec (0,1,1,4)\\(0,0,0,0)\\(0,0,0,0) \evec$ &
\includegraphics*[height=6cm]{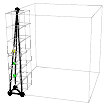} & 12
\\
\hline
(6.3) & 6 & $\mathbb{C}^{4}/\mathbb{Z}_{6}$ & $\bvec (0,1,2,3)\\(0,0,0,0)\\(0,0,0,0) \evec$ &
\includegraphics*[height=6cm]{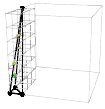} & 24
\\
\hline \hline

\end{tabular}
\end{center}

\caption{Orbifold Actions and corresponding Toric Diagrams for $\mathbb{C}^{4}/\Gamma_N$ orbifolds with order $N=1\dots 6$ \textbf{(Part 7/10)}.}
\label{t7}
\end{table}

%%%%%%%%%%%%%%%%%%%%%%%%%%%%%%%%%%%%%%%%%%%%%%%
%%%%%%%%%%%%%%%%%%%%%%%%%%%%%%%%%%%%%%%%%%%%%%%

\begin{table}[ht]

\begin{center}
\begin{tabular}{m{1cm}|m{0.8cm}|m{2.2cm}|m{3cm}|m{6cm}|m{2cm}}
\hline \hline
\# & $N$ & Orbifold & Orbifold Action & Toric Diagram & Multiplicity\\
\hline \hline
(6.4) & 6 & $\mathbb{C}^{4}/\mathbb{Z}_{6}$ & $\bvec (1,1,5,5)\\(0,0,0,0)\\(0,0,0,0) \evec$ &
\includegraphics*[height=6cm]{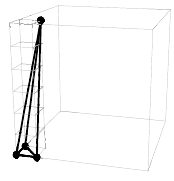} & 3
\\
\hline
(6.5) & 6 & $\mathbb{C}^{4}/\mathbb{Z}_{6}$ & $\bvec (1,1,2,2)\\(0,0,0,0)\\(0,0,0,0) \evec$ &
\includegraphics*[height=6cm]{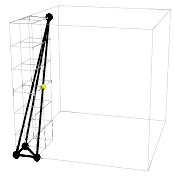} & 12
\\
\hline
(6.6) & 6 & $\mathbb{C}^{4}/\mathbb{Z}_{6}$ & $\bvec (1,3,3,5)\\(0,0,0,0)\\(0,0,0,0) \evec$ &
\includegraphics*[height=6cm]{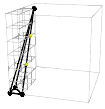} & 6
\\
\hline \hline

\end{tabular}
\end{center}

\caption{Orbifold Actions and corresponding Toric Diagrams for $\mathbb{C}^{4}/\Gamma_N$ orbifolds with order $N=1\dots 6$ \textbf{(Part 8/10)}.}
\label{t7b}
\end{table}

\clearpage

%%%%%%%%%%%%%%%%%%%%%%%%%%%%%%%%%%%%%%%%%%%
%%%%%%%%%%%%%%%%%%% N=6..6 PART II ================
\begin{table}[ht]

\begin{center}
\begin{tabular}{m{1cm}|m{0.8cm}|m{2.2cm}|m{3cm}|m{6cm}|m{2cm}}
\hline \hline
\# & $N$ & Orbifold & Orbifold Action & Toric Diagram & Multiplicity\\
\hline \hline

%%%%%%%%%%%%%%%%%%% N=6 %%%%%%%%%%%%%%%%%%%
(6.7) & 6 & $\mathbb{C}^{4}/\mathbb{Z}_{6}$ & $\bvec (1,3,4,4)\\(0,0,0,0)\\(0,0,0,0) \evec$ &
\includegraphics*[height=6cm]{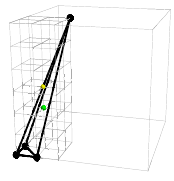} & 12
\\
\hline
(6.8) & 6 & $\mathbb{C}^{4}/\mathbb{Z}_{6}$ & $\bvec (1,1,1,3)\\(0,0,0,0)\\(0,0,0,0) \evec$ &
\includegraphics*[height=6cm]{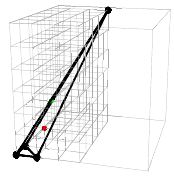} & 4
\\
\hline
(6.9) & 6 & $\mathbb{C}^{4}/\mathbb{Z}_{6}$ & $\bvec (1,2,4,5)\\(0,0,0,0)\\(0,0,0,0) \evec$ &
\includegraphics*[height=6cm]{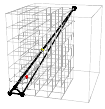} & 6
\\
\hline \hline

\end{tabular}
\end{center}

\caption{Orbifold Actions and corresponding Toric Diagrams for $\mathbb{C}^{4}/\Gamma_N$ orbifolds with order $N=1\dots 6$ \textbf{(Part 9/10)}.}
\label{t8}
\end{table}

\clearpage

%%%%%%%%%%%%%%%%%%% N=6..6 PART II ================
\begin{table}[ht]

\begin{center}
\begin{tabular}{m{1cm}|m{0.8cm}|m{2.2cm}|m{3cm}|m{6cm}|m{2cm}}
\hline \hline
\# & $N$ & Orbifold & Orbifold Action & Toric Diagram & Multiplicity\\
\hline \hline
(6.10) & 6 & $\mathbb{C}^{4}/\mathbb{Z}_{6}$ & $\bvec (2,3,3,4)\\(0,0,0,0)\\(0,0,0,0) \evec$ &\includegraphics*[height=6cm]{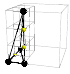} & 6
\\
\hline\hline
\end{tabular}
\end{center}

\caption{Orbifold Actions and corresponding Toric Diagrams for $\mathbb{C}^{4}/\Gamma_N$ orbifolds with order $N=1\dots 6$ \textbf{(Part 10/10)}.}
\label{t8b}
\end{table}

\section{Comments on the counting time}

\begin{figure}[ht!]
\begin{center}
\includegraphics[totalheight=5cm]{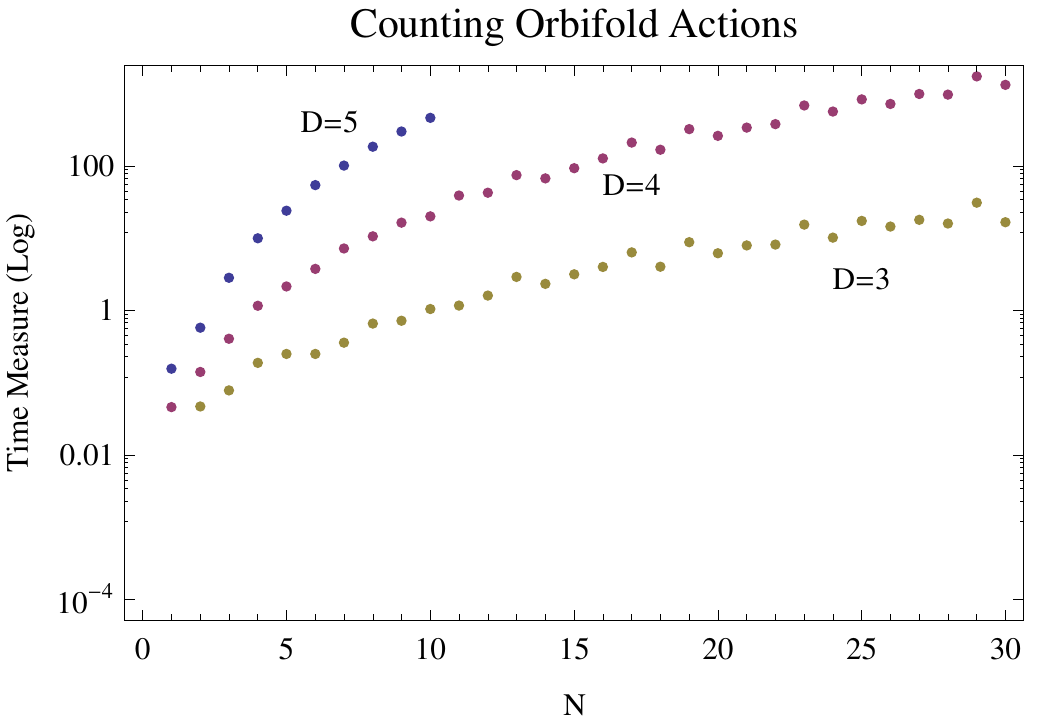}
\includegraphics[totalheight=5cm]{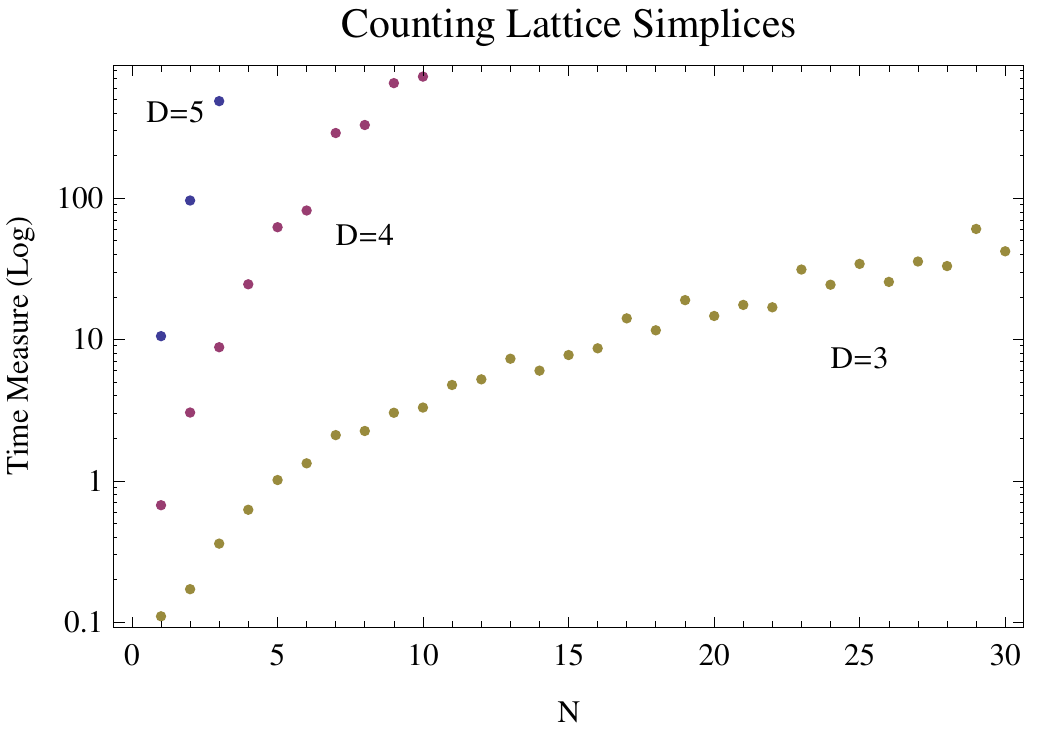}
\caption{Measurement of relative counting time at a given order $N$ of the orbifold $\mathbb{C}^{D}/\Gamma_{N}$. The two plots compare the two different counting methods presented in the current paper.}
  \label{ftime}
 \end{center}
\end{figure}

As a final comment, we compare the counting time required between the two methods of counting inequivalent orbifold actions of $\mathbb{C}^{D}/\Gamma_{N}$ as presented in this paper. The measurement of the counting time on a normal desktop PC is shown in \fref{ftime}. We note that the method of counting inequivalent representations of orbifold actions is computationally less expensive than the counting of inequivalent lattice simplices which requires the geometrical analysis of lattice simplices. For both methods we also note that the counting time increases exponentially with order $N$ and orbifold dimension $D$.\\

%%%%%%%%%%%%%%%%%%%%%%%%%%%%%%%%%%%%%%%%%%%%%%%%%%%%%%%%%%%%%%%%%%%%%%%%
%%%%%%%%%%%%%%%%%%%%%%%%%%%%%%%%%%%%%%%%%%%%%%%%%%%%%%%%%%%%%%%%%%%%%%%%

\end{document}